\documentclass[12pt]{article}

\usepackage[x11names]{xcolor}
\usepackage{tikz}
\usepackage{tikz-cd}
\tikzcdset{scale cd/.style={every label/.append style={scale=#1},    cells={nodes={scale=#1}}}}
\usepackage{cite}
\usepackage{comment}
\usepackage{hyperref}
\usetikzlibrary{decorations.pathreplacing,decorations.markings}
\usepgflibrary{arrows}
\usepgflibrary[arrows]
\usetikzlibrary{arrows}
\usetikzlibrary{positioning}
\usetikzlibrary{cd}
\usepackage{graphicx}
\usepackage{amsmath}
\usepackage{amssymb}
\usepackage{subsupscripts}
\usepackage{braket}
\usepackage{amsthm}
\usepackage{xcolor}
\usepackage{subcaption}
\usepackage{bbm}

\makeatletter

\@addtoreset{equation}{section}
\makeatother
\newcommand{\ba}{\begin{eqnarray}}
\newcommand{\ea}{\end{eqnarray}}

\newcommand{\bpsi}{\bar\psi}

\usetikzlibrary{decorations.pathreplacing,decorations.markings}
\tikzset{
  on each segment/.style={
    decorate,
    decoration={
      show path construction,
      moveto code={},
      lineto code={
        \path [#1]
        (\tikzinputsegmentfirst) -- (\tikzinputsegmentlast);
      },
      curveto code={
        \path [#1] (\tikzinputsegmentfirst)
        .. controls
        (\tikzinputsegmentsupporta) and (\tikzinputsegmentsupportb)
        ..
        (\tikzinputsegmentlast);
      },
      closepath code={
        \path [#1]
        (\tikzinputsegmentfirst) -- (\tikzinputsegmentlast);
      },
    },
  },
  mid arrow/.style={postaction={decorate,decoration={
        markings,
        mark=at position .5 with {\arrow[#1]{stealth}}
      }}},
}

\tikzstyle{decision} = [diamond, draw, fill=blue!20, 
    text width=4.5em, text badly centered, node distance=3cm, inner sep=0pt]
\tikzstyle{block} = [rectangle, draw, fill=blue!20, 
    text width=8em, text centered, rounded corners, minimum height=2em]
\tikzstyle{line} = [draw, -latex']
\tikzstyle{cloud} = [draw, ellipse,fill=red!20, node distance=3cm,
    minimum height=2em, text centered]

\DeclareMathOperator*{\simx}{\sim}
\DeclareMathOperator*{\tox}{\longrightarrow}
\DeclareMathOperator*{\Sym}{\text{Sym}}
\DeclareMathOperator*{\Span}{\text{Span }}
\DeclareMathOperator*{\End}{\text{End}}
\DeclareMathOperator*{\Hom}{\text{Hom}}

\newcommand{\bigslant}[2]{{\raisebox{.2em}{$#1$}\left/\raisebox{-.2em}{$#2$}\right.}}

\newcommand{\dket}[1]{\ket{#1}\!\rangle}
\newcommand{\dbra}[1]{\langle\!\bra{#1}}
\newcommand{\superp}[2]{\genfrac{}{}{0pt}{}{#1}{#2}}

 \def\d{\delta}

 \def\p{\partial}
 
 \def\a{\alpha}
 \def\b{\beta}
 \def\g{\gamma}
 \def\d{\delta}
 \def\e{\varepsilon}
 
 \def\th{\theta}
 
 \def\k{\kappa}
 \def\l{\lambda}

 \def\s{\sigma}
 \def\t{\tau}
 \def\th{\theta}
 \def\z{\zeta }
 
 \def\G{\Gamma}
 \def\D{\Delta}

 \def\O{\Omega}
 \def\o{\omega }
 
\def\CA{{\mathcal{A}}}
\def\CB{{\mathcal{B}}}

\def\CE{{\mathcal{E}}}
\def\CF{{\mathcal{F}}}
\def\CG{{\mathcal{G}}}

\def\CK{{\mathcal{K}}}
\def\CL{{\mathcal{L}}}
\def\CM{{\mathcal{M}}}
\def\CN{{\mathcal{N}}}
\def\CO{{\mathcal{O}}}

\def\CS{{\mathcal{S}}}
\def\CT{{\mathcal{T}}}
\def\CU{{\mathcal{U}}}
\def\CV{{\mathcal{V}}}

\def\CY{{\mathcal{Y}}}

\def\la{\left\langle}
\def\ra{\right\rangle}

\def\bc{{\bar{c}}}

\def\implies{\quad\Rightarrow\quad}

\def\vphi{\varphi}

\def\CS{\mathcal{S}}

\def\bf{\mathfrak{b}}
\def\qf{\mathfrak{q}}
\def\tf{\mathfrak{t}}
\def\CZv{\mathcal{Z}^v}
\def\CZbf{\mathcal{Z}^\text{bf}}
\def\CZf{\mathcal{Z}^{f}}
\def\CZaf{\mathcal{Z}^{\bar{f}}}

\def\CZCS{\mathcal{Z}^{\text{CS}}}
\def\Zv{Z^{v}}

\def\Zf{Z^{f}}
\def\Zaf{Z^{\bar{f}}}

\def\ZCS{Z^{\text{CS}}}

\def\CZinst{\mathcal{Z}_I}

\def\hg{{\hat\gamma}}

\def\bd{{\bar d}}

\def\vac{\emptyset}
\def\res{\mathop{\text{Res}}}

\def\bt{{\bar\tau}}

\def\bl{{\boldsymbol{\lambda}}}
\def\bv{{\boldsymbol{v}}}
\def\bk{{\boldsymbol{k}}}
\def\bnu{{\boldsymbol{\nu}}}

\def\mZ{\mathbb{Z}}
\def\mR{\mathbb{R}}
\def\mC{\mathbb{C}}
\def\mI{\mathbb{I}}
\def\End{\text{End}}
\def\Hom{\text{Hom}}

\def\tpsi{\tilde{\psi}}
\def\tx{\tilde{x}}

\def\gl{\mathfrak{gl}}
\def\sl{\mathfrak{sl}}

\def\Uqsl{U_q(\widehat{\mathfrak{sl}(2)})}

\def\Abox{{\tikz[scale=0.007cm] \draw (0,0) rectangle (1,1);}}
\def\sAbox{{\tikz[scale=0.005cm] \draw (0,0) rectangle (1,1);}}

\def\AboxB{{\tikz[scale=0.007cm] \draw[fill=black] (0,0) rectangle (1,1);}}
\def\sAboxB{{\tikz[scale=0.005cm] \draw[fill=black] (0,0) rectangle (1,1);}}

\def\tPhi{\tilde{\Phi}}

\def\tX{\tilde{X}}
\def\tPsi{\tilde{\Psi}}

\def\bmu{{\boldsymbol{\mu}}}

\def\vrho{\varrho}

\def\bn{\bar n}

\def\tN{\tilde{N}}
\def\bCM{\overline{\mathcal{M}}}
\def\tk{\tilde{\kappa}}

\begin{document}
\begin{titlepage}
% \vspace*{-2cm}
% \marginpar[]{\jobname .pdf\\ \today\\}
% \begin{flushright}
% \jobname .pdf\\ \today\\
% % CQUEST-2021-????
% \end{flushright}
% \vspace*{1cm}
% \vskip 12mm

\begin{center}
{\Huge  Shifted quantum groups and matter multiplets\\
\vskip .5cm
in supersymmetric gauge theories}

\vskip 2cm
{\Large Jean-Emile Bourgine}\\

\vskip 1cm
% {\it ARC Centre of Excellence for Mathematical and Statistical Frontiers (ACEMS)}\\
{\it School of Mathematics and Statistics}\\
{\it University of Melbourne}\\
{\it Parkville, Victoria 3010, Australia}\\

\vskip 1cm    
{\it Center for Quantum Spacetime (CQUeST)}\\
{\it Sogang University}\\
{\it Seoul, 121-742, South Korea}
\vskip 1cm

\texttt{bourgine@kias.re.kr}
\end{center}
\vfill
\begin{abstract}
The notion of \textit{shifted} quantum groups has recently played an important role in algebraic geometry. This subtle modification of the original definition brings more flexibility in the representation theory of quantum groups. The first part of this paper presents new mathematical results for the shifted quantum toroidal $\gl(1)$ and quantum affine $\sl(2)$ algebras (resp. denoted $\ddot{U}_{q_1,q_2}^\bmu(\gl(1))$ and $\dot{U}_q^\bmu(\sl(2))$). It defines several new representations, including finite dimensional highest $\ell$-weight representations for the toroidal algebra, and a vertex representation of $\dot{U}_q^\bmu(\sl(2))$ acting on Hall-Littlewood polynomials. It also explores the relations between representations of $\dot{U}_q^\bmu(\sl(2))$ and $\ddot{U}_{q_1,q_2}^\bmu(\gl(1))$ in the limit $q_1\to\infty$ ($q_2$ fixed), and present the construction of several new intertwiners. These results are used in the second part to construct BPS observables for 5d $\CN=1$ and 3d $\CN=2$ gauge theories. In particular, it is shown that 5d hypermultiplets and 3d chiral multiplets can be introduced in the algebraic engineering framework using shifted representations, and the Higgsing procedure is revisited from this perspective.
\end{abstract}
\vfill
\end{titlepage}

\setcounter{footnote}{0}

\newpage
\tableofcontents
\newpage

% \tikzexternalize

\section{Introduction}
The notion of shifted quantum groups, and more precisely shifted Yangians, has been introduced in \cite{Brundan2004} as a way to represent finite W-algebras. These algebras enter in a finite version of the AGT correspondence \cite{Braverman2010}, and arise in the description of quantum Coulomb branches of 3d $\CN=4$ gauge theories \cite{Braverman2016a,Nakajima2019}. It is also in this setting that the trigonometric version of shifted Yangians, namely the shifted quantum affine algebras, were introduced by Finkelberg and Tsymbaliuk in \cite{Finkelberg2017a}.\footnote{The shifted quantum affine $\sl(2)$ algebra with antidominant shift had appeared before in \cite{DiFrancesco2016} in the context of quantum Q-systems of type A and cluster algebras.}

However, quantum groups were initially introduced by Drinfeld and Jimbo to describe the algebra of symmetries of quantum integrable systems. In this context, Hernandez and Jimbo have defined in \cite{Hernandez2012} a closely related notion of \textit{asymptotic} quantum groups, of which the prefundamental representation entering in the construction of Baxter's Q-operator \cite{BLZ3} is an important realization. The proximity of these two notions prompted the study of the representation theory of shifted quantum affine algebras in \cite{Hernandez2020} and Yangians in \cite{Hernandez2021}, with applications to integrable systems in sight.

In \cite{Bourgine2021d}, the author proposed an algebraic construction of certain BPS observables for a class of 3d $\CN=2$ gauge theories based on the representation theory of the shifted quantum affine $\sl(2)$ algebras. This construction is a specific instance of the \textit{algebraic engineering} which aims to extend the ``CFT techniques'' of vertex operators to supersymmetric gauge theories using the representation theory of quantum groups. It is inspired from the string theory realization of gauge theories as the low energy dynamics of certain NS5-D5-D3 brane systems, with each brane carrying a module of the quantum group \cite{Bourgine2017a} (see also the original papers \cite{AFS,Awata2016a,Mironov2016,Bourgine2017b,BFMZ,Zhu2017,Zenkevich2018,Zenkevich2020,Ghoneim2020}). In the case of 3d $\CN=2$ theories, it involves certain \textit{shifted representations} arising from the presence of $\CN=2$ chiral multiplets. One of the main objectives of this paper is to investigate further the connection between such shifted representations and the presence of matter multiplets in supersymmetric gauge theories. In particular, we will show that 5d $\CN=1$ hypermultiplets can be introduced in this framework using the shifted versions of the quantum toroidal $\gl(1)$ algebra. We also derive several technical results that justify the claims made in \cite{Bourgine2021d}. Finally, we will explore the relation between representations of shifted quantum toroidal $\gl(1)$ and quantum affine $\sl(2)$ algebras, and revisit the Higgsing procedure relating 3d and 5d theories from an algebraic perspective.

It is worth mentioning that shifted quantum groups also appeared very recently in the study of BPS algebras \cite{Galakhov2021,Noshita2021}. It would be interesting to explore more the connections between this application and the results presented in this paper.

\begin{figure}
\begin{center}
\begin{tabular}{|c|c|c|c|c|c|}
\hline
Algebra & Module & Representation & Weights & Description & Brane\\
\hline
$\CU$ & $\CM_N$ & $\vrho_a^{L(M_N)}$ & $a$ & Kirillov-Reshetikhin module (dim. $N+1$) & $-$\\
$\CU^{(0,-1)}$ & $\CL$ & $\vrho_\nu$ & $\nu$ & Prefundamental representation & D3\\
$\CU^{(-1,-1)}$ & $\CV$ & $\vrho_\nu^{(0)}$ & $\nu$ & Vector representation & D3\\
$\CU^{(-\infty,0)}$ & $\CF$ & $\vrho_{u,n}^{(LT)}$ & $u$ & Left twisted Fock representation & NS5\\
$\CU^{(0,-\infty)}$ & $\CF$ & $\vrho_{u,n}^{(RT)}$ & $u$ & Right twisted Fock representation & NS5\\
$\CE$ & $\CF_0$ & $\rho_v^{(0,1)}$ & $v$ & Vertical Fock representation & D5\\
$\CE$ & $\CF_0$ & $\rho_u^{(1,n)}$ & $u$ & Horizontal Fock representation & NS5+$n$D5\\
$\CE$ & $\CV$ & $\rho_v^{(0,0)}$ & $v$ & Vector representation & D3\\
$\CE$ & $\CM(v,K)$ & $\rho_v^{(\CM)}$ & $v$ & MacMahon representation & $-$\\
$\CE^{(-\infty,0)}$ & $\CF$ & $\rho_{u,n}^{(LT)}$ & $u$ & Left twisted Fock representation & $-$\\
$\CE^{(0,-\infty)}$ & $\CF$ & $\rho_{u,n}^{(RT)}$ & $u$ & Right twisted Fock representation & $-$\\

\hline
\end{tabular}
\end{center}
\caption{Summary of the representations encountered in this paper.}
\label{table1}
\end{figure}

\paragraph{Organization of the paper and main results.} The physical description presented in this paper relies on several new mathematical results on shifted quantum groups that could be of general interest. This is why these results have been gathered in the first three sections of the paper, and were made accessible to an audience of both mathematicians and physicists. On the other hand, the sections five and six are intended to physicists and deal with the specific application to supersymmetric gauge theories and string theory.

In details, section II focuses on shifted quantum affine $\sl(2)$ algebras (abbreviated here $\CU^\bmu$). Following \cite{Finkelberg2017a,Hernandez2020}, the notion of shifted representations is introduced, it is then applied to relate the prefundamental representation with the Kirillov-Reshetikhin modules. Furthermore, new representations called \textit{left/right twisted Fock representations} are presented. They can be defined as vertex representations of a Borel subalgebra of the quantum affine $\sl(2)$ algebra acting on the bosonic Fock space, and extended to the ``infinitely shifted'' algebras $\CU^{(-\infty,0)}$ and $\CU^{(0,-\infty)}$ (respectively). Finally, we recall the construction of the intertwining operators acting between a twisted Fock representation and its tensor product with a prefundamental representation \cite{Bourgine2021d}.

The third section presents a study of representations of the shifted quantum toroidal $\gl(1)$ algebras which are denoted here $\CE^\bmu$ for short. The notion of shifted representations is extended to the toroidal case, and used to define \textit{pit representations} as deformations of the ``vertical'' Fock representation \cite{feigin2011quantum}. These representations include a number of finite dimensional highest $\ell$-weight representations for the algebras $\CE^\bmu$ with a dominant shift. They act on a module spanned by states labeled by the sub-Young diagrams of a fixed Young diagram. In addition, this section also extends the definition of \textit{left/right twisted Fock representations} to the toroidal case. We conclude this section by extending the construction of the Awata-Feigin-Shiraishi intertwiners \cite{AFS} to include shifted representations.

The fourth section investigates the relation between representations of the shifted algebras $\CE_{q_1,q_2}^\bmu$ and $\CU^\bmu$ following from the ``crystal'' limit $q_1\to\infty$ with $q_2$ fixed. We study the shifted highest $\ell$-weight representations, and introduce a notion of limit for this class of representations. In this sense, the vertical Fock representation of $\CE$ tends to an infinite sum of (negative) prefundamental representations. We explain how some generators of the left/right twisted Fock representations for $\CU^\bmu$ can be obtained as a limit of generators of $\CE$ in the horizontal Fock representation. For further reference, we summarized the representations considered in this paper for various shifted algebras in the table of figure \ref{table1}.

In the fifth section, we show that the shifted representations of $\CE^\bmu$ can be used to introduce (anti)fundamental matter hypermultiplets in the algebraic engineering of 5d $\CN=1$ gauge theories. We also discuss the connection with another realization of hypermultiplets using ``matter vertex operators'' \cite{Bourgine2017b}. 

The algebraic construction of 3d $\CN=2$ gauge theories presented in \cite{Bourgine2021d} is briefly recalled in the sixth section, and then extended to (A-type) quiver theories by analyzing the case of the gauge group $U(1)\times U(1)$. We also come back to the description of the Higgsing procedure using the results on the limit of representations obtained in the sections VI. Finally, we develop the point of view of the equivariant character of instanton/vortex moduli space in appendix \ref{AppCOHA}.

\section{Shifted quantum affine \texorpdfstring{$\sl(2)$}{sl(2)} algebras}\label{sec_SQA_sl2}
\subsection{Definition and main properties}
The quantum affine algebra of $\sl(2)$ with quantum parameter $q$ is traditionally denoted $\Uqsl$ or $\dot{U}_q(\sl(2))$, it will be abbreviated by $\CU$ here. The shifted quantum affine $\sl(2)$ algebras depend on the shift parameters $\bmu=(\mu_+,\mu_-)\in\mZ\times\mZ$, and we will denote them $\CU^\bmu$. In fact, the algebras $\CU^{\bmu}$ and $\CU^{\bmu+(n,n)}$ are isomorphic and so their properties only depend on $\mu_++\mu_-$ \cite{Finkelberg2017a}.\footnote{Explicitly, we can choose the following isomorphism  $X^+(z)\to z^n X^+(z)$, $X^-(z)\to X^-(z)$ and $\Psi^\pm(z)\to z^n\Psi^\pm(z)$ in terms of the Drinfeld currents defined below.} However. it is useful to keep the two parameters $\bmu=(\mu_+,\mu_-)$ in our application to string theory. It is also convenient to introduce the partial orderings $\bmu\geq\bmu'$ iff $\mu_++\mu_-\geq \mu_+'+\mu_-'$, and $\bmu\geq k$ iff $\mu_++\mu_-\geq k$ for $k\in\mZ$ to state the main properties of the shifted algebras. The shifts satisfying $\bmu\geq0$ are called \textit{dominant} while shifts with $\bmu\leq0$ are called \textit{antidominant}. This notion determines the main properties of the shifted algebras. For instance, Hernandez has shown in \cite{Hernandez2012} that the shifted algebras $\CU^\bmu$ have non-trivial finite dimensional highest $\ell$-weight representations only if $\bmu$ is dominant.

The notion of shifted quantum groups is introduced using the Drinfeld presentation of the algebra in which a set of currents $X_\o^\pm(z), \Psi_\o^\pm(z)$ is assigned to each node $\o$ of the Dynkin diagram of the underlying Lie algebra. The translation between this presentation and the RTT presentation employed in the field of integrable systems can be found in \cite{Ding1993}, while the expression of the Chevalley generators in the Drinfeld-Jimbo presentation is found e.g. in \cite{Khoroshkin2007}. In this section, we focus on the case of the Lie algebra $\sl(2)$ for which the Dynkin diagram has a single node, and so we can drop the index $\o$. Thus, the shifted quantum affine $\sl(2)$ algebra $\CU^\bmu$ is spanned by the modes of the four currents
\begin{equation}\label{Uqsl_currents}
X^\pm(z)=\sum_{k\in\mathbb{Z}}z^{-k}X^\pm_k,\quad \Psi^\pm(z)=\sum_{\pm k\geq -\mu_\pm} z^{-k}\Psi_{k}^\pm,
\end{equation}
and a central element $q^c$. The shifts $\mu_\pm$ only enter in the expansion of the Cartan currents $\Psi^\pm(z)$, and the relations obeyed by the modes remain unchanged, they can be found in \cite{Chari1995}. Once summed over, they produce the following exchange relations for the currents (see, for instance, \cite{Ding2000})
\begin{align}
\begin{split}\label{def_Uqsl2}
&[\Psi^\pm(z),\Psi^\pm(w)]=0,\quad \Psi^+(z)\Psi^-(w)=\dfrac{G(q^cz/w)}{G(q^{-c}z/w)}\Psi^-(w)\Psi^+(z),\\
&\Psi^+(z)X^\pm(w)=G(q^{\pm c/2}z/w)^{\pm1}X^\pm(w)\Psi^+(z),\quad \Psi^-(z)X^\pm(w)=G(q^{\mp c/2}z/w)^{\pm1}X^\pm(w)\Psi^-(z),\\
&(z-q^{\pm2}w)\ X^\pm(z)X^\pm(w)=(q^{\pm2} z-w)\ X^\pm(w)X^\pm(z),\\
% X^\pm(z)X^\pm(w)=G(z/w)^{\pm1}X^\pm(w)X^\pm(z),\\
&[X^+(z),X^-(w)]=\dfrac1{q-q^{-1}}\left(\d(q^{-c}z/w)\Psi^+(q^{c/2}w)-\d(q^c z/w)\Psi^-(q^{-c/2}w)\right),
\end{split}
\end{align}
where we introduced the structure function
\begin{equation}\label{def_G}
G(z)=q^2\dfrac{z-q^{-2}}{z-q^2}.
\end{equation} 
The zero modes $\Psi_{-\mu_+}^+$ and $\Psi_{\mu_-}^-$ of the Cartan currents $\Psi^\pm(z)$ are assumed to be invertible. They are not central, but instead obey the relations
\begin{equation}\label{rel_Psi_X}
\Psi_{-\mu_+}^+ X^\pm_k=q^{\pm2} X^\pm_k\Psi_{-\mu_+}^+,\quad \Psi_{\mu_-}^- X^\pm_k=q^{\mp2} X^\pm_k\Psi_{\mu_-}^-,\quad \Psi_{-\mu_+}^+\Psi_{\mu_-}^-=\Psi_{\mu_-}^-\Psi_{-\mu_+}^+.
\end{equation} 
The product $\Psi_{-\mu_+}^+\Psi_{\mu_-}^-$ is central, it can be set to one upon a rescaling of the generators but we will refrain from doing so here for reasons that will be explained below.

Alternatively, the Cartan currents can also be expanded as
\begin{equation}\label{shifted_Psi}
\Psi^+(z)=z^{\mu_+}\Psi_{-\mu_+}^+ e^{\sum_{k>0}z^{-k}A_{k}},\quad \Psi^-(z)=z^{-\mu_-}\Psi_{\mu_-}^- e^{-\sum_{k>0}z^{k}A_{-k}},
% \Psi^\pm(z)=z^{\pm\mu_\pm}\Psi_{\mp\mu_\pm}^\pm e^{\pm\sum_{k>0}z^{\mp k}A_{\pm k}},
\end{equation}
with the modes $A_k$ satisfying the relations
\begin{equation}\label{com_Ak}
[A_k,A_l]=\dfrac1k(q^{ck}-q^{-ck})(q^{2k}-q^{-2k})\d_{k+l},\quad [A_k,X_l^\pm]=\pm\dfrac1k q^{\mp c|k|/2}(q^{2k}-q^{-2k})X_{l+k}^\pm.
\end{equation} 

\paragraph{Coproduct} The Drinfeld coproduct can be extended to shifted quantum algebras: it becomes the algebra morphism $\D:\CU^{\bmu+\bmu'}\to\CU^{\bmu}\otimes\CU^{\bmu'}$ for any $\bmu,\bmu'\in\mZ\times\mZ$, upon the usual completion of the tensor product \cite{Finkelberg2017a,Hernandez2020}. It is defined by the same formulas as the coproduct of the original algebra $\CU$, i.e.
\begin{align}
\begin{split}\label{def_coprod}
&\D(X^+(z))=X^+(z)\otimes 1+\Psi^-(q^{c_{(1)}/2}z)\otimes X^+(q^{c_{(1)}}z),\\
&\D(X^-(z))=X^-(q^{c_{(2)}}z)\otimes \Psi^+(q^{c_{(2)}/2}z)+1\otimes X^-(z),\\
&\D(\Psi^\pm(z))=\Psi^\pm(q^{\pm c_{(2)}/2}z)\otimes\Psi^\pm(q^{\mp c_{(1)}/2}z),
\end{split}
\end{align}
with the notations $c_{(1)}=c\otimes 1$, $c_{(2)}=1\otimes c$. We note that $\D(\bf_\pm^{\bmu+\bmu'})\subset\bf_\pm^{\bmu}\otimes\bf_\pm^{\bmu'}$ where the Borel subalgebras $\bf^\bmu_\pm\subset\CU^\bmu$ are generated by
\begin{align}
\begin{split}
&\bf^\bmu_+=\la X^+_k,\Psi^-_{-l}, q^{\pm c/2}\diagup (k,l)\in\mZ,\ l\leq \mu_-\ra,\\
&\bf^\bmu_-=\la X^-_k,\Psi^+_{l}, q^{\pm c/2}\diagup (k,l)\in\mZ,\ l\geq -\mu_+\ra.
\end{split}
\end{align}
Furthermore, we have the triangular decomposition $\CU^\bmu=\CU_-^\bmu\otimes\CU_0^\bmu\otimes\CU_+^\bmu$ with\footnote{It is shown in \cite{Finkelberg2017a} in the case $q^c=1$ and the central deformation is obvious.}
\begin{align}
\begin{split}
&\CU^\bmu_+=\la X^+_k\diagup k\in\mZ\ra,\quad \CU^\bmu_-=\la X^-_k\diagup k\in\mZ\ra,\\
&\CU^\bmu_0=\la \Psi^+_{k},\Psi^-_l,(\Psi_{-\mu_+})^{-1},(\Psi_{\mu_-})^{-1}, q^{\pm c/2}\diagup (k,l)\in\mZ,\ k\geq -\mu_+,\ l\leq\mu_-\ra.
\end{split}
\end{align}
Although we will not consider the Drinfeld-Jimbo coproduct, we mention here for completeness the Borel subalgebras of $\CU$ associated to it,
\begin{align}
\begin{split}
&\bf^\perp_+=\la X^+_k, X^-_{k+1} \Psi^+_{k}, q^{\pm c/2}\diagup k\in\mZ^{\geq0}\ra,\\
&\bf^\perp_-=\la X^+_{-k-1}, X^-_{-k}, \Psi^-_{-k}, q^{\pm c/2}\diagup k\in\mZ^{\geq0}\ra.
\end{split}
\end{align}
The Borel subalgebra $\bf_+^\perp$ is denoted $U_q(\bf)$ in \cite{Hernandez2008,Hernandez2020}.

\paragraph{Remark.} We refrain from imposing any condition on the product $\Psi_{-\mu_+}^+\Psi_{\mu_-}^-$ (which is a central element), and we will do the same for the shifted quantum toroidal $\gl(1)$ algebra in the next section. In this case, it is important to keep this degree of freedom when studying the limit $q_1\to\infty$. In fact, it is possible to formulate more precise statements for the algebras $\CU^{\bmu,\O}=\CU^\bmu\diagup<\Psi_{-\mu_+}^+\Psi_{\mu_-}^-=\O>$ with $\O\in\mC^\times$. Upon the rescaling $X^+(z)\to \O^{-1/2}X^+(z)$, $\Psi^\pm(z)\to \O^{-1/2}\Psi^\pm(z)$, the algebra $\CU^{\bmu,\O}$ is isomorphic to $\CU^{\bmu,1}$, but this factor $\O$ may become singular in the limit of the toroidal algebra. The expressions \ref{coproduct} defines a morphism of algebras $\D:\CU^{\bmu,\O}\to \CU^{\bmu^{(1)},\O_1}\otimes\CU^{\bmu^{(2)},\O_2}$ for any shifts $\bmu$, $\bmu^{(1)}$, $\bmu^{(2)}$ such that $\bmu=\bmu^{(1)}+\bmu^{(2)}$ and parameters $\O$, $\O_1$, $\O_2$ such that $\O=q^{-(\mu_+^{(2)}+\mu_-^{(2)})c_{(1)}/2}q^{(\mu_+^{(1)}+\mu_-^{(1)})c_{(2)}/2}\O_1\O_2$. In particular, $\D(\Psi_{\mp\mu_{\pm}}^\pm)=q^{-\mu_\pm^{(2)}c_{(1)}/2}q^{\mu_\pm^{(1)}c_{(2)}/2}\Psi_{\mp\mu_{\pm}}^\pm\otimes\Psi_{\mp\mu_{\pm}}^\pm$.

\paragraph{Asymptotic algebra} The asymptotic algebra $\CU^\text{as}$ has been defined in \cite{Hernandez2012} using the modes of currents $\tX^\pm(z)$, $\tPsi^\pm(z)$ expanded as in \ref{Uqsl_currents} with $\mu_+=\mu_-=0$. The algebraic relations satisfied by these modes can be found in \cite{Hernandez2012}, once summed over they provide the exchange relations
\begin{align}
\begin{split}\label{def_Uqsl2_as}
&[\tPsi^\pm(z),\tPsi^\pm(w)]=[\tPsi^\pm(z),\tPsi^\mp(w)]=0,\quad \k \tX^\pm(z)=q^{\mp 2}\tX^\pm(z)\k,\\
&\tPsi^+(z)\tX^\pm(w)=q^{\mp2}G(z/w)^{\pm1}\tX^\pm(w)\tPsi^+(z),\quad \tPsi^-(z)\tX^\pm(w)=q^{\mp2}G(z/w)^{\pm1}\tX^\pm(w)\tPsi^-(z),\\
&(z-q^{\pm2}w)\ \tX^\pm(z)\tX^\pm(w)=(q^{\pm2} z-w)\ \tX^\pm(w)\tX^\pm(z),\\
&q^{-2}\tX^+(z)\tX^-(w)-\tX^-(w)\tX^-(z)=\dfrac{\d(z/w)}{q-q^{-1}}\left(\tPsi^+(w)-\tPsi^-(w)\right),
\end{split}
\end{align}
for the currents, together with $\tPsi_0^+=1$, $\tPsi_0^-=\k^2$. When $\k$ is invertible, there is an isomorphism of algebras with the quantum affine $\sl(2)$ algebra at $q^c=1$,
\begin{equation}
\bigslant{\CU}{\la q^c=1\ra}\simeq\CU^\text{as}\otimes_{\mC[\k]}\mC[\k,\k^{-1}]
\end{equation} 
obtained by sending $\tX^+(z)\to X^+(z)$,  $\k^{-1}\tX^-(z)\to X^-(z)$, $\k^{-1}\tPsi^\pm(z)\to\Psi^\pm(z)$. Thus, when $\k$ is invertible, $\CU^\text{as}$ is simply the subalgebra of $\CU$ at $q^c=1$ generated by the elements $\{X^+_n,\ k^{-1}X^-_n,\ k^{-1}\Psi_{\pm k},\ k^{-1}\}$ where $k=\Psi_0^+$. It was shown in \cite{Hernandez2020} that the shifted algebras $\CU^\bmu/\la q^c=1\ra$ with antidominant shift $\bmu$ can be obtained from a quotient of the asymptotic algebra $\CU^\text{as}$. More precisely, the algebra $\CU^{(0,-p)}(q^c=1)$ is obtained as the quotient
\begin{equation}
\bigslant{\CU^\text{as}}{\la\k=\tPsi_{-1}^-=\cdots=\tPsi_{-p+1}=0\ra},
\end{equation} 
to which the generators $\Psi_0^+$ and $\Psi_{-p}^-$ are added with the relations \ref{rel_Psi_X}. Sending $p$ to infinity, we can also see the infintely shifted algebra $\CU^{(0,-\infty)}$ (i.e. with $\psi^-(z)\equiv0$) as a quotient of the asymptotic algebra $\CU^\text{as}$.

% 
% The asymptotic quantum affine $\sl(2)$ algebra, denoted here $\CU^\text{as}$, has been introduced in \cite{Hernandez2012}. It can be defined as the subalgebra of $\CU(q^c=1)$ (i.e. without central deformation) generated by the elements $\{X^+_k,\ \k^{-1}X^-_k,\ \k^{-1}\Psi_{\pm k},\ \k^{-1}\}$ where $\k=\Psi_0^+$.

\subsection{Shifted representations}
The Drinfeld coproduct \ref{coproduct} can be used to introduce shifted versions of any representations $\vrho$ of $\CU$ that are representations of $\CU^\bmu$ with dominant $\bmu$. More generally, starting with a representation $\vrho$ of $\CU^\bmu$, we will build two representations of $\CU^{\bmu'}$ with $\bmu'\geq\bmu$ denoted $\iota_P\vrho$ and $\iota^\ast_P\vrho$. They depend on $P(z)$ which is a polynomial or a finite Laurent series of the form
\begin{equation}\label{def_P}
P(z)=\a_Pz^{-\mu_0}\prod_{a=1}^n(1-z/\nu_a)=\dfrac{\a_P}{\prod_{a=1}^n(-\nu_a)}z^{\mu_\infty}\prod_{a=1}^{n}(1-\nu_a/z),
\end{equation}
with $n=\mu_0+\mu_\infty\geq0$, $\nu_a\in\mC^\times$ and $\a_P\in\mC^\times$. From this quantity, we define a one-dimensional representation of $\CU^{\bmu_P}$ as
\begin{equation}
\vrho_P(X^\pm(z))=0,\quad \vrho_P(\Psi^\pm(z))=P(z),\quad \vrho_P(q^c)=1,
\end{equation} 
where $\bmu_P=(\mu_\infty,\mu_0)\geq0$ is fixed by the asymptotics of $P(z)$. Using the coproduct, we can define further the representations
\begin{equation}
\iota_P\vrho=(\vrho_P\otimes\vrho)\D,\qquad\iota_P^\ast\vrho=(\vrho\otimes\vrho_P)\D.
\end{equation} 
If $\vrho$ is a representation of $\CU^\bmu$ on a module $\CM$, then $\iota_P\vrho$ and $\iota_P^\ast\vrho$ are representations of $\CU^{\bmu+\bmu_P}$ on the same module $\CM$ since $\vrho_P$ is one-dimensional. They will be called \textit{shifted representations} of $\vrho$ by $P(z)$, they depend on the $\mu_0+\mu_\infty+1$ extra parameters $\nu_a$ and $\a_P$. In practice, the shifted representation $\iota_P\vrho$ simply corresponds to multiply the actions of $X^+(z)$ by $P(z)$ and $\Psi^\pm(z)$ by $\vrho(P(q^{\pm c/2}z))$. Similarly, the representation $\iota_P^\ast\vrho$ corresponds to multiply the actions of $X^-(x)$ by $P(z)$ and $\Psi^\pm(z)$ by $\vrho(P(q^{\mp c/2}z))$. As we will see, in some cases these shifted representations become reducible on $\CM$. If so, we will slightly abuse the notation, and denote also $\iota_P\vrho$ or $\iota_P^\ast\vrho$ the subrepresentations acting on a submodule $\CM'\subset\CM$.

Mathematically, this construction assigns to the Laurent series $P(z)$ of the form \ref{def_P} the two functors $\iota_P$ and $\iota_P^\ast$ from the category of representations of $\CU^\bmu$ to the category of representations of $\CU^{\bmu+\bmu_P}$.\footnote{More precisely, $\iota_P$ and $\iota_P^\ast$ send representations of $\CU^{\bmu,\O}$ to representations of $\CU^{\bmu',\O'}$ with $\bmu'=\bmu+\bmu_P$ and $\O'=\frac{\a_P^2}{\prod_a(-\nu_a)}\O$. Thus, to preserve the relation $\Psi_{-\mu_+}^+\Psi_{\mu_-}^-=\O$, we would need to require $\a_P^2=\prod_a(-\nu_a)$.} These functors form an abelian semi-group since $\iota_{P_1}\iota_{P_2}=\iota_{P_1P_2}$, $\iota^\ast_{P_1}\iota^\ast_{P_2}=\iota^\ast_{P_1P_2}$, and $\iota_{P_1}\iota^\ast_{P_2}=\iota^\ast_{P_2}\iota_{P_1}$. In fact, as a consequence of the injectivity of the shift homomorphism introduced by Finkelberg and Tsymbaliuk \cite{Finkelberg2017a} (section 10(vii)), the shifted algebras $\CU^\bmu$ are subalgebras of $\CU^{\bmu'}$ if $\bmu\geq\bmu'$. As a result, representations of $\CU^{\bmu'}$ provide representations of $\CU^{\bmu}$, they are constructed explicitly using these functors $\iota_P$ and $\iota_P^\ast$ which generalize those presented in \cite{Hernandez2020} (section 4.5). This construction of shifted representations was briefly mentioned in \cite{Bourgine2021d}, it is similar to the \textit{subcrystal representations} of quiver BPS algebra introduced a little earlier in \cite{Galakhov2021}. In the next section, we will present a particularly useful application to the vertical Fock representation of the quantum toroidal $\gl(1)$ algebra.

We note that these shifted representations can also be defined for quantum affine algebras of higher rank. In this case, a Laurent polynomial $P_\o(z)$ is introduced for each node $\o$ of the Dynkin diagram. The constructions follows again from the existence of the one-dimensional representations
\begin{equation}
\vrho_P(X_\o^\pm(z))=0,\quad \vrho_P(\Psi_\o^\pm(z))=P_\o(z),\quad \vrho_P(q^c)=1.
\end{equation}

\subsubsection{Prefundamental representation}\label{sec_prefund}
The negative prefundamental representation $L_{1,v}^-$ has been introduced in \cite{Hernandez2012} as a representation of a Borel subalgebra of $\CU$. It has been extended to the asymptotic algebra $\CU^\text{as.}$, and also defines a representation of the shifted algebra $\CU^{(0,-1)}$ \cite{Hernandez2020}. It is an infinite dimensional highest $\ell$-weight representation in which the currents act on a module $\CL$ spanned by the states $\dket{k}$ labeled by integers $k\geq0$. It has level zero, i.e. $q^c=1$, and will be denoted by $\vrho_v$ where $v\in\mC^\times$ is the weight. The Drinfeld currents act as follows on the basis $\dket{k}$,
\begin{align}\label{prefund}
\begin{split}
&X^+(z)\dket{k}=\dfrac{q^{2k}}{q-q^{-1}}\d(v q^{2k}/z)\dket{k+1},\\
&X^-(z)\dket{k}=-\dfrac{q}{q-q^{-1}}\d(vq^{2(k-1)}/z)(1-q^{-2k})\dket{k-1},\\
&\Psi^\pm(z)\dket{k}=q^{2k}\left[\dfrac{z(z-vq^{-2})}{(z-vq^{2k})(z-vq^{2(k-1)})}\right]_\pm\dket{k}.
\end{split}
\end{align}
To lighten mathematical expressions, we will often omit to indicate the representation when describing the action of the currents on the basis of the modules, but the choice of representation will always be clear from the context.

As observed in \cite{Hernandez2012}, the prefundamental representation is the formal limit $N\to\infty$ of the highest $\ell$-weight representations of dimension $(N+1)$ of the quantum affine $\sl(2)$ algebra $\CU$ on the Kirillov-Reshetikhin modules $\CM_N=\text{Span}\{\dket{k},0\leq k<N+1\}$. They are known in physics as \textit{spin $N/2$ representations}, with the famous spin $1/2$ representation acting on the states $\ket{+}=\dket{0}$, $\ket{-}=\dket{1}$ at each site of the Heisenberg XXZ spin chain. Alternatively, these finite dimensional representations can be obtained as a shifted prefundamental representation $\iota_{P_N}\vrho_v$ with the polynomial $P_N(z)=q^{-N}(1-v q^{2N}/z)$ in the variable $z^{-1}$. The zero of the polynomial imposes $X^+(z)\dket{N}=0$, and the representation can be restricted to the submodule $\CM_N\subset\CL$ on which the currents act as\footnote{To compare with the standard literature on quantum groups, on may denote the weight $a=v q^N$, replace $q\to q^{-1}$ using $\s_H$, and use the spectral parameter $u=z^{-1}$. Then, the action of the Cartan on the vacuum state takes the familiar form of a ratio of Drinfeld polynomials,
\begin{equation}
\Psi^\pm(u^{-1})\dket{0}=q^{N}\left[\dfrac{P_D(q^{-1}u)}{P_D(qu)}\right]_\pm\dket{0},\quad P_D(u)=\prod_{j=0}^{N-1}(1-aq^{2j+1-N}u).%=q^{N}\left[\dfrac{1-aq^{-N}u}{1-aq^{N}u}\right]_\pm\dket{0}
\end{equation}}
\begin{align}\label{KR}
\begin{split}
&X^+(z)\dket{k}=\d(v q^{2k}/z)\dfrac{q^{2k-N}-q^{N}}{q-q^{-1}}\dket{k+1},\\
&X^-(z)\dket{k}=-\dfrac{q}{q-q^{-1}}\d(v q^{2(k-1)}/z)(1-q^{-2k})\dket{k-1},\\
&\Psi^\pm(z)\dket{k}=q^{2k-N}\left[\dfrac{(z-v q^{2N})(z-v q^{-2})}{(z-v q^{2k})(z-v q^{2(k-1)})}\right]_\pm\dket{k}.
\end{split}
\end{align}
Of course, this observation is consistent with the shift of the algebra since $\vrho_{v}$ is a representation of $\CU^{(0,-1)}$ and $\bmu_{P_N}=(0,1)$ so $\iota_{P_N}\vrho_v$ is indeed a representation of $\CU$.

\paragraph{Contragredient} The algebraic engineering technique \cite{Bourgine2021d} also involves the dual space $\CL^\ast$ spanned by the states $\dbra{k}$ for $k\in\mZ\geq0$ defined using the scalar product $\dbra{k}\!\!\dket{k'}=(n_k^\CL)^{-1}\d_{k,k'}$. Physical quantities do not depend on the choice of the norm $n_k^\CL$ for the states, but the form of the contragredient action does. Under a specific choice of norm $n_k^\CL=\bar n_k^\CL$ with
\begin{equation}
\bn_k^\CL=(-1)^k\dfrac{q^{2k^2-k}}{(q^2;q^2)_k},
\end{equation} 
the contragredient action of the prefundamental representation satisfies the property $\vrho_v(e)^t=\vrho_v(\s_V\cdot e)$ where $\s_V:\CU_q^\bmu\to\CU_{q^{-1}}^\bmu$ is the morphism sending $X^\pm(z)\to -X^{\mp}(z)$ leaving the Cartan currents invariant (i.e. $X^\pm(z)$ acts on $\dbra{k}$ as $-X^\mp(z)$ does on $\dket{k}$). We recall here the definition of the q-Pochhammer symbols used intensively in this paper,
\begin{equation}\label{def_qPoch}
(z;q^2)_k=\prod_{j=0}^{k-1}(1-zq^{2j}),\quad (z;q^2)_\infty=\prod_{j=0}^\infty(1-zq^{2j}).
\end{equation}
Usually, the choice of normalization $n_k^\CL=\bar n_k^\CL$ is made to simplify the expression of the intertwiners defined below \cite{Bourgine2017b,Bourgine2021d}. Here instead, we decided to keep an arbitrary norm $n^\CL_k$ to highlight some subtleties.

% A contragredient action can be defined on the dual states with the scalar product $\dbra{k}\!\!\dket{k'}=n_k^{-1}\d_{k,k'}$. When the (inverse squared) norm $n_k$ takes the form $\bn_k=(-1)^k\dfrac{q^{2k^2-k}}{(q^2;q^2)_k}$, the currents $X^\pm(z)$ acts on dual states as $-X^\mp(z)$.

\subsubsection{Vector representation}\label{sec_vect_CU}
An equivalent of the vector representation of the quantum toroidal $\gl(1)$ algebra can be defined for the shifted quantum affine $\sl(2)$ algebra $\CU^{(-1,-1)}$. This representation of level $c=0$ and weight $v$ will be denoted $\vrho_v^{(0)}$. The action of the currents on the basis $\dket{k}$ of the module $\CV$, with $k\in\mZ$, takes the form
\begin{align}
\begin{split}
&X^+(z)\dket{k}=\dfrac{q^{2k}}{q-q^{-1}}\d(vq^{2k}/z)\dket{k+1},\\
&X^-(z)\dket{k}=\dfrac{q^{3-2k}}{q-q^{-1}}\d(vq^{2k-2}/z)\dket{k-1},\\
&\Psi^\pm(z)\dket{k}=-vq^{2k}\left[\dfrac{z}{(z-vq^{2k})(z-vq^{2k-2})}\right]_\pm\dket{k}.
\end{split}
\end{align}
The prefundamental representation analyzed in the previous subsection can be recovered from this action by the shift $\vrho_v=\iota_P^\ast\vrho^{(0)}_v$ by $P(z)=q^{-2}(1-zq^2/v)$, in agreement with the shift of the algebra $\CU^{(-1,-1)+\bmu_P}\simeq\CU^{(0,-1)}$.

\subsection{Fock representations}
The bosonic Fock space $\CF_0=\mC[J_{-1},J_{-2},\cdots]\ket{\vac}$ is obtained from the action of the negative modes $J_{-k}$ of a Heisenberg algebra $[J_k,J_l]=k\d_{k+l}$ on a vacuum state $\ket{\vac}$ annihilated by positive modes. It is known to be isomorphic to the ring of symmetric polynomials in infinitely many variables under the identification of the vacuum $\ket{\vac}$ with the constant polynomial $1$, the negative modes $J_{-k}$ with the multiplication by the power sum $p_k=\sum_i x_i^k$, and the positive modes with the formal derivative $k\p/\p p_k$. This Fock space can be extended into the (charged) Fock space $\CF=\oplus_{m\in\mZ}e^{mQ}\CF_0$ using the zero modes $[J_0,Q]=1$. In this subsection, we present several representations of (shifted) quantum affine $\sl(2)$ algebras acting on $\CF$, and for which the Drinfeld currents have the form of vertex operators.

\subsubsection{Twisted Fock representations}\label{sec_limit_horiz}
The \textit{twisted Fock representations} have been introduced in the context of algebraic engineering of 3D $\CN=2$ gauge theories \cite{Bourgine2021d}, and we would like to give here a few more details regarding their construction.\footnote{They are unrelated to the twisted vertex representations introduced by Jing in \cite{Jing1990}.} The first representation, denoted $\vrho^{(LT)}_{u,u',n}$, for \textit{left twisted}, is characterized by $\Psi^+(z)=0$. It is a representation of the infinitely shifted algebra $\CU^{(-\infty,0)}$. In fact, combining it with $\iota_P$ or $\iota_P^\ast$, we can represent any $\CU^{(-\infty,k)}$ for $k\geq0$. In contrast with the Frenkel-Jing representations \cite{Frenkel1988} recalled below, the twisted Fock representations can be defined for any value of the central charges. Indeed, replacing $X^+(z)\to X^+(q^{-c}z)$, and $\Psi^\pm(z)\to\Psi^\pm(q^{\pm c/2}z)$ in the algebraic relations \ref{def_Uqsl2}, we observe that the central element $\textcolor{blue}{q^c}$ only appears in relations involving $\Psi^+(z)$,
\begin{align}
\begin{split}\label{twisted_Uqsl2}
&[\Psi^\pm(z),\Psi^\pm(w)]=0,\quad \Psi^+(z)\Psi^-(w)=\dfrac{G(\textcolor{blue}{q^{2c}}z/w)}{G(z/w)}\Psi^-(w)\Psi^+(z),\\
&\Psi^+(z)X^+(w)=G(\textcolor{blue}{q^{2c}}z/w)X^+(w)\Psi^+(z),\quad \Psi^+(z)X^-(w)=G(z/w)^{-1}X^-(w)\Psi^+(z),\\
&\Psi^-(z)X^\pm(w)=G(q^{\mp c/2}z/w)^{\pm1}X^\pm(w)\Psi^-(z),\quad (z-q^{\pm2}w)\ X^\pm(z)X^\pm(w)=(q^{\pm2} z-w)\ X^\pm(w)X^\pm(z),\\
&[X^+(z),X^-(w)]=\dfrac1{q-q^{-1}}\left(\d(\textcolor{blue}{q^{-2c}}z/w)\Psi^+(w)-\d(z/w)\Psi^-(z)\right).
\end{split}
\end{align}
Taking $\Psi^+(z)\equiv0$, the dependence in $q^{c}$ drops, and the commutator $[X^+,X^-]$ simplifies into 
\begin{equation}
[X^+(z),X^-(w)]=-\dfrac{\d(z/w)}{q-q^{-1}}\Psi^-(z).
\end{equation} 

The left twisted Fock representation of $\CU^{(-\infty,0)}$ will be denoted $\vrho^{(LT)}_{u,u',n}$, it depends on the complex weights $u,u'$ and integer $n$. The Drinfeld currents have the expression
\begin{align}
\begin{split}\label{def_vrho_LT}
&X^+(z)=-\dfrac{uq^{-nc}}{q-q^{-1}}z^{-n}e^Qe^{\sum_{k>0}\frac{z^k}{k}q^{ck}(1-q^{2k})J_{-k}}e^{-\sum_{k>0}\frac{z^{-k}}{k}q^{-ck}J_k}q^{-2J_0},\\
&X^-(z)=(u')^{-1}\dfrac{q^{-1}}{q-q^{-1}}z^{n}e^{-Q}e^{-\sum_{k>0}\frac{z^k}{k}(1-q^{-2k})J_{-k}}e^{\sum_{k>0}\frac{z^{-k}}{k}J_k},\\
&\Psi^+(z)=0,\quad \Psi^-(z)=(u/u')e^{-\sum_{k>0}\frac{z^k}{k}q^{ck/2}(q^{2k}-q^{-2k})J_{-k}}q^{-2J_0},
\end{split}
\end{align}
Note that these weights are in fact redundant because of the property $\vrho^{(LT)}_{\a u,\a'u',n+p}=\iota_{\a z^{-p}}\iota_{(\a')^{-1} z^p}^\ast\vrho^{(LT)}_{u,u',n}$ under shifts of representations by monomials. In the following, we will restrict ourselves to representations with $u=u'$ that we will denote simply $\vrho^{(LT)}_{u,n}$. 

The right twisted representation $\vrho^{(RT)}_{u,u',n}$ of $\CU^{(0,-\infty)}$ on $\CF$ proceeds from a different twist, namely $X^+(z)\to X^+(q^c z)$ and $\Psi^\pm(z)\to\Psi^\pm(q^{\pm c/2}z)$. It is such that $\Psi^-(z)=0$, and the Drinfeld currents have the form
\begin{align}
\begin{split}\label{def_vrho_RT}
&X^+(z)=-\dfrac{uq^{2-nc}}{q-q^{-1}}z^{n}e^{-Q} e^{-\sum_{k>0}\frac{z^k}{k}q^{-kc}J_{-k}} e^{\sum_{k>0}\frac{z^{-k}}{k}q^{kc}(1-q^{2k})J_k}q^{-2J_0},\\
&X^-(z)=(u')^{-1}\dfrac{q^{-1}}{q-q^{-1}}z^{-n}e^{Q}e^{\sum_{k>0}\frac{z^k}{k}J_{-k}}e^{-\sum_{k>0}\frac{z^{-k}}{k}(1-q^{-2k})J_k},\\
&\Psi^+(z)=(u/u')e^{-\sum_{k>0}\frac{z^{-k}}{k}q^{kc/2}(q^{2k}-q^{-2k})J_k}q^{-2J_0},\quad \Psi^-(z)=0.
\end{split}
\end{align}
The two twisted representations are related through $\vrho^{(LT)}_{u,u',n}(e)^\dagger=\vrho^{(RT)}_{u,u',n}(\s_H\cdot e)$ for $e\in\CU^{(-\infty,0)}$. The left twisted Fock representation defines a representation of the Borel subalgebra $\CB_+\subset\CU$  spanned by the modes of $X^+(z)$, $\Psi^-(z)$ and $q^c$, while the right twisted representation gives a representation of the Borel subalgebra $\CB_-\subset\CU$ spanned by the modes of $X^-(z)$, $\Psi^+(z)$ and $q^c$.

\subsubsection{Frenkel-Jing vertex representation}
For the sake of completeness, we would like to mention here the vertex representation introduced by Frenkel and Jing in \cite{Frenkel1988} although they it not be used in our application to brane systems and supersymmetric gauge theories. Specializing their construction to the case of $\sl(2)$, we find an action $\vrho^{(\text{FJ})}_u$ on the Fock space $\CF$ in which the currents are represented as vertex operators\footnote{We made a minor modification of the zero modes to ease the comparison with the twisted Fock representations of the previous section.},
\begin{align}\label{def_FJ}
\begin{split}
&X^+(z)=ue^Qe^{\sum_{k>0}\frac{z^k}{k}(q^k+q^{-k})J_{-k}}e^{-\sum_{k>0}\frac{z^{-k}}{k}q^{-k}J_k}z^{2J_0+1},\\
&X^-(z)=u^{-1}e^{-Q}e^{-\sum_{k>0}\frac{z^k}{k}(1+q^{2k})J_{-k}}e^{\sum_{k>0}\frac{z^{-k}}{k}J_k}z^{-2J_0+1},\\
&\Psi^+(z)=e^{\sum_{k>0}\frac{z^{-k}}{k}q^{-k/2}(q^k-q^{-k})J_{k}}q^{2J_0},\quad \Psi^-(z)=e^{-\sum_{k>0}\frac{z^k}{k}q^{k/2}(q^{2k}-q^{-2k})J_{-k}}q^{-2J_0}.
\end{split}
\end{align}
This is a representation of level one and weight $u\in\mC^\times$, and we note that it is different from the twisted Fock representations of the previous subsection. In fact, these two types of vertex representations seem to come from different limits of the horizontal Fock representation of the quantum toroidal $\gl(1)$ algebra (see the definition of the limits LI/LII in the next section). Unfortunately, it appears difficult to define rigorously the limit in the case of the Frenkel-Jing representation and we will not attempt to do so here.

\subsection{Intertwining operators}\label{sec_def_P_nu}\label{sec_3d_intw}
In \cite{Jimbo1994}, vertex operators intertwining between representations of the quantum algebra $\CU$ are introduced to compute correlation functions of the XXZ spin chain. These operators intertwine the Frenkel-Jing representation with its tensor product with a spin 1/2 representation. Here, we introduce a similar object, replacing the Frenkel-Jing representation with a twisted Fock representation, and the spin 1/2 representation with the prefundamental representation. More precisely, we construct several operators intertwining between the action of a shifted algebra $\CU^{(-\infty,n)}$ on the Fock space $\CF$, and the action of the product $\CU^{(0,-1)}\otimes\CU^{(-\infty,n+1)}$ on $\CL\otimes\CF$ obtained with the generalized Drinfeld coproduct \ref{def_coprod}. Similar constructions can be performed for the shifted algebra $\CU^{(n,-\infty)}$ using right-twisted Fock representations instead of left-twisted ones.

The first intertwiners have appeared previously in \cite{Bourgine2021d}, they satisfy the intertwining relations
\begin{align}\label{intw_Phi_3d}
\begin{split}
&\Phi[u,\nu,n]:\CL\otimes\CF\to\CF,\quad \vrho^{(LT)}_{u,n}(e)\Phi=\Phi\left(\vrho_\nu\otimes\iota_{P_\nu}^\ast \vrho^{(LT)}_{u,n}\D(e)\right),\quad e\in \CU^{(-\infty,0)},\\
&\Phi^\ast[u,\nu,n]:\CF\to\CL\otimes\CF,\quad \left(\vrho_\nu\otimes \vrho^{(LT)}_{u,n}\D'(e)\right)\Phi^\ast=\Phi^\ast\iota_{P_\nu^\ast} \vrho^{(LT)}_{-u\nu,n+1}(e),\quad e\in \CU^{(-\infty,-1)},\\
&\text{with}\quad P_\nu(z)=1-\nu/z,\quad P_\nu^\ast(z)=-\nu^{-1}z(1-q^2z/\nu).
\end{split}
\end{align}
These intertwining operators can be expanded in the basis of the prefundamental module,
\begin{equation}\label{Phi3D}
\Phi[u,\nu,n]=\sum_{k=0}^\infty n_k^\CL\ \dbra{k}\otimes\Phi_k[u,\nu,n],\quad \Phi^\ast[u,\nu,n]=\sum_{k=0}^\infty n_k^\CL \dket{k}\otimes \Phi_k^\ast[u,\nu,n],
\end{equation} 
and the corresponding components are vertex operator acting on the Fock space $\CF$,
\begin{align}\label{expr_Phi3D}
\begin{split}
&\Phi_k[u,\nu,n]=\t_k[u,\nu,n] e^{kQ} e^{-\sum_{l>0}\frac1l\nu^l q^{2kl}J_{-l}}e^{-\sum_{l>0}\frac1l\nu^{-l}\frac{1-q^{-2kl}}{1-q^{-2l}}J_l}q^{-2kJ_0},\\
&\Phi_k^\ast[u,\nu,n]=\t_k^\ast[u,\nu,n]e^{-kQ}e^{\sum_{l>0}\frac{\nu^l}{l}q^{-2l}(1-q^{2kl})J_{-l}}e^{-\sum_{l>0}\frac{\nu^{-l}}{l}\frac{q^{-2kl}}{1-q^{-2l}}J_l},\\
\text{and}\quad &\t_k[u,\nu,n]=(-u\nu^2)^k\prod_{j=1}^k(\nu q^{2j-2})^{-n-2},\quad \t_k^\ast[u,\nu,n]=q^ku^{-k}\dfrac{\bar n_k^\CL}{n_k^\CL}\prod_{j=1}^k(\nu q^{2j-2})^{n}.
\end{split}
\end{align}
% with the normalizations
% \begin{equation}
% \t_k[\nu,n]=(-)^k\nu^{-nk}q^{-2k-(n+2)k(k-1)},\quad \t_k^\ast[\nu,n]=\dfrac{\bar n_k}{n_k}\nu^{nk}q^{k+nk(k-1)}.
% \end{equation} 

It is possible to define another set of intertwiners which solves the algebraic constraints 
\begin{align}
\begin{split}\label{intw_tPhis}
&\iota_{P_\nu^\ast}\vrho_{-u\nu,n-1}^{(LT)}(e)\ \tPhi = \tPhi\ \left(\vrho_\nu\otimes\vrho_{u,n}^{(LT)}\ \D(e)\right),\quad e\in\CU^{(-\infty,-1)}.\\
&\left(\vrho_{\nu}\otimes\vrho_{u,n}^{(LT)}\ \D'(e)\right)\tPhi^\ast=\tPhi^\ast\iota_{P_{\nu}^\ast}^\ast\vrho_{-uq^2/\nu,n-1}^{(LT)}(e),\quad e\in\CU^{(-\infty,-1)}.
\end{split}
\end{align}
They are also expressed by decomposing in the prefundamental states as in \ref{Phi3D} but each component involves a different vertex operator:
\begin{align}
\begin{split}\label{def_tPhi}
&\tPhi_k[u,\nu,n]=\tilde{\t}_k[u,\nu,n]e^{kQ}e^{-\sum_{l>0}\frac{\nu^l}{l}q^{2kl}J_{-l}}e^{\sum_{l>0}\frac{\nu^{-l}}{l}\frac{q^{-2kl}}{1-q^{-2l}}J_l}q^{-2kJ_0},\\
&\tPhi_k^\ast[u,\nu,n]=\tilde{\t}_k^\ast[u,\nu,n]e^{-kQ}e^{-\sum_{l>0}\frac{\nu^l}{l}q^{2(k-1)l}J_{-l}}e^{-\sum_{l>0}\frac{\nu^{-l}}{l}\frac{q^{-2kl}}{1-q^{-2l}}J_l},\\
\text{and}\quad&\tilde{\t}_k[u,\nu,n]=(q^2;q^2)_k^{-1}\t_k[u,\nu,n],\quad \tilde{\t}_k^\ast[u,\nu,n]=(-\nu q^{-2})^k(q^2;q^2)_k\prod_{j=1}^k(\nu q^{2j-2})^{-1}\t_k^\ast[u,\nu,n].
\end{split}
\end{align}
Finally, intertwining relations of the type \ref{intw_Phi_3d} or \ref{intw_tPhis}, but involving more general shifted algebras $\CU^{(-\infty,n)}$ can be solved by a modification of the factors $\t_k$, $\t_k^\ast$ (or $\tilde{\t}_k$, $\tilde{\t}_k^\ast$). We postpone this discussion to the section \ref{sec_shift_intw} since the notion of shifted intertwiners is slightly simpler in the toroidal setting.

% \paragraph{Remark} The algebraic properties of the intertwiners can also be characterized using the right twisted Fock representation $\vrho^{(RT)}_{u,n}$. For instance, for the intertwiner $\Phi^\ast$, we have
% \begin{equation}\label{intw_Phis}
% \left(\vrho_{\nu}\otimes\iota_{P_\nu}\vrho^{(RT)}_{u,n}\ \D'(e)\right)\Phi^\ast=\Phi^\ast\vrho^{(RT)}_{-u\nu,n+1}(e),\quad e\in \CU^{(0,-\infty)},
% \end{equation} 
% with $P_\nu(z)=1-\nu/z$. Note, however, that in order to couple horizontally $\Phi$ and $\Phi^\ast$, the intermediate representation should match, and we are forced to make a consistent choice of left/right twist for all the representations.

\section{Shifted quantum toroidal \texorpdfstring{$\gl(1)$}{gl(1)} algebras}\label{sec_SQTA}
Shifted quantum toroidal algebras have been briefly evoked in \cite{Finkelberg2017a}, but a general study of their main properties is still lacking. Fortunately, we will need only elementary results that can be inferred from the studies of shifted quantum affine algebras \cite{Hernandez2020}. After recalling the definition of the shifted quantum toroidal $\gl(1)$ algebra and presenting some of its properties, we construct the shifted versions of the vector, vertical and horizontal representations entering in the algebraic engineering formalism. In addition, two new twisted horizontal Fock representations are also introduced. In the last subsection, we study the behavior of intertwiners under the shift of representations.

\subsection{Definition of the shifted algebras}
The shifted quantum toroidal $\gl(1)$ algebras depend on two complex parameters $q_1,q_2\in\mC^\times$ and two integers defining the shift $\bmu=(\mu_+,\mu_-)\in\mZ\times\mZ$. Instead of the standard notation $\ddot{U}^\bmu_{q_1,q_2}(\gl(1))$, we will use here the shortcut $\CE_{q_1,q_2}^{\bmu}$ and often omit to indicate the dependence in the parameters $q_1,q_2$. This algebra has an $S_3$-invariance under permutation of the parameters $q_1,q_2,q_3=q_1^{-1}q_2^{-1}$. In the Drinfeld presentation, it is generated by the modes of the currents 
\begin{equation}\label{DIM_currents}
x^\pm(z)=\sum_{k\in\mathbb{Z}}z^{-k}x^\pm_k,\quad \psi^\pm(z)=\sum_{\pm k\geq -\mu_\pm} z^{-k}\psi_{k}^\pm,
\end{equation} 
and a central element $\hg$. Their algebraic relations follow from the expansion of the current relations
\begin{align}
\begin{split}\label{def_DIM}
&[\psi^\pm(z),\psi^\pm(w)]=0,\quad \psi^+(z)\psi^-(w)=\dfrac{g(\hg z/w)}{g(\hg^{-1}z/w)}\psi^-(w)\psi^+(z),\\
&\psi^+(z)x^\pm(w)=g(\hg^{\pm1/2}z/w)^{\pm1}x^\pm(w)\psi^+(z),\quad \psi^-(z)x^\pm(w)=g(\hg^{\mp1/2}z/w)^{\pm1}x^\pm(w)\psi^-(z),\\
&\prod_{\a=1,2,3}(z-q_\a^{\pm1}w)\ x^\pm(z)x^\pm(w)=\prod_{\a=1,2,3}(z-q_\a^{\mp1}w)\ x^\pm(w)x^\pm(z),\\
% x^\pm(z)x^\pm(w)=g(z/w)^{\pm1}x^\pm(w)x^\pm(z),\\
&[x^+(z),x^-(w)]=\k\left(\d(\hg^{-1}z/w)\psi^+(\hg^{1/2}w)-\d(\hg z/w)\psi^-(\hg^{-1/2}w)\right),\\
&\Sym_{z_1,z_2,z_3}\dfrac{z_2}{z_3}[x^\pm(z_1),[x^\pm(z_2),x^\pm(z_3)]]=0\quad \text{(Serre relations)},
\end{split}
\end{align}
where $\k$ is a $\mathbb{C}$-number and $g(z)$ a rational function, both depending on the parameters $q_1,q_2$ , 
\begin{equation}\label{def_g}
\k=\prod_{\a=1,2,3}(1-q_\a),\quad g(z)=\prod_{\a=1,2,3}\dfrac{1-q_\a z}{1-q_\a^{-1}z}.
\end{equation}
In the $\psi-\psi$ and $\psi-x$ relations, the rational functions $g(\a z/w)$ are expanded in positive/negative powers of $z/w$, while the $x-x$ relations lead to quadratic relations between the modes $x_k^\pm$ when the currents are replaced by their expansions \ref{DIM_currents}.

By definition, the zero modes $\psi_{\mp\mu_\pm}^\pm$ are assumed to be invertible, and the relations \ref{def_DIM} imply that they are central. We note that the shift $\bmu$ does not enter in the algebraic relations \ref{def_DIM}, but instead it defines the asymptotic expansions \ref{DIM_currents} of the Cartan currents $\psi^\pm(z)$. When $\bmu=(0,0)$, the shifted algebra $\CE^\bmu$ reduces to the original quantum toroidal $\gl(1)$ algebra $\CE$, also called `Ding-Iohara-Miki algebra' \cite{Ding1997,Miki2007}, provided that we relax the condition $\psi_0^+\psi_0^-=1$ in the definition of $\CE$.\footnote{Comparing the presentation \ref{def_DIM} with the definition of the algebra $\CE$ used in the work of Feigin, Jimbo, Miwa and Mukhin (e.g. \cite{FJMM1}), we note that the Cartan currents have been twisted by the central element $\hg$ (denoted $C$ in \cite{FJMM1}) as $\psi^\pm(z)\to \psi^\pm(\hg^{\pm1/2}z)$, and the currents have also been rescaled which gives a different value for $\k$. Besides, we do not include the grading elements in the definition of the algebra. These small differences should be kept in mind when comparing the papers.} Like in the case of shifted quantum affine algebras, the algebras $\CE^{\bmu}$ and $\CE^{\bmu+(n,n)}$ and their properties only depend on $\mu_++\mu_-$. 

It is sometimes useful to introduce another mode expansion for the Drinfeld currents, namely
\begin{equation}
\psi^+(z)=z^{\mu_+}\psi_{-\mu_+}^+ e^{\sum_{k>0}z^{-k}a_{k}},\quad \psi^-(z)=z^{-\mu_-}\psi_{\mu_-}^- e^{-\sum_{k>0}z^{k}a_{-k}}.
% \psi^\pm(z)=z^{\pm\mu_\pm}\psi_{\mp\mu_\pm}^\pm e^{\pm\sum_{\pm k>0}z^{-k}a_{k}}.
\end{equation} 
Then, the modes $a_k$ and $x_k^\pm$ obey the following commutation relations
\begin{align}
\begin{split}\label{com_ak}
&[a_k,a_l]=(\hg^k-\hg^{-k})c_k\d_{k+l},\quad [a_k,x_l^\pm]=\pm\hg^{\mp |k|/2}c_k x_{l+k}^\pm,\\
&\text{with}\quad [g(z)]_{\pm}=\exp\left(\pm\sum_{k>0}z^{\mp k}c_{\pm k}\right),\quad c_k=-\dfrac1{k}\prod_{\a=1,2,3}(1-q_\a^k),
\end{split}
\end{align}
where the subscripts $[\cdots]_\pm$ denote the expansions in powers of $z^{\mp1}$. It follows from the property $g(z^{-1})=g(z)^{-1}$ that the coefficients $c_k$ are invariant under the sign flip $k\to-k$.

\begin{figure}
\begin{center}
\begin{tabular}{|c|c|c|c|}
\hline
Morphism $F$ & $F(x^\pm(z))$ & $F(\psi^\pm(z))$ & $F(\hg)$ \\
\hline
$\CT:\CE_{q_1,q_2}^\bmu\to \CE_{q_1,q_2}^\bmu$ & $z^{\mp1}x^\pm(z)$ & $\hg^{\mp1}\psi^\pm(z)$ & $\hg$ \\
$\CS^2:\CE_{q_1,q_2}^{(\mu_+,\mu_-)}\to \CE_{q_1,q_2}^{(\mu_-,\mu_+)}$ & $-x^\mp(z^{-1})$ & $\psi^\mp(z^{-1})$ & $\hg^{-1}$ \\
$\s_V:\CE_{q_1,q_2}^\bmu\to\CE_{q_1^{-1},q_2^{-1}}^\bmu$ & $-x^\mp(z)$ & $\psi^\pm(z)$ & $\hg^{-1}$ \\
$\s_H:\CE_{q_1,q_2}^{(\mu_+,\mu_-)}\to\CE_{q_1^{-1},q_2^{-1}}^{(\mu_-,\mu_+)}$ & $x^\pm(z^{-1})$ & $\psi^\mp(z^{-1})$ & $\hg$ \\
$\t_\o:\CE_{q_1,q_2}^\bmu\to \CE_{q_1,q_2}^\bmu$ & $x^\pm(\o z)$ & $\psi^\pm(\o z)$ & $\hg$ \\
$\bt_\o:\CE_{q_1,q_2}^\bmu\to \CE_{q_1,q_2}^\bmu$ & $\o^{\pm1}x^\pm(z)$ & $\psi^\pm(z)$ & $\hg$ \\
\hline
\end{tabular}
\end{center}
\caption{Brief summary of the main morphisms between shifted algebras}
\label{table_morph}
\end{figure}

\paragraph{Morphisms of algebras} The original algebra $\CE$ has a group $SL(2,\mZ)$ of automorphisms generated by Miki's automorphism $\CS$ \cite{Miki2007}, and the automorphism $\CT$ that sends $x^\pm(z)\to z^{\mp1}x^\pm(z)$ and $\psi^\pm(z)\to \hg^{\mp1}\psi^\pm(z)$ (with $\hg$ fixed). While $\CT$ remains an automorphism of $\CE^{\bmu}$, $\CS$ appears no longer well-defined. On the other hand, $\CS^2$ can still be defined as the algebra homomorphism $\CE^{(\mu_+,\mu_-)}\to \CE^{(\mu_-,\mu_+)}$ that sends $x^\pm(z)\to -x^\mp(z^{-1})$, $\psi^\pm(z)\to \psi^\mp(z^{-1})$ and $\hg\to \hg^{-1}$. In terms of modes, $x^\pm_k\to -x_{-k}^\mp$, $a_k\to -a_{-k}$, and $\psi_{-\mu_+}^+\leftrightarrow\psi_{\mu_-}^-$.

It is also instructive to investigate the shifted version of the algebra morphisms $\s_V,\s_H:\CE_{q_1,q_2}\to\CE_{q_1^{-1},q_2^{-1}}$ introduced in \cite{BFMZ}. We observe that $\s_V$ can be extended to the algebra morphisms $\CE_{q_1,q_2}^\bmu\to\CE_{q_1^{-1},q_2^{-1}}^\bmu$ sending $x^\pm(z)\to -x^\mp(z)$, $\psi^\pm(z)$ fixed and $\hg\to\hg^{-1}$. In the same way, $\s_H$ extends to the algebra morphisms $\CE_{q_1,q_2}^{(\mu_+,\mu_-)}\to\CE_{q_1^{-1},q_2^{-1}}^{(\mu_-,\mu_+)}$ sending $x^\pm(z)\to x^\pm(z^{-1})$, $\psi^\pm(z)\to\psi^\mp(z^{-1})$ and $\hg$ fixed. The extended morphisms still satisfy $\CS^2=\s_V\s_H=\s_H\s_V$.

Finally, the shifted algebra $\CE^\bmu$ is compatible with the two grading elements $d$ and $\bd$ acting on the generators as
\begin{equation}
[d,x_k^\pm]=-kx_k^\pm,\quad [d,\psi_k^\pm]=-k\psi_k^\pm,\quad [\bd,x_k^\pm]=\pm x_k^\pm,\quad [\bd,\psi_k^\pm]=0,
\end{equation}
and $[d,\hg]=[\bd,\hg]=0$. We note that the zero modes of the Cartan $\psi_{\mp\mu_\pm}^\pm$ which are central in $\CE^\bmu$ do not commute with the grading $d$ when $\bmu\neq(0,0)$. The adjoint action of these gradings define the automorphisms $\t_\o:e\to \o^d e\o^{-d}$ (resp. $\bt_\o:e\to \o^\bd e\o^{-\bd}$) that generates the rescaling $x^\pm(z)\to x^\pm(\o z)$, $\psi^\pm(z)\to\psi^\pm(\o z)$ (resp. $x^\pm(z)\to \o^{\pm1}x^\pm(z)$, $\psi^\pm(z)\to\psi^\pm(z)$). A brief summary of the various morphisms defined here is given in the table \ref{table_morph}. In fact, the definitions of these morphisms also apply to the shifted quantum affine $\sl(2)$ algebras by simply replacing $x^\pm(z),\psi^\pm(z),\hg$ with $X^\pm(z),\Psi^\pm(z),q^c$.

\paragraph{Coproduct} Like before, the Drinfeld coproduct can be extended into algebra morphisms $\D:\CE^{\bmu+\bmu'}\to\CE^{\bmu}\otimes\CE^{\bmu'}$ for any $\bmu,\bmu'\in\mZ\times\mZ$,
\begin{align}
\begin{split}\label{coproduct}
&\D(x^+(z))=x^+(z)\otimes 1+\psi^-(\hg_{(1)}^{1/2}z)\otimes x^+(\hg_{(1)}z),\\
&\D(x^-(z))=x^-(\hg_{(2)} z)\otimes \psi^+(\hg_{(2)}^{1/2}z)+1\otimes x^-(z),\\
&\D(\psi^\pm(z))=\psi^\pm(\hg_{(2)}^{\pm1/2}z)\otimes\psi^\pm(\hg_{(1)}^{\mp1/2}z),%\quad \D(\hg)=\hg\otimes\hg,\\%\quad \D(a_k)=a_k\otimes \hg^{-|k|/2}+\hg^{|k|/2}\otimes a_k,\\
% &,\quad \D(c)=c\otimes 1+1\otimes c,%\quad \D(\bc)=\bc\otimes1+1\otimes \bc.
\end{split}
\end{align}
with $\hg_{(1)}=\hg\otimes 1$, $\hg_{(2)}=1\otimes\hg$ and $\D(\hg)=\hg_{(1)}\hg_{(2)}$. Defining the Borel subalgebras $\CB^\bmu_\pm\subset\CE^\bmu$ generated by
\begin{align}
\begin{split}
&\CB^\bmu_+=\la x^+_k,\psi^-_{-l}, \hg^{\pm 1/2}\diagup (k,l)\in\mZ,\ l\leq \mu_-\ra,\\
&\CB^\bmu_-=\la x^-_k,\psi^+_{l}, \hg^{\pm 1/2}\diagup (k,l)\in\mZ,\ l\geq -\mu_+\ra,
\end{split}
\end{align}
we note that $\D(\CB_\pm^{\bmu+\bmu'})\subset\CB^{\bmu}\otimes\CB^{\bmu'}$. Using Miki's automorphism, it is possible to define another decomposition of the algebra $\CE$ into Borel subalgebras $\CB^\perp_\pm$,\footnote{Note that we use here a different notation than in \cite{FJMM2}.}
\begin{align}
\begin{split}
&\CB^\perp_+=\la x^+_k, x^-_{k+1} \psi^+_{k}, \hg^{\pm 1/2}\diagup k\in\mZ^{\geq0}\ra,\\
&\CB^\perp_-=\la x^+_{-k-1}, x^-_{-k}, \psi^-_{-k}, \hg^{\pm 1/2}\diagup k\in\mZ^{\geq0}\ra.
\end{split}
\end{align}
Since there is no definition of an equivalent to Miki's automorphism for shifted algebra, it appears difficult to extend this notion to the shifted algebras $\CE^\bmu$. Finally, it is easy to show that the triangular decomposition $\CE^\bmu=\CE_-^\bmu\otimes\CE_0^\bmu\otimes\CE_+^\bmu$ holds with
\begin{align}
\begin{split}
&\CE^\bmu_\pm=\la x^\pm_k\diagup k\in\mZ\ra,\quad \CE^\bmu_0=\la \psi^+_{k},\psi^-_l,(\psi_{-\mu_+})^{-1},(\psi_{\mu_-})^{-1}, \hg^{\pm c/2}\diagup (k,l)\in\mZ,\ k\geq -\mu_+,\ l\leq\mu_-\ra.
\end{split}
\end{align}

\paragraph{Asymptotic algebra} Consider the quantum toroidal $\gl(1)$ algebra without the central deformation $\hg$, i.e. $\CE_0=\CE\diagup\la\psi_0^+\psi_0^-=1,\hg=1\ra$. It is possible to mimic the construction of the asymptotic algebra $\CU^\text{as}$ and introduce the quantum toroidal algebra $\CE^\text{as}$ generated by the twisted Drinfeld currents $\tx^+(z)=x^+(z)$, $\tx^-(z)=\bar\g x^-(z)$, $\tpsi^\pm(z)=\bar\g\psi^\pm(z)$ with $\bar\g=(\psi_0^+)^{-1}$. As a result, $\tpsi_0^+=1$ and $\tpsi_0^-=\bar\g^2$. Since $\bar\g$ is central, the twisted currents still obey the relations \ref{def_DIM}. When $\bar\g$ is invertible, this algebra is isomorphic to $\CE_0$. On the other hand, when $\bar\g=0$ it realizes the shifted algebras $\CE^\bmu$ with antidominant shift, as e.g.
\begin{equation}
\CE^{(0,-p)}\simeq \CE^\text{as}\diagup\la \bar\g=\tpsi_{-1}^-=\cdots=\tpsi_{-p+1}^-=0\ra.
\end{equation} 

\subsection{Shifted representations}
The definition of shifted representation of the previous section extends trivially to the shifted quantum toroidal algebras $\CE^\bmu$. Hence, we can construct the shifted representations $\iota_P\rho$ and $\iota_P^\ast\rho$ of $\CE^{\bmu+\bmu_P}$ from a representation $\rho$ of $\CE^\bmu$ and a Laurent polynomial $P(z)$ of the form \ref{def_P}. The representation $\iota_P\rho$ (resp. $\iota_P^\ast\rho$) correspond to multiply the action of the current $x^+(z)$ (resp. $x^-(z)$) by $P(z)$ and the action of the Cartan currents $\psi^\pm(z)$ by $\rho(P(\hg^{\pm1/2}z))$ (resp $\rho(P(\hg^{\mp1/2}z))$). Note that, to help distinguish the representations of the algebra $\CE^\bmu$ and $\CU^\bmu$, we use the letter $\rho$ for the former and $\vrho$ for the latter. 

In the remainder of this subsection, we study the shifted representations obtained from the representations of $\CE$ entering in the algebraic engineering formalism, namely the vector, vertical Fock and horizontal Fock representations. For these representations, the central elements are parameterized by the operators $(c,\bc)$ as $\hg=q_3^{c/2}$ and $\psi_0^\pm=q_3^{\mp\bc/2}$, which breaks the $S_3$-invariance under the permutation of $(q_1,q_2,q_3)$ into an $S_2$-invariance under the exchange of $q_1,q_2$. A representation $\rho$ is said to be of level $(n,m)$ if $\rho((c,\bc))=(n,m)$, it will be denoted $\rho_v^{(n,m)}$ where $v\in\mC^\times$ is the weight. The three representations mentioned above correspond to $\rho_v^{(0,0)}$, $\rho_v^{(0,1)}$ and $\rho_u^{(1,n)}$ respectively.

\subsubsection{Vector representation}
The vector representation has been introduced in \cite{feigin2011quantum} and describes the action of the algebra $\CE$ on a module $\CV$ spanned by the states $\dket{k}$ parameterized by an integer $k\in\mZ$. This representation is not invariant under the exchange of the parameters $q_1$ and $q_2$. In string theory, it has been associated to D3-branes wrapping either $\mC_{\e_1}\times S^1$ or $\mC_{\e_2}\times S^1$ \cite{Zenkevich2018}. We will focus here on $q_2$, keeping in mind that another representation can be obtained by the exchange $q_1\leftrightarrow q_2$. The action of the Drinfeld currents on the basis $\dket{k}$ in the shifted representations $\iota_P\rho_v^{(0,0)}$ reads,\footnote{The factors $q_2^{k}$ and $q_2^{1-k}$ in the matrix elements of $x^+(z)$ and $x^-(z)$ resp. can be eliminated by the rescaling $\dket{k}\to q_2^{-k(k-1)/2}\dket{k}$ of the states. We kept these factors here to simplify the comparison with other representations.}
\begin{align}
\begin{split}
&x^+(z)\dket{k}=(1-q_1)(1-q_3)q_2^{k+1}\d(vq_2^k/z)P(vq_2^k)\dket{k+1},\\
&x^-(z)\dket{k}=(1-q_1)(1-q_3)q_2^{1-k}\d(vq_2^{k-1}/z)\dket{k-1},\\
&\psi^\pm(z)\dket{k}=P(z)\left[\dfrac{(z-vq_1q_2^k)(z-vq_1^{-1}q_2^{k-1})}{(z-vq_2^k)(z-vq_2^{k-1})}\right]_\pm\dket{k},
\end{split}
\end{align}
and for $\iota_P^\ast\rho_v^{(0,0)}$,
\begin{align}
\begin{split}
&x^+(z)\dket{k}=(1-q_1)(1-q_3)q_2^{k+1}\d(vq_2^k/z)\dket{k+1},\\
&x^-(z)\dket{k}=(1-q_1)(1-q_3)q_2^{1-k}P(vq_2^{k-1})\d(vq_2^{k-1}/z)\dket{k-1},\\
&\psi^\pm(z)\dket{k}=P(z)\left[\dfrac{(z-vq_1q_2^k)(z-vq_1^{-1}q_2^{k-1})}{(z-vq_2^k)(z-vq_2^{k-1})}\right]_\pm\dket{k}.
\end{split}
\end{align}

The vector representation is not a highest $\ell$-weight representation, and neither are the shifted representations in general. However, they reduce to highest/lowest weight representations for certain choices of $P(z)$. For instance, choosing $P(z)=1-vq_2^{-1}/z$, we observe that $\iota_P^\ast\rho_v(x^-(z))$ annihilates the state $\dket{0}$. As a result, $\iota_P^\ast\rho_v^{(0,0)}$ is now reducible, and it defines a lowest weight subrepresentation of $\CE^{(0,1)}$ on the module $\CL$ spanned by the states $\dket{k}$ with $k\geq0$ on which the Drinfeld currents act as follows,\footnote{Choosing more generally $P(z)=\a_Pz^k(1-vq_2^{-1}/z)$, we would find a lowest weight representation of $\CE^{(k,1-k)}$ on $\CL$.} 
\begin{align}
\begin{split}
&x^+(z)\dket{k}=(1-q_1)(1-q_3)q_2^{k+1}\d(vq_2^k/z)\dket{k+1},\\
&x^-(z)\dket{k}=(1-q_1)(1-q_3)q_2^{1-k}(1-q_2^{-k})\d(vq_2^{k-1}/z)\dket{k-1},\\
&\psi^\pm(z)\dket{k}=\left[\dfrac{(z-vq_2^{-1})(z-vq_1q_2^k)(z-vq_1^{-1}q_2^{k-1})}{z(z-vq_2^k)(z-vq_2^{k-1})}\right]_\pm\dket{k}.
\end{split}
\end{align}
This action can be interpreted as a two parameters version of the prefundamental representation defined in the previous section. In the same way, taking instead $P(z)=1-v/z$, $\iota_P\rho_v^{(0,0)}(x^+(z))$ annihilates $\dket{0}$, and $\iota_P\rho_v^{(0,0)}$ defines a highest weight subrepresentation of $\CE^{(0,1)}$ on the module $\bar\CL$ spanned by the states $\dket{k}$ with $k<0$.

Combining $\iota_{P_1}$ and $\iota_{P_2}^\ast$, it is also possible to define a finite dimensional representation for any shifted algebras $\CE^\bmu$ with $\bmu\geq2$. Since $\bmu\neq(0,0)$, it does not contradict the fact that $\CE$ has no non-trivial finite dimensional representation. For instance, choosing $P_2(z)=1-vq_2^{-1}/z$ and $P_1(z)=1-vq_2^N/z$, we find that the representation $\iota_{P_1}\iota_{P_2}^\ast\rho_v^{(0,0)}$ of $\CE^{(0,2)}$ has a subrepresentation of dimension $N+1$ on the module $\CM_N$ spanned by $\dket{k}$ with $k=0,1,\cdots N$,
\begin{align}
\begin{split}
&x^+(z)\dket{k}=(1-q_1^{-1})(1-q_3^{-1})q_2^k(1-q_2^{N-k})\d(vq_2^k/z)\dket{k+1},\\
&x^-(z)\dket{k}=-(1-q_1)(1-q_3)q_2^{1-k}(1-q_2^{-k})\d(vq_2^{k-1}/z)\dket{k-1},\\
&\psi^\pm(z)\dket{k}=(1-vq_2^{N}/z)(1-vq_2^{-1}/z)\left[\dfrac{(z-vq_1q_2^k)(z-vq_1^{-1}q_2^{k-1})}{(z-vq_2^k)(z-vq_2^{k-1})}\right]_\pm\dket{k}.
\end{split}
\end{align}
Rewriting the action of the Cartan currents on the vacuum as
\begin{equation}
\psi^\pm(u^{-1})\dket{0}=\left[(1-q_1vu)(1-q_3vu)\dfrac{P_D(q_2^{1/2}u)}{P_D(q_2^{-1/2}u)}\right]_\pm\dket{0},\quad P_D(u)=\prod_{j=0}^{N-1}(1-v q_2^{j+1/2}u)
\end{equation}
it is tempting to interpret $\iota_{P_1}\iota_{P_2}^\ast\rho_v^{(0,0)}$ as a two parameters version of the spin $N/2$ representations of the quantum affine $\sl(2)$ algebra acting on the Kirillov-Reshetikhin modules $\CM_N$. And indeed, we will show in the next section that the extra factors $(1-q_1vu)(1-q_3vu)$ drop in a certain limit, and the action reduces to the spin $N/2$ representation. These representations are special cases of the finite dimensional representations constructed in the next subsection.

\paragraph{Contragredient} The contragredient vector representation acts on the dual space $\CV^\ast$ spanned by the states $\dbra{k}$ for $k\in\mZ$ defined using the scalar product $\dbra{k}\!\!\!\dket{k'}=(n_k)^{-1}\d_{k,k'}$. Under the specific choice of norm $n_k=\bar n_k$ with $\bn_{k}=(-)^kq_2^{k^2}$, the contragredient action of the unshifted vector representation satisfies the property $\rho_v^{(0,0)}(e)^t=\rho_v^{(0,0)}(\s_V\cdot e)$ for $e\in \CE$, i.e. $x^\pm(z)$ acts on $\dbra{k}$ as $-x^\mp(z)$ does on $\dket{k}$. In the presence of shifts, we find two more remarkable values, 
\begin{equation}\label{def_Pik}
\bn_{k}^P=(-)^kq_2^{k^2}\Pi_k^P(v),\quad\bn_{k}^{P\ast}=(-)^kq_2^{k^2}\Pi_k^P(v)^{-1},\quad \text{with}\quad
\Pi_k^P(v)=
\begin{cases}
\prod_{j=1}^kP(vq_2^{j-1}), & k\geq0,\\
% 1, & k=0,\\
\prod_{j=1}^{-k}P(vq_2^{-j})^{-1}, & k<0.
\end{cases}
% \prod_{j=1}^kP(vq_2^{j-1}),\quad P_{-k<0}(v)=\prod_{j=1}^kP(vq_2^{-j})^{-1},
\end{equation} 
Depending on the choice of the scalar product, we observe different properties,
\begin{align}
\begin{split}
&n_k=\bn_k^P\implies \iota_P\rho_v^{(0,0)}(e)^t=\iota_P\rho_v^{(0,0)}(\s_V\cdot e),\\
&n_k=\bn_k^{P\ast}\implies \iota_P^\ast\rho_v^{(0,0)}(e)^t= \iota_P^\ast\rho_v^{(0,0)}(\s_V\cdot e),\\
&n_k=\bn_k\implies \iota_P\rho_v^{(0,0)}(e)^t=\iota_P^\ast\rho_v^{(0,0)}(\s_V\cdot e),\\
&n_k=\bn_k\implies \iota_P^\ast\rho_v^{(0,0)}(e)^t= \iota_P\rho_v^{(0,0)}(\s_V\cdot e).
\end{split}
\end{align}
We note that these particular values of $n_k$ depend on the weight of the representation, as well as the parameters $\nu_a$ and $\a_P$ entering in $P(z)$. From the point of view of string theory, this dependence is very natural since the modules are associated to different branes which are characterized by their position, encoded in the weight $v$, and the position of other branes ending on them, entering through the parameters $\nu_a$.

% For the application to algebraic engineering, we need to define a contragredient action on the dual states $\dbra{k}$. In this context, it is natural to choose a scalar product that depends on the weight of the representation and the parameters $\nu_a$ entering in $P(z)$. Indeed, since modules are associated to branes, they will depend on their position which is encoded in the weight $v$. They also depend on the position of other branes that end on them, which enter through the parameters $\nu_a$. Denoting the scalar product $\dbra{k}\!\!\!\dket{k'}=(n_k^P)^{-1}\d_{k,k'}$ for an $\iota_P\rho_v^{(0,0)}$ module (or $\dbra{k}\!\!\!\dket{k'}=(n_k^{P\ast})^{-1}\d_{k,k'}$ for an $\iota_P^\ast\rho_v^{(0,0)}$ module), we observe that when $n_k^P$ (resp. $n_k^{P\ast}$) coincide with the expression 
% then $x^\pm(z)$ act on $\dbra{k}$ as $-x^\mp(z)$ does on $\dket{k}$. This choice simplifies the definition of the intertwiners, and was used in the case of vertical Fock representations in \cite{Bourgine2017b}.

\subsubsection{Vertical Fock representation}
The quantum toroidal $\gl(1)$ algebra $\CE$ has two isomorphic representations on the free boson Fock space, called vertical and horizontal Fock representations. The equivalence between these two representations follows from Miki's automorphism \cite{Miki2007,Bourgine2018a}, and so it is expected to no longer hold for the shifted algebras. To put it differently, Miki's automorphism does not commute with the shift of the representations, and it is then possible to introduce inequivalent vertical and horizontal shifts. In this section, we focus on the shifted vertical Fock representations based on the Fock representation constructed in \cite{feigin2011quantum}. In this representation, the currents act on a basis $\dket{\l}$, labeled by the Young diagrams $\l$, of the Fock module $\CF_0$. In this absence of shift, this basis is identified with the Macdonald polynomials $P_\l$ (up to a normalization factor) using the isomorphism between the bosonic Fock space and the ring of symmetric polynomials in infinitely many variables \cite{Feigin2009a} (see also \cite{Bourgine2018a} for explicit formulas).

To each box $\Abox\in\l$ of coordinate $(i,j)$ we associate a complex parameter $\chi_\sAbox=vq_1^{i-1}q_2^{j-1}\in\mathbb{C}^\times$, called the \textit{box content}, with $v\in\mC^\times$ the weight of the representation. We also denote $A(\l)$ (resp. $R(\l)$) the set of boxes that can be added to $\l$ (or removed from $\l$). They define the rational functions
\begin{equation}\label{def_CYY}
\psi_{\l}(z)=\frac{\CY_{\l}(q_3^{-1}z)}{\CY_{\l}(z)},\quad \CY_{\l}(z)=\frac{\prod_{\sAbox\in A(\l)}1-\chi_\sAbox/z}{\prod_{\sAbox\in R(\l)}1-\chi_\sAbox/(q_3z)}.
\end{equation} 
Then, the action of the Drinfeld currents of $\CE^{\bmu_P}$ under the representation $\iota_P\rho_v^{(0,1)}$ reads
\begin{align}\label{def_vert_rep}
\begin{split}
&x^+(z)\dket{\l}=\k_+\sum_{\sAbox\in A(\l)}\delta(z/\chi_\sAbox)P(\chi_\sAbox)\res_{w=\chi_\sAbox}\dfrac1{w\CY_{\lambda}(w)}\dket{\l+\Abox},\\
&x^-(z)\dket{\l}=q_3^{-1/2}\k_- \sum_{\sAbox\in R(\l)}\delta(z/\chi_\sAbox)\res_{w=\chi_\sAbox}w^{-1}\CY_{\l}(q_3^{-1}w)\dket{\l-\Abox},\\
&\psi^\pm(z)\dket{\l}=q_3^{-1/2}\left[P(z)\psi_{\l}(z)\right]_\pm\dket{\l}.
\end{split}
\end{align}
Comparing with our previous conventions \cite{Bourgine2017b,Bourgine2018a}, we have introduced here the factors $\k_\pm$ to absorb the modification of the factor $\k$ in the definition of the algebra. They are such that $\k_+\k_-=(1-q_1q_2)\k/((1-q_1)(1-q_2))$, and we will take $\k_+=(1-q_3)q_3^{-1/2}$ and $\k_-=(1-q_3^{-1})q_3^{1/2}$ for definiteness.

The module $\CF_0$ is built from the lowest weight state $\dket{\vac}$ by the action of the modes $x_n^+$. In general, it is infinite dimensional, but finite dimensional modules can be constructed for shifted algebras $\CE^\bmu$ with $\bmu\geq2$. Indeed, when a zero $\nu_a$ of the polynomial $P(z)$ coincides with the content $\chi_\sAboxB=vq_1^{I-1}q_2^{J-1}$ for a box $\AboxB$ of coordinates $(I,J)$, the matrix elements $\dbra{\l+\AboxB}x^+_n\dket{\l}$ vanish. Thus, states $\dket{\l}$ with $\AboxB\in\l$ cannot be created by the action of products and sums of generators on the vacuum. A ``pit'', or a forbidden box, is introduced in the representation and it becomes reducible, with a subrepresentation on a module spanned by the states $\dket{\l}$ with $\AboxB\notin\l$. This construction bears some similarity with the one introduced in \cite{Bershtein2018} for the MacMahon modules, this point will be discussed in the next subsection. One should also emphasize that the parameters $q_1,q_2$ remain generic here. Once again, we will abuse the notation and keep the notation $\iota_P\rho_v^{(0,1)}$ for this subrepresentation.

For instance, choosing $P(z)=q_3^{1/2}(1-vq_1/z)$, a pit of coordinates $(2,1)$ is introduced. As a result, the representation can be restricted to a submodule spanned by states $\dket{\l}$ where the Young diagrams $\l$ consists of a single column. Such states can be identified with the states $\dket{k}\in\CL$ where $k$ is the height of the column. In fact, this subrepresentation coincides with the shifted vector representation $\iota_P^\ast\rho_v^{(0,0)}$ on $\CL$ defined in the previous subsection (up to a rescaling of the states by a power of $q_2$). This fact comes as no surprise given the construction of the Fock representation as a tensor product of vector representations presented in \cite{feigin2011quantum}.

In order to obtain a finite dimensional representation, we need to introduce at least two pits. Taking for instance the zeros of $P(z)$ at $\nu_1=vq_1^I$ and $\nu_2=vq_2^J$, we find a subrepresentation acting on the module of dimension $\left(\superp{I+J}{I}\right)$ spanned by the states $\dket{\l}$ where the Young diagrams are restricted to fit in the rectangle $I\times J$. The most general finite dimensional modules $\CM_\mu$ constructed in this way are labeled by a fixed Young diagram $\mu$ and consists of all Young diagrams $\l$ fitting inside $\mu$ (i.e. the columns height must satisfy $\l_i\leq\mu_i$), $\CM_\mu=\text{Span}\{\dket{\l},\l\subseteq\mu\}$. A representation of $\CE^{(k,|A(\mu)|-k)}$ on $\CM_\mu$ can be obtained from the vertical one $\rho_v^{(0,1)}$ as the shift $\iota_{P_\mu}\rho_v^{(0,1)}$ with 
\begin{equation}
P_\mu(z)=\a_\mu z^{k}\prod_{\sAboxB\in A(\mu)}(1-\chi_\sAboxB/z),
\end{equation} 
for any $\a_\mu\in\mC^\times$ and $k\in\mZ$. The action of the Drinfeld currents in the example of $\mu=21$ can be found in the appendix \ref{AppA}. These finite dimensional representations can be constructed only for the shifted algebras $\CE^\bmu$ with $\bmu\geq2$. In the case of the shifted quantum affine $\sl(2)$ algebras $\CU^\bmu$, the existence of finite dimensional representations requires $\bmu\geq0$ \cite{Hernandez2020}. Thus, it is tempting to conjecture that $\bmu\geq2$ is indeed a necessary condition for the existence of finite dimensional highest $\ell$-weight representations in the toroidal case.

In \cite{FJMM1}, a set of Bethe equations associated to the quantum toroidal $\gl(1)$ symmetry have been derived using the q-character technique. In the rational case, similar equations have been obtained, they describe the spectrum of the transfer matrices constructed in \cite{Litvinov2020} for the Intermediate Long Wave hierarchy. It would be particularly interesting to extend these two constructions to the case of finite dimensional shifted representations. In this way, we could define a class of spin chains with the spins degrees of freedom taking values in the finite modules $\CM_\mu$. For such representations, the R-matrix could be obtained using the method described in \cite{Hernandez2019} and based on the Maulik-Okounkov stable maps \cite{MO2012}.

The shifted vertical representations $\iota_P^\ast\rho_v^{(0,1)}$ of $\CE^{\bmu_P}$ will play a less important role here, but they can be analysed along the same lines. The action of the currents reads
\begin{align}\label{def_vert_rep_2}
\begin{split}
&\iota_P^\ast\rho_v^{(0,1)}(x^+(z))\dket{\l}=\k_+\sum_{\sAbox\in A(\l)}\delta(z/\chi_\sAbox )\res_{w=\chi_\sAbox}\dfrac1{w\CY_{\l}(w)}\dket{\l+\Abox},\\
&\iota_P^\ast\rho_v^{(0,1)}(x^-(z))\dket{\l}=q_3^{-1/2}\k_- \sum_{\sAbox\in R(\l)}\delta(z/\chi_\sAbox)P(\chi_\sAbox)\res_{w=\chi_\sAbox}w^{-1}\CY_{\l}(q_3^{-1}w)\dket{\l-\Abox},\\
&\iota_P^\ast\rho_v^{(0,1)}(\psi^\pm(z))\dket{\l}=q_3^{-1/2}\left[P(z)\psi_{\l}(z)\right]_\pm\dket{\l}.
\end{split}
\end{align}
For specific values of the zeros $P(z)$, $x^-(z)$ annihilates certain states of the module. Hence, when $P(z)$ is of the form 
\begin{equation}
P_\mu(z)=\a_\mu z^{k}\prod_{\sAboxB\in R(\mu)}(1-\chi_\sAboxB/z),
\end{equation} 
for a fixed Young diagram $\mu$, the representation can be reduced to a submodule spanned by states $\dket{\l}$ where $\l$ is restricted to contain $\mu$ as a sub-Young diagram.

\paragraph{Contragredient} Like in the case of vector representations, we also introduce the dual space $\CF_0^\ast$ spanned by $\dbra{\l}$, and the scalar product $\dbra{\l}\!\!\!\dket{\mu}=(n_\l)^{-1}\d_{\l,\mu}$. For the specific choice $n_\l=\bn_\l$ with  
\begin{equation}\label{def_bn_l}
\bn_\l=(-q_3^{1/2}v)^{-|\l|}\CN_{\l,\l}(1)^{-1}\prod_{\sAbox\in\l}\chi_\sAbox,
\end{equation} 
where $\CN_{\l,\mu}(\a)$ is the Nekrasov factor defined in equ. \ref{def_CN} below, the contragredient representation satisfies $\rho_v^{(0,1)}(e)^t=\rho_v^{(0,1)}(\s_V\cdot e)$, which leads to some simplification. When considering shifted representations, we find two more remarkable values,
\begin{equation}
\bn_\l^P=\Pi_\l^P(v)\bn_\l,\quad \bn_\l^{P\ast}=\Pi_\l^P(v)^{-1}\bn_\l,\quad \text{with}\quad \Pi_\l^P(v)=\prod_{\sAbox\in\l}P(\chi_\sAbox).
\end{equation} 
Then, we observe the following properties:
\begin{align}
\begin{split}
&n_\l=\bn_\l^P\implies \iota_P\rho_v^{(0,1)}(e)^t=\iota_P\rho_v^{(0,1)}(\s_V\cdot e),\\
&n_\l=\bn_\l^{P\ast}\implies \iota_P^\ast\rho_v^{(0,1)}(e)^t= \iota_P^\ast\rho_v^{(0,1)}(\s_V\cdot e),\\
&n_\l=\bn_\l\implies \iota_P\rho_v^{(0,1)}(e)^t=\iota_P^\ast\rho_v^{(0,1)}(\s_V\cdot e),\\
&n_\l=\bn_\l\implies \iota_P^\ast\rho_v^{(0,1)}(e)^t= \iota_P\rho_v^{(0,1)}(\s_V\cdot e).
\end{split}
\end{align}

% Like before, the normalization of the dual states $\dket{\l}$ follows from the choice of a scalar product, and we will denote $\dbra{\l}\!\!\!\dket{\mu}=(n_\l^P)^{-1}\d_{\l,\mu}$ for $\iota_P\rho_v^{(0,1)}$ and $\dbra{\l}\!\!\!\dket{\mu}=(n_\l^{P\ast})^{-1}\d_{\l,\mu}$ for $\iota_P^\ast\rho_v^{(0,1)}$. In \cite{Bourgine2017b}, the norm was chosen such that $x^\pm(z)$ act on $\dbra{\l}$ as $\s_V(x^\pm(z))=-x^\mp(z)$ does on $\dket{\l}$ (while the action of the Cartan remains the same). In the case of shifted representations, this condition remain satisfied if we choose 
% \begin{equation}
% \bn_\l^P=(-q_3^{1/2}v)^{-|\l|}\CN_{\l,\l}(1)\prod_{x\in\l}\chi_\sAbox  P(\chi_\sAbox ),\quad \bn_\l^{P\ast}=(-q_3^{1/2} v)^{-|\l|}\CN_{\l,\l}(1)\prod_{x\in\l}\chi_\sAbox  P(\chi_\sAbox )^{-1}.
% \end{equation} 
% where the expression of the . It introduces an extra dependence in the parameters $\nu_a$ for the module which is natural from the point of view of string theory.

% \cmt{Recaling of $x^\pm$ might have introduced some extract factor in $n_\l^P$... What if we use instead $\dbra{\l}\!\!\!\dket{\mu}=\d_{\l,\mu}$? Would we better understand what happens when $P(\chi_\sAbox)$ vanishes?}

\subsubsection{MacMahon module and prefundamental representation}
It is instructive to compare the shifted vertical Fock representations and the quantum toroidal $\gl(1)$ equivalents of the representations discussed in the previous section. The latter follow from the MacMahon modules $\CM(v,K)$ on which is defined a representation $\rho^{(\CM)}_v$ of $\CE$ of highest $\ell$-weight
\begin{equation}
\psi^\CM_\vac(z)=K^{1/2}\dfrac{1-K^{-1}v/z}{1-v/z},\quad\text{i.e.}\quad \psi^\pm(z)\dket{\vac}=\left[\psi^\CM_\vac(z)\right]_\pm\dket{\vac}.
\end{equation}
The module $\CM(v,K)$ is spanned by states labeled by plane partitions, $v$ denotes the weight of the representation and $\rho^{(\CM)}_v(\psi^\pm_0)=K^{\pm1/2}$ defines the action of the central elements together with $\rho^{(\CM)}_v(\hg)=1$ \cite{FJMM3d}. When $K=q_3^N$ with $N\in\mZ^{>0}$ (or, in fact, any $q_\a^N$ upon exchanging the quantum group parameters), the MacMahon representation becomes reducible, and defines a subrepresentation acting on $N$-tuple Young diagrams which is isomorphic to a tensor product of $N$ vertical Fock representations. These representations are the quantum toroidal analog of the Kirillov-Reshetikhin representations of quantum affine algebras. In particular, the vertical Fock module is the smallest highest $\ell$-weight $\CE$-module, it is the analog of the spin $1/2$ representation of the algebra $\CU$. It is also the analog of the fundamental representation defined in \cite{Hernandez2008} since the highest $\ell$-weight can be written as a ratio of Drinfeld polynomials of degree one,
\begin{equation}
\psi_\vac(u^{-1})=q_3^{-1/2}\dfrac{P(q_3^{1/2}u)}{P(q_3^{-1/2}u)},\quad P(u)=1-ua,
\end{equation}
with the rescaled weight $a=q_3^{1/2}v$.

These representations can be seen as pit representations, with the pit located at the position $(0,0,N)$ in the plane partition \cite{Bershtein2018}. More general pit representations can also be defined as subrepresentation after specializing to $K=q_1^{i-1}q_2^{j-1}q^{k-1}$ with $(i,j,k)\in\mZ^{\geq0}\times\mZ^{\geq0}\times\mZ^{\geq0}$ corresponding to the location of the pit. In parallel, the construction of shifted representations applies to the MacMahon representation as well, and it gives a different construction of pit representations. A clear advantage of the shifted representations is the possibility to introduce more than one pit, which is also necessary in order to define finite dimensional representations.

The negative prefundamental representation $\rho_v^{(\CL^-)}$ of the quantum toroidal $\gl(1)$ algebra is introduced in \cite{FJMM2} as a representation of the Borel subalgebra $\CB^\perp_+$. It can be obtained as the limit $K\to\infty$ of the representation $\iota_{K^{-1/2}}\rho^\CM$ of $\CB^\perp_+\subset\CE$ and it is characterized by the $\ell$-weight
\begin{equation}
\lim_{K\to\infty}K^{-1/2}\psi^\CM_\vac(z)=\dfrac1{1-v/z}.
\end{equation}
From the explicit form of the action of the Drinfeld currents in the MacMahon representation given in \cite{FJMM2}, it is easy to show that this prefundamental representation can be extended to the asymptotic algebra $\CE^\text{as}$, and to the shifted algebra $\CE^{(0,-1)}$. The prefundamental module, like the MacMahon module, is spanned by states labeled by plane partitions. In fact, the MacMahon representation can also be obtained from the prefundamental one as the shifted representation $\rho^{(\CM)}_v=\iota_P\rho_v^{\CL^-}$ with $P(z)=K^{1/2}(1-K^{-1}v/z)$.\footnote{To be really precise and recover the formulas given in \cite{FJMM2} for the matrix elements of the Drinfeld currents, we would need to take $\iota_{P_1}\iota_{P_2}^\ast\rho_v^{\CL^-}$ with $P_1(z)=1-K^{-1}v/z$ and $P_2(z)=K^{1/2}$ but these algebras are isomorphic when $K$ is invertible.} The vertical Fock representation is obtained as a subrepresentation after specialization to $K=q_3$. In this way, the notion of shifted representation provide a unifying point of view on the construction of pit representations.

\subsubsection{Horizontal Fock representation}
In the horizontal Fock representation \cite{Feigin2009a}, the Drinfeld currents' action is defined in terms of the vertex operators,\footnote{It is possible to use different normalizations of the Heisenberg algebra, sending $J_k\to r_kJ_k$ with $r_kr_{-k}=1$. The choice taken here identifies the elementary symmetric polynomials $p_k$ with $q_3^{k/2}(1-q_2^k)J_{-k}$ so that the zero mode $\eta_0^+$ in the expansion $\eta^+(z)=\sum_{k\in\mC}z^{-k}\eta_k^+$ acts as the Macdonald operator \cite{Feigin2009a}.}
\begin{align}
\begin{split}\label{def_eta_vphi}
&\eta^+(z)=e^{\sum_{k>0}\frac{z^k}{k}q_3^{k/2}(1-q_1^k)(1-q_2^k)J_{-k}}e^{-\sum_{k>0}\frac{z^{-k}}{k}q_3^{-k/2}J_k},\\
&\eta^-(z)=e^{-\sum_{k>0}\frac{z^k}{k}(1-q_1^{-k})(1-q_2^{-k})J_{-k}}e^{\sum_{k>0}\frac{z^{-k}}{k}J_k},\\
&\vphi^+(z)=e^{\sum_{k>0}\frac{z^{-k}}{k}q_3^{k/4}(1-q_3^{-k})J_k},\quad \vphi^-(z)=e^{-\sum_{k>0}\frac{z^k}{k}q_3^{k/4}c_kJ_{-k}},
\end{split}
\end{align}
with the coefficients $c_k$ given in \ref{com_ak}. The shifted representations $\iota_P\rho_u^{(1,n)}$ and $\iota_P^\ast\rho_u^{(1,n)}$ read simply
\begin{align}
\begin{split}
&\iota_P\rho_u^{(1,n)}:\quad x^+(z)=\k_+ uz^{-n}P(z)\eta^+(z),\quad x^-(z)=\k_-u^{-1}z^n\eta^-(z),\quad \psi^\pm(z)=q_3^{\mp n/2}P(q_3^{\pm1/4}z)\vphi^\pm(z),\\
&\iota_P^\ast\rho_u^{(1,n)}:\quad x^+(z)=\k_+ uz^{-n}\eta^+(z),\quad x^-(z)=\k_-u^{-1}z^nP(z)\eta^-(z),\quad \psi^\pm(z)=q_3^{\mp n/2}P(q_3^{\mp1/4}z)\vphi^\pm(z).
\end{split}
\end{align}

\subsubsection{Twisted Fock representations} 
The construction of the twisted Fock representation can also be extended to the shifted algebras $\CE^\bmu$. Indeed, under the replacement $x^+(z)\to x^+(\hg^{-1}z)$, and $\psi^\pm(z)\to\psi^\pm(\hg^{\pm1/2}z)$ in the algebraic relations \ref{def_DIM}, we observe that the central element $\textcolor{blue}{\hg}$ only appears in relations involving $\psi^+(z)$,
\begin{align}\label{twisted_DIM}
\begin{split}
&[\psi^\pm(z),\psi^\pm(w)]=0,\quad \psi^+(z)\psi^-(w)=\dfrac{g(\textcolor{blue}{\hg^2} z/w)}{g(z/w)}\psi^-(w)\psi^+(z),\\
&\psi^+(z)x^+(w)=g(\textcolor{blue}{\hg^2}z/w)x^+(w)\psi^+(z),\quad\psi^+(z)x^-(w)=g(z/w)^{-1}x^-(w)\psi^+(z),\\
&\psi^-(z)x^\pm(w)=g(z/w)^{\pm1}x^\pm(w)\psi^-(z),\quad [x^+(z),x^-(w)]=\k\left(\d(\textcolor{blue}{\hg^{-2}}z/w)\psi^+(w)-\d(z/w)\psi^-(z)\right),\\
&\prod_{\a=1,2,3}(z-q_\a^{\pm1}w)\ x^\pm(z)x^\pm(w)=\prod_{\a=1,2,3}(z-q_\a^{\mp1}w)\ x^\pm(w)x^\pm(z),\\
&\Sym_{z_1,z_2,z_3}\dfrac{z_2}{z_3}[x^\pm(z_1),[x^\pm(z_2),x^\pm(z_3)]]=0\quad \text{(Serre relations)},
\end{split}
\end{align}
Just like before, the $\hg$-dependence drops if we take $\psi^+(z)=0$.

The left twisted representation $\rho_{u,u',n}^{(LT)}$ of $\CE^{(-\infty,0)}$ acts on the charged Fock space $\CF$, it has two weights $u,u'\in\mC^\times$ and also depend on an integer $n$, and the Drinfeld currents read
\begin{align}
\begin{split}
&x^+(z)=ue^Qe^{\sum_{k>0}\frac{z^k}{k}\hg^k(1-q_1^k-q_2^k-q_3^k)J_{-k}}e^{-\sum_{k>0}\frac{z^{-k}}{k}\hg^{-k}J_k}(\hg z)^{-2J_0-n-1},\\
&x^-(z)=(u')^{-1}e^{-Q}e^{-\sum_{k>0}\frac{z^k}{k}(1-q_1^{-k}-q_2^{-k}-q_3^{-k})J_{-k}}e^{\sum_{k>0}\frac{z^{-k}}{k}J_k}(\hg z)^{2J_0+n-1},\\
&\psi^+(z)=0,\quad \psi^-(z)=(u/u')e^{-\sum_{k>0}z^k\hg^{k/2}c_kJ_{-k}}.
\end{split}
\end{align}
We note that $\psi_0^-=u/u'$ is indeed central. Under the trivial shift $\iota_P^\ast$ with $P(z)=u'/u$, this representation is equivalent to $\rho_{u,u,n}^{(LT)}$ which we denote simply $\rho_{u,n}^{(LT)}$. Note also that under the shift $Q\to Q+\a$, the weights transform as $u,u'\to e^\a u,e^\a u'$.

The right twisted representation $\rho_{u,u',n}^{(RT)}$ of $\CE^{(0,-\infty)}$ is such that $\psi^-(z)=0$, and the Drinfeld currents have the form
\begin{align}
\begin{split}
&x^+(z)=ue^{-Q}e^{-\sum_{k>0}\frac{z^k}{k}\hg^{-k}J_{-k}}e^{\sum_{k>0}\frac{z^{-k}}{k}\hg^{k}(1-q_1^k-q_2^k-q_3^k)J_k}(\hg^{-1}z)^{2J_0+n-1},\\
&x^-(z)=(u')^{-1}e^{Q}e^{\sum_{k>0}\frac{z^k}{k}J_{-k}}e^{-\sum_{k>0}\frac{z^{-k}}{k}(1-q_1^{-k}-q_2^{-k}-q_3^{-k})J_k}(\hg^{-1}z)^{-2J_0-n-1},\\
&\psi^+(z)=(u/u')e^{-\sum_{k>0}z^{-k}\hg^{k/2}c_kJ_{k}}.
\end{split}
\end{align}
The two representations are related by the property $\rho_{u,u',n}^{(LT)}(e)^\dagger=\rho_{u,u',n}^{(RT)}(\s_H\cdot e)$ for all $e\in\CE_{q_1,q_2}^{(-\infty,0)}$ with the usual definition $J_k^\dagger=J_{-k}$, $Q^\dagger=-Q$.

We note that the algebra $\CE^{(-\infty,0)}$ contains the Borel subalgebra $\CB_+\subset\CE$ spanned by the modes of $x^+(z)$, $\psi^-(z)$ and $\hg$. In the same way $\CE^{(0,-\infty)}$ contains the Borel subalgebra $\CB_-\subset\CE$ spanned by the modes of $x^-(z)$, $\psi^+(z)$ and $\hg$. Thus, the left (resp. right) twisted Fock representations define representations of the Borel subalgebra $\CB^+$ (resp. $\CB^-$). It would be interesting to study further these representations, and in particular the possible connections with shuffle algebras, symmetric polynomials or W-algebras.

\subsection{Shifted intertwiners} \label{sec_shift_intw}
In \cite{AFS}, Awata, Feigin and Shiraishi (AFS) introduced the intertwiners $\Phi$ and $\Phi^\ast$ of the algebra $\CE$ between a horizontal Fock representation and the tensor product of a vertical and a horizontal representation. We keep here the same notation $\Phi$ and $\Phi^\ast$ as in the previous section for the intertwiners, but it will be clear from the context to which algebra they refer. The matrix elements of these intertwiners were identified with the refined topological vertex, which serves as the starting point of the algebraic engineering formalism. These objects are uniquely determined (up to an overall normalization factor) by the following intertwining properties for $e\in\CE$,
\begin{align}\label{prop_intw}
\begin{split}
&\Phi\left(\rho_{v}^{(0,1)}\otimes\rho_u^{(1,n)}\ \D(e)\right)=\rho_{u'}^{(1,n+1)}(e)\Phi\\
&\Phi^\ast\rho_{u'}^{(1,n+1)}(e)=\left(\rho_{v}^{(0,1)}\otimes\rho_u^{(1,n)}\ \D'(e)\right)\Phi^\ast,
\end{split}
\end{align}
with the constraint $u'=-q_3^{1/2}uv$ and $\D'$ the opposite coproduct. The solution of these equations can be expanded in the vertical Macdonald basis as follows,
\begin{equation}\label{def_Phi}
\Phi[u,v,n]=\sum_{\l}n_\l\dbra{\l}\otimes \Phi_{\l}[u,v,n],\quad \Phi^\ast[u,v,n]=\sum_{\l}n_\l\dket{\l}\otimes\Phi_{\l}^\ast[u,v,n],
\end{equation} 
with the (arbitrary) coefficients $n_\l$ introduced earlier as a normalization of the scalar product $\dbra{\l}\!\!\!\dket{\mu}=n_\l^{-1}\d_{\l,\mu}$. The vertical components are vertex operators that can be built as normal-ordered products of the vertex operators $\eta^\pm(z)$ defined in \ref{def_eta_vphi},
\begin{align}\label{def_AFS}
\begin{split}
&\Phi_\l[u,v,n]=t_{\l}[u,v,n]\ :\Phi_\vac(v)\prod_{\sAbox\in\l}\eta^+(\chi_\sAbox ):,\quad \Phi_{\l}^\ast[u,v,n]=t_{\l}^\ast[u,v,n]\ :\Phi_\vac^\ast(v)\prod_{\sAbox\in\l}\eta^-(\chi_\sAbox ):,\\
\text{with:}\quad &t_\l[u,v,n]=(-q_3^{1/2}uv)^{|\l|}\prod_{\sAbox\in\l}\chi_\sAbox^{-n-1},\quad t_{\l}^\ast[u,v,n]= \frac{\bn_\l}{n_\l}u^{-|\l|}q_3^{-|\l|/2}\prod_{\sAbox\in\l}\chi_\sAbox^{n},\\
&\Phi_\vac(v)=e^{-\sum_{k>0}\frac{v^k}{k}q_3^{k/2}J_{-k}}e^{\sum_{k>0}\frac{v^{-k}q_3^{-k/2}}{k(1-q_1^{-k})(1-q_2^{-k})}J_k},\quad
\Phi_\vac^\ast(v)=e^{\sum_{k>0}\frac{v^k}{k}q_3^{k}J_{-k}}e^{-\sum_{k>0}\frac{v^{-k}}{k(1-q_1^{-k})(1-q_2^{-k})}J_k},
\end{split}
\end{align}
and $\bn_\l$ given in \ref{def_bn_l}. We note that the expression of $t_\l^\ast[u,v,n]$ does indeed simplify for the choice $n_\l=\bn_\l$.

As it turns out, the expressions \ref{def_AFS} can also be used to define intertwiners between shifted representations. Indeed,
the intertwiner $\Phi$ intertwines the representations $\iota_P \rho_v^{(0,1)}\otimes \rho_u^{(1,n)}$ and $\iota_P\rho_{u'}^{(1,n+1)}$ of the shifted algebras $\CE^{\bmu_P}\otimes\CE$ and $\CE^{\bmu_P}$. Furthermore, it is also an intertwiner between the representations $\rho^{(0,1)}_v\otimes\iota_P^\ast\rho_u^{(1,n)}$ and $\iota_P^\ast\rho_{u'}^{(1,n+1)}$ of $\CE\otimes\CE^{\bmu_P}$ and $\CE^{\bmu_P}$. The proof is done by explicit calculation, tracking carefully the extra factors $P(z)$ inside the intertwining equations \ref{prop_intw} expanded in the vertical basis. In the same way, one can show that $\Phi^\ast$ intertwines the representations $\iota_P^\ast\rho_{u'}^{(1,n+1)}$ and $\iota_P^\ast\rho_v^{(0,1)}\otimes\rho_u^{(1,n)}$ of $\CE^{\bmu_P}$ and $\CE^{\bmu_P}\otimes\CE$, but also the representations $\iota_P\rho_{u'}^{(1,n+1)}$ and $\rho_v^{(0,1)}\otimes\iota_P\rho_u^{(1,n)}$ of $\CE^{\bmu_P}$ and $\CE\otimes\CE^{\bmu_P}$. All these properties hold provided that the weights obey the constraint $u'=-q_3^{1/2} uv$.

New intertwining relations can also be produced if we allow for the modification of the factors $t_\l$ in the vertical components. We define the shifted intertwiners $\Phi^P$ and $\Phi^{P\ast}$ by the same formulas as in \ref{def_AFS} with $t_\l$ and $t_\l^\ast$ respectively replaced with
\begin{equation}\label{shift_tl}
t_\l^P[u,v,n]=t_{\l}[u,v,n]\prod_{\sAbox\in\l}P(\chi_\sAbox),\quad \text{and}\quad t_\l^{P\ast}[u,v,n]= t_\l^{\ast}[u,v,n]\prod_{\sAbox\in\l}P(\chi_\sAbox).
\end{equation} 
Then, $\Phi^P$ intertwines between the representations $\rho_v^{(0,1)}\otimes\iota_P\rho_u^{(1,n)}$ and $\iota_P\rho_{u'}^{(1,n+1)}$ of $\CE\otimes\CE^{\bmu_P}$ and $\CE^{\bmu_P}$, while $\Phi^{P\ast}$ intertwines between $\iota_P^\ast\rho_{u'}^{(1,n+1)}$ and $\rho_v^{(0,1)}\otimes\iota_P^\ast\rho_u^{(1,n)}$ of $\CE^{\bmu_P}$ and $\CE\otimes\CE^{\bmu_P}$. By the same procedure, we can also show that the new intertwiner $\Phi^P$ intertwines between the representations $\iota_P^\ast\rho_v^{(0,1)}\otimes\rho_u^{(1,n)}$ and $\iota_P\rho_{u'}^{(1,n+1)}$ of $\CE^{\bmu_P}\otimes\CE$ and $\CE^{\bmu_P}$, and $\Phi^{P\ast}$ intertwines between $\iota_P^\ast\rho_{u'}^{(1,n+1)}$ and $\iota_P\rho_v^{(0,1)}\otimes\rho_u^{(1,n)}$ of $\CE^{\bmu_P}$ and $\CE^{\bmu_P}\otimes\CE$. All these results are summarized in the table of figure \ref{table2}. We note that, by composing various shifts, it possible replace $\CE$ by a shifted version $\CE^\bmu$ with $\bmu\geq0$.

\begin{figure}
\begin{center}
\begin{tabular}{|c|c|c|}
\hline
Representations & Algebras & Intertwiner\\
\hline
$\iota_P \rho_v^{(0,1)}\otimes \rho_u^{(1,n)}\to\iota_P\rho_{u'}^{(1,n+1)}$ & $\CE^{\bmu_P}\otimes\CE\to\CE^{\bmu_P}$ & $\Phi$\\
$\iota_P^\ast \rho_v^{(0,1)}\otimes \rho_u^{(1,n)}\to\iota_P\rho_{u'}^{(1,n+1)}$ & $\CE^{\bmu_P}\otimes\CE\to\CE^{\bmu_P}$ & $\Phi^P$\\
$\rho^{(0,1)}_v\otimes\iota_P^\ast\rho_u^{(1,n)}\to\iota_P^\ast\rho_{u'}^{(1,n+1)}$ & $\CE\otimes\CE^{\bmu_P}\to\CE^{\bmu_P}$ & $\Phi$\\
$\rho_v^{(0,1)}\otimes\iota_P\rho_u^{(1,n)}\to\iota_P\rho_{u'}^{(1,n+1)}$ & $\CE\otimes\CE^{\bmu_P}\to\CE^{\bmu_P}$ & $\Phi^P$\\
$\iota_P^\ast\rho_{u'}^{(1,n+1)}\to\iota_P^\ast\rho_v^{(0,1)}\otimes\rho_u^{(1,n)}$ & $\CE^{\bmu_P}\to\CE^{\bmu_P}\otimes\CE$ & $\Phi^\ast$\\
$\iota_P^\ast\rho_{u'}^{(1,n+1)}\to\iota_P\rho_v^{(0,1)}\otimes\rho_u^{(1,n)}$ & $\CE^{\bmu_P}\to\CE^{\bmu_P}\otimes\CE$ & $\Phi^{P\ast}$\\
$\iota_P\rho_{u'}^{(1,n+1)}\to\rho_v^{(0,1)}\otimes\iota_P\rho_u^{(1,n)}$ & $\CE^{\bmu_P}\to\CE\otimes\CE^{\bmu_P}$ & $\Phi^\ast$\\
$\iota_P^\ast\rho_{u'}^{(1,n+1)}\to\rho_v^{(0,1)}\otimes\iota_P^\ast\rho_u^{(1,n)}$ & $\CE^{\bmu_P}\to\CE\otimes\CE^{\bmu_P}$ & $\Phi^{P\ast}$\\
\hline
\end{tabular}
\end{center}
\caption{Summary of the various intertwining relations for $\CE^\bmu$}
\label{table2}
\end{figure}

\paragraph{Back to quantum affine $\sl(2)$} The construction of shifted intertwiners also applies to the intertwiners defined in section \ref{sec_3d_intw}. Shifted intertwiners $\Phi^P$ and $\Phi^{P\ast}$ are defined by replacing the factors $\t_k[\nu,n]$ and $\t_k^\ast[\nu,n]$ entering in the expression \ref{expr_Phi3D} of the intertwiners by\footnote{We keep here the same notation as in \ref{def_Pik} since the quantities coincide for $k\geq0$ if we replace $q_2=\to q^2$.}
\begin{equation}
\t_k^P[u,\nu,n]=\Pi^P_k(\nu)\t_k[u,\nu,n],\quad \t_k^{P\ast}[\nu,n]=\Pi^P_k(\nu)\t_k^{\ast}[u,\nu,n],\quad \Pi^P_k(\nu)=\prod_{j=1}^k P(\nu q^{2j-2}).
\end{equation} 
The rules for choosing which intertwiner to insert between two given shifted representations are the same as in the toroidal case, they have been summarized in the table of figure \ref{table3}.

\begin{figure}
\begin{center}
\begin{tabular}{|c|c|c|}
\hline
Representations & Algebras & Intertwiner\\
\hline
$\iota_P \vrho_\nu\otimes \iota_{P_\nu}^\ast\vrho_{u,n}^{(LT)}\to\iota_P\vrho_{u,n}^{(LT)}$ & $\CU^{(0,-1)+\bmu_P}\otimes\CU^{(-\infty,1)}\to\CU^{(-\infty,0)+\bmu_P}$ & $\Phi$\\
$\iota_P^\ast \vrho_\nu\otimes \iota_{P_\nu}^\ast\vrho_{u,n}^{(LT)}\to\iota_P\vrho_{u,n}^{(LT)}$ & $\CU^{(0,-1)+\bmu_P}\otimes\CU^{(-\infty,1)}\to\CU^{(-\infty,0)+\bmu_P}$ & $\Phi^P$\\
$\vrho_\nu\otimes\iota_{P_\nu P}^\ast\vrho_{u,n}^{(LT)}\to\iota_P^\ast\vrho_{u,n}^{(LT)}$ & $\CU^{(0,-1)}\otimes\CU^{(-\infty,1)+\bmu_P}\to\CU^{(-\infty,0)+\bmu_P}$ & $\Phi$\\
$\vrho_\nu\otimes\iota_P\iota_{P_\nu}^\ast\vrho_{u,n}^{(LT)}\to\iota_P\vrho_{u,n}^{(LT)}$ & $\CU^{(0,-1)}\otimes\CU^{(-\infty,1)+\bmu_P}\to\CU^{(-\infty,0)+\bmu_P}$ & $\Phi^P$\\
$\iota_P^\ast\iota_{P_\nu^\ast}\vrho_{-u\nu,n+1}^{(LT)}\to\iota_P^\ast\vrho_\nu\otimes\vrho_{u,n}^{(LT)}$ & $\CU^{(-\infty,-1)+\bmu_P}\to\CU^{(0,-1)+\bmu_P}\otimes\CU^{(-\infty,0)}$ & $\Phi^\ast$\\
$\iota_P^\ast\iota_{P_\nu^\ast}\vrho_{-u\nu,n+1}^{(LT)}\to\iota_P\vrho_\nu\otimes\vrho_{u,n}^{(LT)}$ & $\CU^{(-\infty,-1)+\bmu_P}\to\CU^{(0,-1)+\bmu_P}\otimes\CU^{(-\infty,0)}$ & $\Phi^{P\ast}$\\
$\iota_{PP_\nu^\ast}\vrho_{-u\nu,n+1}^{(LT)}\to\vrho_\nu\otimes\iota_P\vrho_{u,n}^{(LT)}$ & $\CU^{(-\infty,-1)+\bmu_P}\to\CU^{(0,-1)}\otimes\CU^{(-\infty,0)+\bmu_P}$ & $\Phi^\ast$\\
$\iota_P^\ast\iota_{P_\nu^\ast}\vrho_{-u\nu,n+1}^{(LT)}\to\vrho_\nu\otimes\iota_P^\ast\vrho_{u,n}^{(LT)}$ & $\CU^{(-\infty,-1)+\bmu_P}\to\CU^{(0,-1)}\otimes\CU^{(-\infty,0)+\bmu_P}$ & $\Phi^{P\ast}$\\
\hline
\end{tabular}
\end{center}
\caption{Summary of the various intertwining relations for $\CU^\bmu$}
\label{table3}
\end{figure}

\section{Limit of the quantum toroidal representations}
% In this section, we investigate the relation between the quantum affine $\sl(2)$ algebra and the quantum toroidal $\gl(1)$ algebra in the limit $q_1\to\infty$. Since it is difficult to define this limiting procedure rigorously at the level of the algebra, we will focus instead on the limit of certain representations. As we will see, representations of shifted algebras appear naturally in this context.

% \subsection{Limit of the quantum toroidal \texorpdfstring{$\gl(1)$}{gl(1)} algebraic relations}
The algebra $\CE$ exhibits an $S_3$ symmetry under the permutation of the variables $q_\a$, but Fock representations break this symmetry because of the choice of parameterization of the central elements $\hg=q_3^{c/2}$ and $\psi_0^\pm=q_3^{\mp\bc/2}$. Thus, two inequivalent limits $q_1\to\infty$ can be considered, differing on whether central elements are kept fixed or not,
\begin{itemize}
 \item \textbf{Limit LI:} Send $q_1\to\infty$, hold $q_3$ fixed (and so $q_2\to0$).
 \item \textbf{Limit LII:} Send $q_1\to\infty$, hold $q_2$ fixed (and so $q_3\to0$).
\end{itemize}
The limit relevant to our application to 5d gauge theories on the omega-deformed background $\mC_{\e_1}\times\mC_{\e_2}\times S^1$ is the limit LII which corresponds to $\e_1\to-\infty$ with $\e_2$ fixed, this will be our main focus in this section.

In both cases (due to the $S_3$ symmetry), the structure function $g(z)$ entering in the currents relations \ref{def_DIM} reduces to the structure function $G(z)$ of the quantum affine $\sl(2)$ algebra defined in \ref{def_G}, with either $q^2=q_3$ (LI) or $q^2=q_2$ (LII). On the other hand, the parameter $\k$ entering in \ref{def_DIM} behaves as $q_1^{-1}\k\to -(1-q^2)$. Thus, assuming for now that $\hg$ tends to $q^c$, the algebraic relations between the Drinfeld currents of $\CE$ reduce in both limits to the relations \ref{def_Uqsl2} between the currents of the quantum affine $\sl(2)$ algebra, up to a harmless rescaling of the currents $x^\pm(z)$. To be explicit, sending in the limit LII $\psi^\pm(z)\to\Psi^\pm(z)$, $\hg\to q^c$ and $\a^\pm q_1^{\b^\pm}x^\pm(z)\to X^\pm(z)$ with $\a^\pm\in\mC^\times$, $\b^\pm\in\mZ$ such that $\a^+\a^-=q(1-q^2)^{-2}$ and $\b^++\b^-=-1$, the relations \ref{def_DIM} reproduce the relations \ref{def_Uqsl2}. We note that in this limit, the Serre relations simply follow from the $X^\pm-X^\pm$ exchange relations, and can be omitted in the definition of the algebra. It is readily observed that the Drinfeld coproduct \ref{coproduct}, the algebra morphisms of figure \ref{table_morph}, the shift of Cartan currents and the functors $\iota_P$ and $\iota_P^\ast$ all have a well-defined equivalent for the quantum affine $\sl(2)$ algebra obtained by taking the limit.

Unfortunately, our previous observation is a little crude since the limits only hold for the currents and not at the level of the modes. To put it otherwise, and as we will see by studying specific representations, the limits do not commute with the modes expansion of the Drinfeld currents. Hence, the zero modes $\psi_0^\pm$ are central in $\CE$, while the zero modes $\Psi_0^\pm$ are not.\footnote{This is also one of the reasons why Miki's automorphism \cite{Miki2007} cannot be extended to the quantum affine algebra in a simple way.} It might be possible to define rigorously this limit in the context of (q-deformed) vertex operator algebras or using category theory, but it is beyond the scope of this paper. On the other hand, it should be easier to formulate the limit as a functor between categories of representations. In this section, we take a first step in this direction by analysing the limit of specific representation. This analysis is important for the application to brane systems discussed in the next section.

\paragraph{Remark 1.} It is important to emphasize that the condition $\Psi_0^+\Psi_0^-=1$ for the Cartan modes of the algebra $\CU$ does not follow from the algebraic relations of the toroidal algebra in the limits LI/LII, and so in general we expect $\CE$ to relate to a \textbf{shifted} quantum affine $\sl(2)$ algebra $\CU^\bmu$.

\paragraph{Remark 2.} The limits LI and LII studied here are equivalent to the \textit{crystal limits} $q_1\to0$ (with either $q_2$ or $q_3$ fixed) under the algebra isomorphism $\s_V:\CE^\bmu_{q_1,q_2}\to\CE^\bmu_{q_1^{-1},q_2^{-1}}$ defined previously, and its quantum affine $\sl(2)$ counterpart. Indeed, at the level of the currents' exchange relations, the following diagram is commutative
% \tikzexternaldisable
\begin{equation}
\begin{tikzcd}
\CE_{q_1,q_2} \arrow{r}{\s_V} \arrow[swap]{d}{q_1\to\infty} & \CE_{q_1^{-1},q_2^{-1}} \arrow{d}{q_1^{-1}\to0} \\
\CU_q^\text{as} \arrow{r}{\s_V} & \CU_{q^{-1}}^\text{as},
\end{tikzcd}
\end{equation} 
% \tikzexternalenable
and the limits $q_1\to0$ coincide with the limits $q_1\to\infty$ upon the replacement $q\to q^{-1}$.

\paragraph{Generalization to higher ranks} This relation between quantum toroidal and quantum affine algebras is expected to extend to the quantum toroidal $\gl(p)$ and quantum affine $\sl(p+1)$ algebras. Indeed, the exchange relations of the Drinfeld currents defining the quantum toroidal algebras are written using a set of structure functions $g_{\o,\o'}(z)$ indexed by two nodes $\o,\o'\in\{0,1,\cdots,p-1\}$ of the affine Dynkin diagram (see e.g. the appendix A in \cite{Bourgine:2019phm}). These structure functions depend on two quantum group parameters $q,\k$,\footnote{Note that we use here a standard Kronecker delta, instead of the Kronecker delta modulo $p$ used in \cite{Bourgine:2019phm}.}
\begin{align}
\begin{split}
g_{\o,\o'}(z)&=\left(q^2\dfrac{1-q^2z}{1-q^{-2}z}\right)^{\d_{\o,\o'}}\left(q\dfrac{1-q^{-1}\k z}{1-q\k z}\right)^{\d_{\o,\o'-1}}\left(q\dfrac{1-q^{-1}\k^{-1} z}{1-q\k^{-1}}\right)^{\d_{\o,\o'+1}}\\
&\times\left(q\dfrac{1-q^{-1}\k z}{1-q\k z}\right)^{\d_{\o,p-1}\d_{\o',0}}\left(q\dfrac{1-q^{-1}\k^{-1} z}{1-q\k^{-1}}\right)^{\d_{\o,0}\d_{\o',p-1}}
\end{split}
\end{align}
Introducing a twist of the Drinfeld currents as $x_\o^\pm(z)\to x_\o^\pm(\k^\o z)$, $\psi_\o^\pm(z)\to\psi^\pm_\o(\k^\o z)$, the structure functions become
\begin{align}
\begin{split}
g_{\o,\o'}(\k^{\o-\o'}z)&=\left(q^2\dfrac{1-q^2z}{1-q^{-2}z}\right)^{\d_{\o,\o'}}\left(q\dfrac{1-q^{-1}z}{1-qz}\right)^{\d_{\o,\o'-1}}\left(q\dfrac{1-q^{-1} z}{1-qz}\right)^{\d_{\o,\o'+1}}\\
&\times\left(q\dfrac{1-q^{-1}\k^p z}{1-q\k^p z}\right)^{\d_{\o,p-1}\d_{\o',0}}\left(q\dfrac{1-q^{-1}\k^{-p} z}{1-q\k^{-p}z}\right)^{\d_{\o,0}\d_{\o',p-1}}
\end{split}
\end{align}
Further sending $\k\to\infty$ (or $\k\to0$), we thus recover the structure functions of the quantum affine $\sl(p+1)$ algebra up to the extra factor $q^{-\d_{\o,p-1}\d_{\o',0}+\d_{\o,0}\d_{\o',p-1}}$ (resp. $q^{\d_{\o,p-1}\d_{\o',0}-\d_{\o,0}\d_{\o',p-1}}$). It is also likely that the relation can be interpreted in the language of instanton and vortex moduli spaces of supersymmetric gauge theories. In fact, the algebraic engineering has been extended in \cite{Bourgine:2019phm} to 5D $\CN=1$ gauge theories on orbifolds $\left(\mC_{\e_1}\times\mC_{\e_2}\times S_1\right)\diagup\mZ_p$. In addition to the parameter $p$, these algebras are labeled on two integers $(\nu_1,\nu_2)\in\mZ_p\times\mZ_p$ associated to the $\mZ_p$-orbifold action on the spacetime coordinates
\begin{equation}
(z_1,z_2)\in\mC_{\e_1}\times\mC_{\e_2}\to (e^{2i\pi\frac{\nu_1}{p}}z_1,e^{2i\pi\frac{\nu_2}{p}}z_2).
\end{equation}
In the case $(\nu_1,\nu_2)=(1,0)$ corresponding to a surface defect, the structure functions of these algebra reduce to those of the quantum toroidal $\gl(p)$ algebra (up to an extra power of $q_2$) with the identification of the parameters $q_2=q^2$, $q_1=q^{-1}\k$. Thus, the limit $\k\to\infty$ corresponds to send $q_1\to\infty$ with $q_2$ fixed, it coincides with the one studied here. It would be interesting to analyse further this limit for specific representations, but we will keep this study for a future work and instead focus here on the case of the quantum toroidal $\gl(1)$ algebra.

\subsection{Limit of the vertical Fock representation}
It has been observed in the previous section that the shifted vertical Fock representation $\iota_P\rho_v^{(0,1)}$ by $P(z)=\a_P(1-z/(vq_1))$ can be reduced to the module $\CL$, with the identification of states $\dket{k}$ with single column Young diagrams of $k$ boxes. For such Young diagrams, the functions $\CY_\l(z)$ and $\psi_\l(z)$ simplify into
\begin{equation}\label{CY_k}
\CY_k(z)=\dfrac{(z-vq_1)(z-vq_2^k)}{z(z-vq_1q_2^k)},\quad \psi_k(z)=\dfrac{(z-vq_1q_2^k)(z-vq_1^{-1}q_2^{k-1})(z-vq_2^{-1})}{(z-vq_1)(z-vq_2^k)(z-vq_2^{k-1})}.
\end{equation}
As a result, the action of the Drinfeld currents on the states $\dket{k}$ can be written explicitly as
\begin{align}\label{act_ip_vert}
\begin{split}
&\iota_P\rho_v^{(0,1)}(x^+(z))\dket{k}=\a_Pq_3^{-1/2}(1-q_1^{-1})(1-q_3)q_2^k\d(vq_2^k/z)\dket{k+1},\\
&\iota_P\rho_v^{(0,1)}(x^-(z))\dket{k}=(1-q_1^{-1})(1-q_3^{-1})(1-q_2^{-k})\d(vq_2^{k-1}/z)\dket{k-1},\\
&\iota_P\rho_v^{(0,1)}(\psi^\pm(z))\dket{k}=\dfrac{\a_P}{-vq_1} q_3^{-1/2}\left[\dfrac{(z-vq_1q_2^k)(z-vq_1^{-1}q_2^{k-1})(z-vq_2^{-1})}{(z-vq_2^k)(z-vq_2^{k-1})}\right]_\pm\dket{k}.
\end{split}
\end{align}
We need to choose $\a_P=q_3^{1/2}$ to counteract the fact that $\psi_0^-=q_3^{-1/2}$ is divergent in the limit LII. Then, the action \ref{act_ip_vert} simply reduces to the prefundamental representation \ref{prefund} with
\begin{align}
\begin{split}
&\dfrac{q}{1-q^2}\iota_P\rho_v^{(0,1)}(x^+(z))\tox_{LII}\vrho_v(X^+(z)),\\
&-\dfrac{q_1^{-1}}{1-q^2}\iota_P\rho_v^{(0,1)}(x^-(z))\tox_{LII}\vrho_v(X^-(z)),\\
&\iota_P\rho_v^{(0,1)}(\psi^\pm(z))\tox_{LII} \vrho_v(\Psi^\pm(z)).
\end{split}
\end{align}
There is an important subtlety with the order of limits here as we have to take first the limit LII before performing the expansion of the rational fraction between the brackets $\left[\cdots\right]_\pm$ that define the action of $\psi_\pm(z)$. This concur with the previous observation that the limits are defined only for the currents and not for the modes. For instance, the modes $a_n$ in the shifted representation act as
\begin{equation}
\iota_P\rho_v^{(0,1)}(a_n)\dket{k}=\dfrac{v^n}{n}(1-q_3^n)\left[(1+q_2^{-n}-q_1^n-q_3^n)q_2^{kn}-q_2^{-n}\right]\dket{k},
\end{equation} 
and this expression diverges in the limit LII.

\paragraph{Performing the limit without shift} The prefundamental representation belongs to the category of modules $\CO_\bmu$ of the shifted algebra $\CU^\bmu$ defined in \cite{Hernandez2020}, with $\bmu=(0,-1)$. Modules in this category are characterized by the fact that they decompose into a direct sum of finite dimensional weight spaces. It was proven in \cite{Hernandez2020} that the simple modules in this category are fully characterized by their $\ell$-weight, i.e. the eigenvalue of the Cartan currents $\psi_\a^\pm(z)$ on the highest weight state $\dket{\vac}$. The same notion also exists for quantum toroidal algebras \cite{Hernandez2008}, and a classification of simple modules in terms of highest $\ell$-weight can probably be established for the shifted quantum toroidal algebras as well. This observation prompts us to propose a formal notion of limit. Let $\CM$ be a highest $\ell$-weight module in the category $\CO_{\CA}$ of a quantum group $\CA$ on which it acts by the representation $\rho_{\CA}$ such that $\rho_{\CA}(\psi_\a^\pm(z))\dket{\vac}=\left[\Psi_\a^{(\CA)\pm}(z)\right]_\pm\dket{\vac}$ with rational functions $\Psi_\a^{(\CA)\pm}(z)$. We say that the highest $\ell$-weight representation $\rho_{\CB}$ of the quantum group $\CB$ is the formal limit of $\rho_{\CA}$ if $\CB$ acts on $\CM$ with $\rho_{\CB}$, and if its highest $\ell$-weights are obtained by taking the limit of the rational functions $\Psi_\a^{(\CA)\pm}(z)$,
\begin{equation}
\rho_{\CB}(\psi_\a^\pm(z))\dket{\vac}=\left[\Psi_\a^{(\CB)\pm}(z)\right]_\pm\dket{\vac},\quad \Psi_\a^{(\CB)\pm}(z)=\lim\Psi_\a^{(\CA)\pm}(z).
\end{equation} 
We note that $\rho_\CB$ is not a simple representation in general, even if $\rho_\CA$ is. In practice, we are simply proposing to reverse the order of the limits, taking first the limit of the rational functions and then the mode expansions in the spectral parameter $z$.

This notion of limit is also natural from the point of view of cohomological Hall algebras. In this context, the $\ell$-weight are ratios of the Y-observables, a set of rational functions that encodes the variation of the equivariant character (see appendix \ref{AppCOHA}). These functions also determine the matrix elements of the currents $x^\pm(z)$ which are expressed in terms of their residues (as in \ref{def_vert_rep} or \ref{COHA_vertex}). With this definition, the formal limit corresponds to the limit of the equivariant character which translates into a limit of the Y-observables.

The limit LII of the vertical Fock representation $\rho_v^{(0,1)}$ on the Fock space $\CF_0$ can be analyzed along these lines. The limit of the Y-observables and $\ell$-weights is
\begin{align}
\begin{split}
&\CY_\l(z)\tox_{LII}\dfrac{z-vq^{2\l_1}}{zq^{2\l_1}},\quad \CY_\l(q_3^{-1}z)\tox_{LII}q^{-2\l_2}\dfrac{z-vq^{2\l_2-2}}{z-vq^{2\l_1-2}},\\
&\psi_\l(z)=\dfrac{\CY_\l(q_3^{-1}z)}{\CY_\l(z)}\tox_{LII}q^{2(\l_1-\l_2)}\dfrac{z(z-vq^{2\l_2-2})}{(z-vq^{2\l_1})(z-vq^{2\l_1-2})},
\end{split}
\end{align}
where $\l_1$ and $\l_2$ are the first two columns of the Young diagram $\l$. We recognize in the limit of $\psi_\l(z)$ the expression of the eigenvalue of the Cartan $\Psi^\pm(z)$ acting on the state $\dket{k=\l_1-\l_2}$ in the prefundamental representation of weight $v q^{2\l_2}$. Decomposing the Young diagram labeling the basis $\dket{\l}$ of the Fock space $\CF_0$ as $\l=(\l_1,\l_2,\mu)$ where $\mu$ is a Young diagram with at most $\l_2$ boxes in the first column, we can identify the state $\dket{\l}$ with a state $\dket{k=\l_1-\l_2}$ in a copy $\CL^{(\l_2,\mu)}$ of the module $\CL$. It leads to the decomposition of $\CF_0$ into an infinite direct sum of prefundamental modules
\begin{equation}
\CF_0=\bigoplus_{\l_2\geq0} \bigoplus_{\mu\diagup\mu_1\leq\l_2}\CL^{(\l_2,\mu)}.
\end{equation}
Using this decomposition, we can define an action of $\CU^\bmu$ on $\CF_0$. The limit of the $\ell$-weight implies
\begin{equation}
\iota_{q_3^{1/2}}\rho_v^{(0,1)}\tox_{LII}\bigoplus_{\l_2\geq0}\bigoplus_{\mu\diagup\mu_1\leq\l_2}\vrho_{vq^{2\l_2}},
\end{equation} 
where the shift $\iota_{q_3^{1/2}}$ removes the divergence of $\rho_v^{(0,1)}(\psi_0^+)=q_3^{-1/2}$ while ensuring that $\iota_{q_3^{1/2}}\rho_v^{(0,1)}(\psi_0^-)=q_3\to0$. The examination of the structure of poles and zeros for the Y-observables confirm this analysis. Indeed, the zero of $\CY_\l(z)$ at $z=vq^{2\l_1}$ corresponds to the possibility for $X^+(z)$ to add a box in the first column. When $\l_1=\l_2$, $\CY_\l(q_3^{-1}z)\tox_{LII}q_2^{-\l_2}$ has no pole and thus $X^-(z)$ annihilates the corresponding state. On the other hand, if $\l_1>\l_2$, $\CY_\l(q_3^{-1}z)$ has a unique pole at $z=vq_2^{\l_1-1}$ in the limit LII which corresponds to the possibility of $X^-(z)$ removing the top box in the first column. 

\paragraph{Limit LI} The formal limit LI of the vertical Fock representation can be analyzed in the same way. Decomposing the set of boxes that can be added to the Young diagram into $A(\l)=A_-(\l)\cup A_0(\l)\cup A_+(\l)$ according to their position below/on/above the diagonal, and similarly for the boxes in $R(\l)$ that can be removed, we can write the limit of the $\ell$-weights as\footnote{This expression follows from the limit of the functions $\CY_\l(z)$,
\begin{equation}
\CY_\l(z)\simx_{LI}\ z^{|R_-(\l)|-|A_-(\l)|}y_\l \dfrac{\prod_{\sAbox\in A_0(\l)}(1-\chi_\sAbox/z)}{\prod_{\sAbox\in R_0(\l)}(1-\chi_\sAbox/(q_3z))},\quad y_\l=\dfrac{\prod_{\sAbox\in A_-(\l)}(-\chi_\sAbox)}{\prod_{\sAbox\in R_-(\l)}(-\chi_x/q_3)}.
\end{equation}
We also note that $|A_-(\l)|-|R_-(\l)|$ is one if $|A_0(\l)|=0$ and zero otherwise.}
\begin{equation}
\psi_\l(z)\tox_{LI} q_3^{|A_-(\l)|-|R_-(\l)|} \prod_{\sAbox\in A_0(\l)}\dfrac{1-q_3\chi_\sAbox/z}{1-\chi_\sAbox/z}\prod_{\sAbox\in R_0(\l)}\dfrac{1-q_3^{-1}\chi_\sAbox/z}{1-\chi_\sAbox/z}.
\end{equation} 
Thus, in this limit the operators $x^\pm(z)$ are only able to add/remove boxes on the diagonal of the Young diagram.
Since $|A_0(\l)|+|R_0(\l)|\leq 1$, Young diagrams $\l$ fall into three categories. In the first category, $|A_0(\l)|=|R_0(\l)|=0$ and no box can be added nor removed from the Young diagram, both $x^\pm(z)$ annihilates the corresponding states, they form a trivial one-dimensional representation for the quantum affine $\sl(2)$ algebra. Diagrams in the second category satisfy $|A_0(\l)|=1$ and $|R_0(\l)|=0$, a box can be added on the diagonal but none can be removed. Thus, they are annihilated by $x^-(z)$, they form the vacuum state of a two-dimensional representation. The excited state is obtained by the action of $x^+(z)$, the corresponding partition obviously belong to the third category, namely $|A_0(\l)|=0$ and $|R_0(\l)|=1$ since a box can be removed from the diagonal but none can be added.

To identify the representation, we focus on a pair of diagrams $(\l,\mu)$ such that $\mu=\l+\Abox$ with $\Abox\in A_0(\l)$. In the limit LI, we find
\begin{equation}
\psi_\l(z)\tox_{LI} \dfrac{z-vq^{-2d(\l)+2}}{z-vq^{-2d(\l)}},\quad \psi_\mu(z)\tox_{LI} q^2\dfrac{z-vq^{-2d(\l)-2}}{z-vq^{-2d(\l)}},
\end{equation} 
where $d(\l)$ is the number of boxes on the diagonal of $\l$. Including the prefactor $q_3^{-1/2}=q^{-1}$ in the action of the Cartan in \ref{def_vert_rep}, we recover the $\ell$-weight of the representation \ref{KR} with $N=1$, i.e. the spin 1/2 representation of $\CU$, with weight $vq^{-2d(\l)}$. Thus, ignoring trivial representations, the vertical Fock representation $\rho_v^{(0,1)}$ of $\CE$ tends in the formal limit LI to an infinite sum of spin 1/2 representations of $\CU$.
% \begin{equation}
% \rho_v^{(0,1)}\tox_{LI}\bigoplus_{d=0}^\infty \bigoplus_{\l_d^+,\l_d^-} \vrho_{vq^{-2d}}^{L(M_2)}
% \end{equation} 
% where $\l_d^+$ (resp. $\l_d^-$) are Young diagrams with at most $d$ columns (rows). They enter in the decomposition of $\l$ into a $d\times d$ rectangle and the two extra Young diagrams $\l_d^+,\l_d^-$.

\subsection{Limit of the vector representation}
\begin{figure}
% \tikzexternaldisable
% \begin{equation}
% \begin{tikzcd}[scale cd=.8, xsep=1pt]
% \rho_v^{(0,1)}(\CE)\acts \CF  \arrow{rd}{P(z)\propto 1-vq_1/z} & & \rho^{(0,0)}_v(\CE)\acts\CV \arrow{ld}{P'(z)\propto 1-vq_2^{-1}/z} \arrow{rd}{\text{LII}} & \\
%  & \Superp{\iota_P\rho_v^{(0,1)}(\CE^{(0,1)})\acts \CL}{\simeq\iota_{P'}^\ast\rho_v^{(0,0)}(\CE^{(0,1)})\acts \CL} \arrow{rd}{\text{LII}}  & & \vrho_v^{(0)}(\CU^{(-1,-1)})\acts\CV \arrow{ld}{P''(z)\propto 1-zq^2/v}\\
%  & & \Superp{\vrho_v(\CU^{(0,-1)})\acts\CL}{\simeq\iota_{P''}^\ast\vrho^{(0)}_v(\CU^{(0,-1)})\acts\CL } & \\
% \end{tikzcd}
% \end{equation}
% \tikzexternalenable
\begin{center}
\includegraphics[width=170mm]{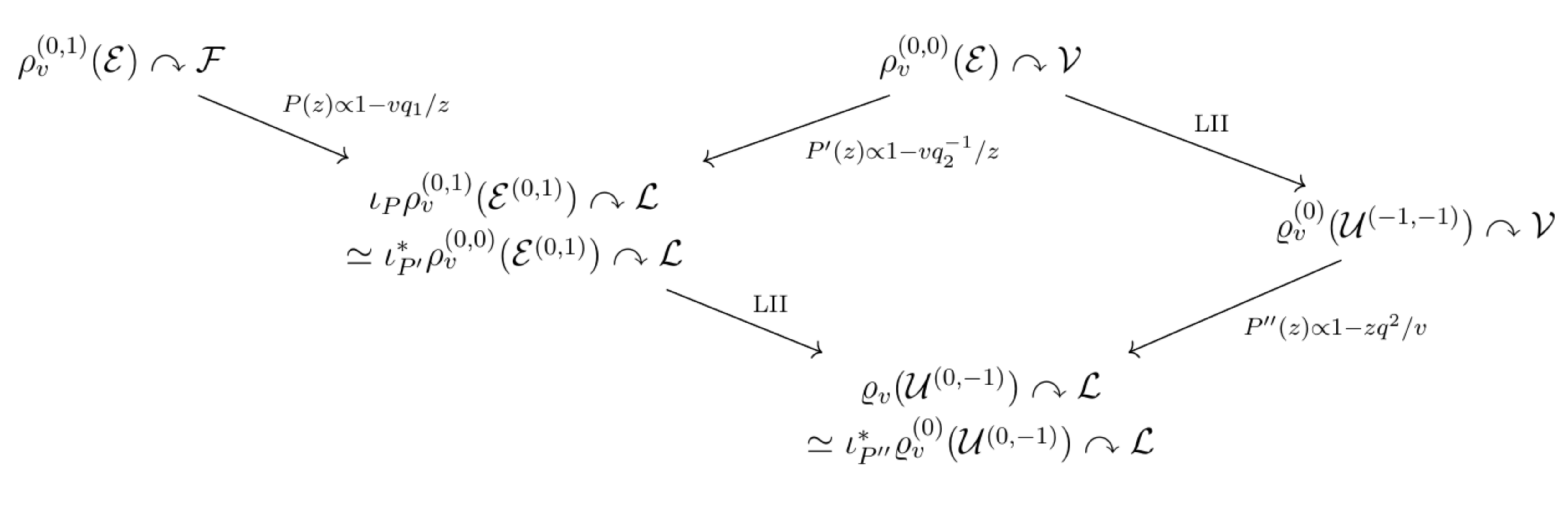}
\end{center}
\caption{Interplay between different shifted representations}
\label{fig_limit}
\end{figure}

The vector representation $\vrho_v^{(0)}$ of $\CU^{(-1,-1)}$ defined in section \ref{sec_vect_CU} can be obtained as the limit LII of the vector representation $\iota_P\rho_v^{(0,0)}$ of $\CE$ with a trivial shift $P(z)=q_1^{-1}$ to absorb the divergence of the $\ell$-weights.
\begin{align}
\begin{split}
&\dfrac{q^{-1}}{1-q^2}\iota_{q_1^{-1}}\rho_v^{(0,0)}(x^+(z))\tox_{LII}\vrho_v^{(0)}(X^+(z)),\\
&\dfrac{q_1^{-1}q^2}{1-q^2}\iota_{q_1^{-1}}\rho_v^{(0,0)}(x^-(z))\tox_{LII}\vrho_v^{(0)}(X^-(z)),\\
&\iota_{q_1^{-1}}\rho_v^{(0,0)}(\psi^\pm(z))\tox_{LII} \vrho_v^{(0)}(\Psi^\pm(z)).
\end{split}
\end{align}
We note that the prefundamental representation analyzed in the previous subsection can also be recovered from this action by the shift $\vrho_v=\iota_P^\ast\vrho^{(0)}_v$ by $P(z)=q^{-2}(1-zq^2/v)$. The figure \ref{fig_limit} summarizes the interplay between shifted vertical Fock, vector and prefundamental representations. It is observed that the shift $\mu_++\mu_-\in\mZ$ decreases by two in the limit LII.

\paragraph{Remark.} It is easy to take the formal limit LI of the vector representation $\rho_v^{(0,0)}$ but it is not very interesting. We observe that the eigenvalues of the Cartan currents $\psi^\pm(z)$ on the states $\dket{k}$ tend to one, unless $k=0$ or $k=1$, in which case
\begin{equation}
\psi^\pm(z)\dket{0}=q^{-2}\dfrac{z-vq^2}{z-v}\dket{0},\quad \psi^\pm(z)\dket{1}=\dfrac{z-vq^{-2}}{z-v}\dket{1}.
\end{equation} 
Comparing with the formula \ref{KR}, it suggests that the vector representation tend to the direct sum of the spin 1/2 representation of $\CU$ (with $\CM_1=\Span\{\dket{0},\dket{1}\}$) and an infinite sum of trivial one-dimensional representations.

\subsection{Limit of the horizontal Fock representation}
The twisted Fock representations of the shifted algebras $\CU^{(-\infty,0)}$ can be interpreted as a certain limit of the horizontal Fock representation $\rho_u^{(1,n)}$ for the toroidal algebra $\CE$. To remove the divergences due to the central element $\rho^{(1,n)}(\hg^{-1})=q_3^{-1/2}$, we examine the action of the left twisted currents $x^+(\hg^{-1}z)$, $x^-(z)$ and $\psi^-(\hg^{-1/2}z)$, omitting $\psi^+(z)$. The limit LII of the current $x^-(z)$ is easy to perform, and we find $\vrho_{u,u,n}^{(LT)}(X^-(z))$ up to a simple prefactor reflecting the normalization of algebraic relations. On the other hand, the limit of the current 
$x^+(\hg^{-1}z)$ appears ill-defined at first. In fact, we need to introduce the decomposition 
\begin{align}
\begin{split}
&x^+(q_3^{-1/2}z)=\tk uq_3^{n/2}z^{-n}\eta_A(z)\eta_B(z),\quad \eta_A(z)=e^{-\sum_{k>0}\frac{z^k}{k}q_1^k(1-q_2^k)J_{-k}},\\
&\eta_B(z)=e^{\sum_{k>0}\frac{z^k}{k}(1-q_2^k)J_{-k}}e^{-\sum_{k>0}\frac{z^{-k}}{k}J_k}.%\to e^{\sum_{k>0}\frac{z^k}{k}(1-q^{2k})J_{-k}}e^{-\sum_{k>0}\frac{z^{-k}}{k}J_k}.
\end{split}
\end{align}
The limit LII of the vertex operator $\eta_B(z)$ is well-defined, and it reproduces the exponential part of the expression \ref{def_vrho_LT} for $X^+(q^{-c}z)$. To define the limit of the vertex operator $\eta_A(z)$, we examine the exchange relation with a generic vertex operator $V_\a(z)$ of charge $\a\in\mC$,
\begin{equation}
V_\a(z)\eta_A(w)=\left(\dfrac{1-q_1q_2w/z}{1-q_1w/z}\right)^\a \eta_A(w)V_\a(z),\quad V_\a(z)=e^{\a Q}e^{\a\sum_{k>0}\frac{z^k}{k}J_{-k}}e^{-\a\sum_{k>0}\frac{z^{-k}}{k}J_k}z^{\a J_0}.
\end{equation}
This relation has a well-defined limit LII in which $\eta_A(w)$ is replaced by the operator $\hat\eta_A$, independent of the coordinate $w$, and satisfying the relation $V_\a(z)\hat\eta_A=q^{2\a} V_\a(z)\hat\eta_A$. This operator is realized on the Fock space $\CF$ as $\hat\eta_A=q^{-2J_0}$. The set of vertex operators $V_\a(z)$ contains as a particular case the fermionic fields $\psi(z)=V_{-1}(z)$ and $\bpsi(z)=V_1(z)$, which modes define the Schur basis of the Fock space $\CF$ through the bosonization formulas. This fact justifies the replacement of $\eta_A(z)$ by $\hat\eta_A$ in the limit LII.\footnote{This argument works also at the level of modes. For instance, the exchange relation for $\bpsi(z)$ reads in terms modes $\eta_{A,n+1}\bpsi_m-q_1\eta_{A,n}\bpsi_{m+1}=\bpsi_m\eta_{A,n+1}-q_1q_2\bpsi_{m+1}\eta_{A,n}$, and gives in the limit LII $\hat\eta_{A,n}\bpsi_{m+1}=q^2\bpsi_{m+1}\hat\eta_{A,n}$ which is independent of $n$.} The current $\psi^-(\hg^{-1/2}z)\propto:x^+(\hg^{-1}z)x^-(z):$ contains a similar seemingly divergent part for which the same procedure applies. As a result, we find indeed\footnote{In the first line, an extra factor $q^{-2}$ arise from changing the position of $\hat\eta_A$ in the normal ordering, it compensates the limit $-q_1^{-1}\tk\to q^2$.}
\begin{align}
\begin{split}
&q_1^{-3/2}q_3^{-n/2}\rho_u^{(1,n)}(x^+(\hg^{-1}z))\tox_{LII}-q^{nc}(1-q^2)\vrho_{u',u',n}^{(LT)}(X^+(q^{-c}z)),\\
&q_1^{1/2}\rho_u^{(1,n)}(x^-(z))\tox_{LII}-q^{-1}(1-q^2)\vrho_{u',u',n}^{(LT)}(X^-(z)),\\
&q_3^{-n/2}\rho_u^{(1,n)}(\psi^-(\hg^{-1/2}z))\tox_{LII}\vrho_{u',u',n}^{(LT)}(\Psi^-(q^{-c/2}z)),
\end{split}
\end{align}
after introducing a zero mode dependence through the weight $u'=e^{Q}u$. Alternatively, it should be possible to define this limit in the framework of shuffle algebras \cite{Feigin2009a,Garbali2021}, but this is beyond the scope of this paper.

% A similar limiting procedure can be define for the right twisted currents, producing instead the representation $\vrho_{u,u,n}^{(RT)}$ of $\CU^{(0,-\infty)}$. 

\paragraph{Symmetric polynomials} The horizontal Fock representation $\rho_u^{(1,0)}$ is known to have a deep connection with Macdonald polynomials \cite{Feigin2009a}. The identification of the modes $q_3^{k/2}(1-q_2^k)J_{-k}$ with the elementary power sum $p_k$, and the vacuum state $\ket{\vac}$ with the constant $1$, defines an isomorphism between the Fock space $\CF_0$ and the ring of symmetric polynomials. The Drinfeld currents $x^\pm(z)$ act on this ring as the vertex operators,
\begin{align}
\begin{split}
&x^+(z)=ue^{\sum_{k>0}\frac{z^k}{k}(1-q_1^k)p_k}e^{-\sum_{k>0}z^{-k}(1-q_2^{k})\frac{\p}{\p p_k}},\\
&x^-(z)=u^{-1}e^{-\sum_{k>0}\frac{z^k}{k}q_3^{k/2}(1-q_1^k)p_k}e^{\sum_{k>0}z^{-k}q_3^{k/2}(1-q_2^{k})\frac{\p}{\p p_k}},
\end{split}
\end{align}
and their zero modes are diagonal in the basis of Macdonald polynomials $P_\l(x,\qf,\tf)$ under the identification $(q_1,q_2)=(\tf^{-1},\qf)$,
\begin{align}
\begin{split}
&x_0^+ P_\l(x,\qf,\tf)=u\left(1-(1-\qf)(1-\tf^{-1})\sum_{(i,j)\in\l}\tf^{1-i}\qf^{j-1}\right)P_\l(x,\qf,\tf),\\
&x_0^- P_\l(x,\qf,\tf)=u^{-1}\left(1-(1-\qf)(1-\tf^{-1})\sum_{(i,j)\in\l}\tf^{i}\qf^{-j}\right)P_\l(x,\qf,\tf).
\end{split}
\end{align}

In the limit $\qf\to0$ with $\tf$ fixed, i.e. $q_2\to0$ with $q_1$ fixed, the Macdonald polynomials reduce to the Hall-Littlewood polynomials $H_\l(x,\tf)$. In this limit, the eigenvalue relation for $x_0^-$ diverges, but the one for $x_0^+$ remains finite. The Drinfeld current $x^+(z)$ tends to the t-fermionic field defined by Jing in \cite{Jing1991},
\begin{equation}\label{def_H}
x^+(z)\to uh(z),\quad h(z)=e^{\sum_{k>0}\frac{z^k}{k}(1-\tf^{-k})p_k}e^{-\sum_{k>0}z^{-k}\frac{\p}{\p p_k}}.
\end{equation} 
Its zero mode $h_0$ is diagonal, but highly degenerate, on the Hall-Littlewood basis, with $h_0H_\l(x,\tf)=\tf^{-\ell(\l)}H_\l(x,\tf)$ where $\ell(\l)$ is the number of columns in the Young diagram $\l$.

The limit LII we are investigating here is closely related to the Hall-Littlewood limit. Using the duality relation of Macdonald polynomial $\o_{\qf,\tf}P_\l(x,\qf,\tf)=Q_{\l^t}(x,\tf,\qf)$ (c.f. \cite{Macdonald} VI.5), we deduce that Macdonald polynomials tend to the dual Hall-Littlewood polynomials $H_\l^\ast(x,\qf)$ in the limit $\tf\to 0$ (also called $\qf$-Whittaker symmetric functions \cite{Gerasimov2008}), with
\begin{equation}\label{def_Hl_ast}
H_\l^\ast(x,\qf)=b_{\l^t}(\qf)\o_{\qf}^{-1}H_{\l^t}(x,\qf),\quad b_{\l^t}(\qf)=\prod_{j=1}^{\l_1}(1-\qf^{\l_{\l^t_j}-j+1}),
\end{equation} 
and the morphism $\o_{\qf}$ acting on elementary power sum as $\o_\qf p_k=(-)^{k-1}(1-\qf^k)p_k$. In this limit, it is the eigenvalue equation of $x^-_0$ which is well-defined, giving $h_0^\ast H_\l^\ast(x,\qf)=\qf^{-\l_1}H_\l^\ast(x,\qf)$ where $h_0^\ast$ is the zero mode of the current
\begin{equation}
h^\ast(z)=e^{\sum_{k>0}\frac{z^k}{k}q^{-2k}p_k}e^{\sum_{k>0}z^{-k}(1-q^{2k})\frac{\p}{\p p_k}},\quad x^-_0(q_3^{1/2}z)\tox_{LII} u^{-1}h^\ast(z).
\end{equation} 
This is consistent with the previous relation for Hall-Littlewood polynomials since $h(z)=\o_\qf h^\ast(-z)\o_{\qf}^{-1}$ with $\qf=q^2$ (and $\tf$ replaced by $\qf$ in equ. \ref{def_H}). The modes of the operator $H^\ast(z)$ obey the same $t$-commutation relations as Jing's t-fermions, and it coincides with $\vrho_{u',u',n}^{(LT)}(X^-(z))$ upon the identification $p_k=(1-q^{2k})J_{-k}$ (up to zero modes). We deduce the action of the zero mode
\begin{equation}
X_0^- H_\l^\ast(x,q^2)=-u^{-1}e^{-Q}\dfrac{q^{-2\l_1}}{1-q^2}H_\l^\ast(x,q^2).
\end{equation}
It would be interesting to compare these results with those obtained in \cite{Ohkubo2015,Ohkubo2017} and based on the crystal limit of the q-deformed Virasoro algebra \cite{Awata1996}. However, their approach leads to the introduction of a different limit for positive and negative modes of the currents, with a rescaling $x_n^\pm\to q_3^{-|n|/2}x_n^\pm$, and so it appears very different from ours.

% Let's take $c=0$ for simplicity in the twisted left representation $\vrho^{(LT)}$. Interestingly, the commutation relation 
% \begin{equation}
% [X_n^+,X_m^-]=-\dfrac1{q-q^{-1}}\Psi_{n+m},\quad n+m\leq0
% \end{equation} 
% imply that $X_n^+$ commutes with $X_0^-\propto H_0^\ast$ for $n>0$. It defines an infinite set of diagonal operator on the dual Hall-Littlewood polynomials $H_\l^\ast(x,\qf)$. Taking the conjugate by $\o_\qf$, it implies that the operator $X_n$ defined by the expansion of
% \begin{equation}
% X(z)=e^{-\sum_{k>0}\frac{z^k}{k}(1-\tf^{k})p_k}e^{\sum_{k>0}z^{-k}\frac{\p}{\p p_k}}=\sum_{n\in\mZ} z^{-n} X_n
% \end{equation} 
% act diagonally on the Hall-Littlewood polynomials. \cmt{I should check all this explicitly...} 

\section{Algebraic engineering of 5d \texorpdfstring{$\CN=1$}{N=1} matter theories}
In this section, we show that the shifted representations of the quantum toroidal $\gl(1)$ algebra can be used to describe the (anti)fundamental hypermultiplets in the algebraic engineering formalism. After recalling the expression of instanton partition functions for 5d $\CN=1$ gauge theories, we examine several key examples of gauge theories with matter hypermultiplets and present the corresponding algebraic construction. A general treatment of A-type quiver theories with $U(N)$ gauge groups can be easily inferred from our analysis of these examples. We note that the algebraic construction is not unique, and the equivalence between different choices of representations is expected to correspond to different possible brane realizations of gauge theories.

\subsection{5d \texorpdfstring{$\CN=1$}{N=1} partition functions and \texorpdfstring{$(p,q)$}{(p,q)}-brane webs} 
In this section, we review the algebraic construction of the partition functions of 5d $\CN=1$ gauge theories in the omega-deformed background $\mC_{\e_1}\times\mC_{\e_2}\times S^1$. Our presentation will be brief, and we refer the reader to \cite{Bourgine2017b} for further information. These theories preserve eight supercharges, they are specified by a choice of gauge group $G$ for the vector multiplet, and of representations for the hypermultiplets. We restrict ourselves to A-type quiver theories with a unitary gauge group at each node, i.e. $G=\times_\a U(N_\a)$, and hypermultiplets in fundamental ($f$), antifundamental ($\bar{f}$) or bifundamental (bf) representations. The partition function of these theories have been computed by localization on the Coulomb branch in \cite{Nekrasov2002}, they factorize into perturbative, one-loop and instanton contributions, and we focus on the latter. 

The instanton contribution is a sum over instanton configurations labeled by $N_\a$-tuple Young diagrams $\bl^{(\a)}=(\l^{(\a,1)},\cdots,\l^{(\a,N_\a)})$ at each node $\a$ of the Dynkin diagram. Each configuration $\bl^{(\a)}$ is weighted by the total number of boxes $|\bl^{(\a)}|$ with a fugacity $\qf_\a$ corresponding to a (renormalized) exponentiated gauge coupling. The summands decompose further into the contributions of the vector multiplets, the fundamental/antifundamental/bifundamental hypermultiplets, and the Chern-Simons term,
\begin{align}
\begin{split}\label{def_CZI}
\CZinst[G,R]=\sum_{\{\bl^{(\a)}\}}&\prod_{\a\in\G}\qf_\a^{|\bl^{(\a)}|}\CZv(\bl^{(\a)},\bv^{(\a)})\CZCS(\bl^{(\a)},\bv^{(\a)},\k_\a)\CZf(\bl^{(\a)},\bv^{(\a)},\bmu^{(\a)})\CZaf(\bl^{(\a)},\bv^{(\a)},\bar{\bmu}^{(\a)})\\
&\times\prod_{\a\rightarrow\b}\CZbf(\bl^{(\a)},\bv^{(\a)},\bl^{(\b)},\bv^{(\b)}|\mu_{\a\b}).
\end{split}
\end{align}
In addition to the background parameters $(q_1,q_2)=(e^{-R\e_1},e^{-R\e_2})$, all factors depend on the vectors $\bv^{(\a)}=(v_1^{(\a)},\cdots,v_{N_\a}^{(\a)})$ of exponentiated Coulomb branch vev (i.e. the vacuum expectation value of the adjoint scalar field in the vector multiplets). The fundamental/antifundamental contributions also depend on the vectors of (exponentiated) masses $\bmu^{(\a)}=(\mu_1^{(\a)},\cdots,\mu^{(\a)}_{N_\a^f})$ and $\bar\bmu^{(\a)}=(\bar\mu_1^{(\a)},\cdots,\bar\mu_{N_\a^{\bar{f}}}^{(a)})$. Finally, we denote the Chern-Simons levels $\k_\a$, and the bifundamental masses $\mu_{\a\b}$. It is noted that in the case of A-type quivers, bifundamental masses can be set to $\mu_{\a\b}=q_3^{-1/2}$ by a rescaling of the parameters. The expression of the vector and bifundamental contributions involve the Nekrasov factor $\CN_{\l,\mu}(\a)$,
\begin{align}
\begin{split}\label{def_CN}
&\CZv(\bl,\bv)=\prod_{l,l'=1}^N \CN_{\l^{(l)},\l^{(l')}}(v_l/v_{l'})^{-1},\quad \CZbf(\bl,\bv,\bl',\bv'|\mu)=\prod_{l=1}^N\prod_{l'=1}^{N'} \CN_{\l^{(l)},\l^{\prime(l')}}(v_l/(\mu v'_{l'})),\\
&\CN_{\l,\mu}(\a)=\prod_{\superp{(i,j)\in\l}{(i',j')\in\mu}}\dfrac{(1-\a q_1^{i-i'+1}q_2^{j-j'})(1-\a q_1^{i-i'}q_2^{j-j'+1})}{(1-\a q_1^{i-i'}q_2^{j-j'})(1-\a q_1^{i-i'+1}q_2^{j-j'+1})}\times\prod_{(i,j)\in\l}(1-\a q_1^iq_2^j)\times\prod_{(i',j')\in\mu}(1-\a q_1^{1-i'}q_2^{1-j'}).
\end{split}
\end{align}
The other contributions are simple products over the boxes $\Abox\in\bl$,%\footnote{To simplify formula, we have introduced some harmless shift of the (anti)fundamental masses $\bmu\to q_3^{\pm1/2}\bmu$.}
\begin{equation}\label{CZ_f_af}
\CZCS(\bl,\bv,\k)=\prod_{\sAbox\in\bl}\chi_\sAbox^\k,\quad \CZf(\bl,\bv,\bmu)=\prod_{\sAbox\in\bl} P_\bmu^f(\chi_\sAbox ),\quad 
\CZaf(\bl,\bv,\bar{\bmu})=\prod_{\sAbox\in\bl}P_{\bar\bmu}^{\bar{f}}(\chi_\sAbox ),
\end{equation} 
where we associated the box content $\chi_\sAbox =v_lq_1^{i-1}q_2^{j-1}$ to the box $\Abox=(l,i,j)$ with $(i,j)\in\l^{(l)}$, and we have introduced the following polynomials in $z^{\pm1}$,
\begin{equation}\label{def_Pmu}
P_\bmu^f(z)=\prod_{a=1}^{N^f}(1-q_1q_2z/\mu_a),\quad  P_{\bar{\bmu}}^{\bar{f}}(z)=\prod_{a=1}^{N^{\bar{f}}}(1-\bar{\mu}_a/z).
\end{equation} 
This expression of the (anti)fundamental contributions using polynomials of the form \ref{def_P} strongly suggests the possibility of using shifted intertwiners with the modified factors $t_\l^P$ and $t_\l^{P\ast}$ given in \ref{shift_tl} to introduce hypermultiplets in the algebraic formalism. We will show below that it is indeed the case.

\paragraph{Brane realization} This class of supersymmetric gauge theories can be realized as the low-energy description of the dynamics of brane systems called $(p,q)$-brane webs \cite{Aharony1997,Aharony1997a}. The $(p,q)$-branes are 5-branes in type IIB string theory corresponding to bound states of $q$ NS5 and $p$ D5 branes. Considering that the NS5-branes fill up the directions $012345$ and the D5-branes the directions $012346$, the $(p,q)$-branes fill up $01234$, and are segments in the $(56)$-plane tilted at an angle $\th_{p,q}$ with $\tan\th_{p,q}=p/q$. All the branes overlap in the directions $01234$ which coincide with the 5d spacetime $\mC_{\e_1}\times\mC_{\e_2}\times S^1$, and they form a trivalent lattice in the $(56)$-plane that encodes the field content and the couplings of the gauge theory. For instance, pure $U(N)$ gauge theories are obtained by suspending $N$ D5-branes between two dressed NS5-branes. We call here \textit{dressed NS5-branes} the $(n,1)$-branes corresponding to the bound states of a single NS5-brane and $n$ D5-branes. Additional matter hypermultiplets can be introduced in three different ways: either as semi-infinite D5-branes ending on the dressed NS5-branes, or finite D5-branes ending on a transverse D7-brane and a dressed NS5-brane, or, after Hanany-Witten transition, as a free transverse D7-brane lying in between the two dressed NS5-branes. The transverse D7-branes extend in the directions $01234789$.

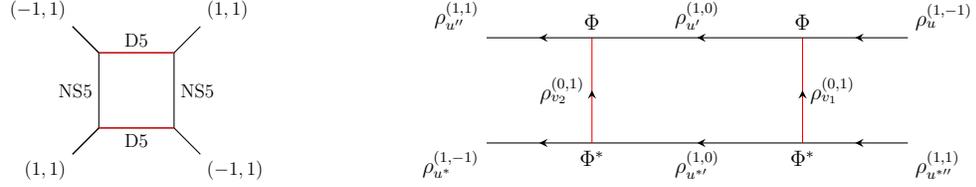
\begin{figure}
\begin{center}
\begin{subfigure}[c]{0.3\textwidth}
\begin{tikzpicture}[scale=.5]
\draw (-.7,-.7) -- (0,0) -- (2,0) -- (2.7,-.7);
\draw (-.7,-.7) -- (0,0) -- (0,2) -- (-.7,2.7);
\draw (-.7,2.7) -- (0,2) -- (2,2) -- (2.7,2.7);
\draw (2,0) -- (2,2);
\draw[red] (0,0) -- (2,0);
\draw[red] (0,2) -- (2,2);
\node[scale=.6,left] at (0,1) {NS5};
\node[scale=.6,below] at (1,0) {D5};
\node[scale=.6,right] at (2,1) {NS5};
\node[scale=.6,above] at (1,2) {D5};
\node[scale=.6,above right] at (2.7,2.7) {$(1,1)$};
\node[scale=.6,below left] at (-.7,-.7) {$(1,1)$};
\node[scale=.6,above left] at (-.7,2.7) {$(-1,1)$};
\node[scale=.6,below right] at (2.7,-.7) {$(-1,1)$};
\end{tikzpicture}
\end{subfigure}
% \hspace{15mm}
\begin{subfigure}[c]{0.3\textwidth}
\begin{tikzpicture}[scale=.7]
\draw[postaction={on each segment={mid arrow=black}}] (8,0) -- (6,0) -- (2,0) -- (0,0);
\draw[red, postaction={on each segment={mid arrow=black}}] (2,0) -- (2,2);
\draw[red, postaction={on each segment={mid arrow=black}}] (6,0) -- (6,2);
\draw[postaction={on each segment={mid arrow=black}}] (8,2) -- (6,2) -- (2,2) -- (0,2);
% \draw[blue, thick, postaction={on each segment={mid arrow=blue}}] (8,0) -- (6,0) -- (6,2) -- (2,2) -- (0,2);
\node[scale=.7,left] at (2,1) {$\rho_{v_2}^{(0,1)}$};
\node[scale=.7,right] at (6,1) {$\rho_{v_1}^{(0,1)}$};
\node[scale=.7,below right] at (8,0) {$\rho_{u^{\ast\prime\prime}}^{(1,1)}$};
\node[scale=.7,below] at (4,0) {$\rho_{u^{\ast\prime}}^{(1,0)}$};
\node[scale=.7,below left] at (0,0) {$\rho_{u^\ast}^{(1,-1)}$};
\node[scale=.7,above right] at (8,2) {$\rho_u^{(1,-1)}$};
\node[scale=.7,above] at (4,2) {$\rho_{u'}^{(1,0)}$};
\node[scale=.7,above left] at (0,2) {$\rho_{u''}^{(1,1)}$};
\node[scale=.7,below] at (2,0) {$\Phi^\ast$};
\node[scale=.7,above] at (2,2) {$\Phi$};
\node[scale=.7,below] at (6,0) {$\Phi^\ast$};
\node[scale=.7,above] at (6,2) {$\Phi$};
\end{tikzpicture}
\end{subfigure}
\end{center}
\caption{$(p,q)$-brane web and the corresponding network of representations for a 5d $\CN=1$ pure $U(2)$ gauge theory (D5-branes and the corresponding representations are in red)}
\label{pureU2}
\end{figure}

In the algebraic engineering of 5d $\CN=1$ gauge theories, a \textit{network of representations} for the quantum group $\CE$ is associated to the brane web in the $(56)$-plane. A network of representations is a set of modules that can be connected with intertwining operators. It gives the prescription to glue these operators together to form an operator $\CT$ from which we obtain the BPS observables of the gauge theory. In the case of 5d $\CN=1$ gauge theories, this network can be described as a rotated and flattened version of the $(p,q)$-brane web where all dressed NS5-branes are drawn horizontally, while (pure) D5-branes are drawn vertically \cite{Bourgine2017a}. As an example, the network of representations associated to the pure $U(2)$ gauge theory is drawn on figure \ref{pureU2}.\footnote{Alternatively, it can be seen as the brane web realizing the 4D $\CN=2$ gauge theory obtained in the compactification limit $R\to0$. This theory is realized in type IIA string theory by suspending D4-branes (drawn vertically) between NS5-branes (drawn horizontally), while matter hypermultiplets are inserted using D6-branes transversal to the $(56)$-plane. However, in this case, the algebraic engineering is based on the affine Yangian double of $\gl(1)$ that is obtained from $\CE$ in the limit $R\to0$ (see \cite{Bourgine2018}).}

D5-branes at position $\sim\log v$ correspond to vertical representations $\rho_v^{(0,1)}$ and $(n,1)$ dressed NS5-branes at position $\sim\log u$ are associated to horizontal representations $\rho_u^{(1,n)}$. The junctions of three $(p,q)$-branes correspond to one of the AFS intertwining operators \ref{prop_intw}, either $\Phi$ or $\Phi^\ast$ depending on the orientation. We choose to orient D5-branes from left to right, and NS5-branes from top to bottom. Junctions with two incoming and one outgoing brane leads to the insertion of the intertwiner $\Phi$, while junctions with one incoming and two outgoing branes leads to insertions of $\Phi^\ast$. These trivalent operators are glued along the modules attached to their common leg. The gluing is realized either by a product of vertex operators (in horizontal representations) or a scalar product (in vertical representations). Glued intertwiners define the operator $\CT$ which intertwines between the tensor product of representations associated to incoming and outgoing external edges. The partition function of the gauge theory is obtained by taking the vacuum expectation value of this operator, i.e. by projection on the vacuum states in the modules associated to the external edges.

\subsection{Engineering fundamental hypermultiplets}
In \cite{Bourgine2017b}, hypermultiplets in the (anti)fundamental representations have been introduced using a specific matter vertex operator. We present here a different construction based on the shifted representations for the algebras $\CE^\bmu$ defined in section \ref{sec_SQTA}. In fact, the shift of the vertical representation is natural from the point of view of the ADHM construction of the instanton moduli space. Indeed, we show in appendix \ref{AppCOHA} that the modification of the equivariant character introduced by the presence of antifundamental hypermultiplets leads to a shift of the cohomological Hall algebra, i.e. the vertical representation, by $\iota_{P}$ with $P=P_{\bar{\bmu}}^{\bar{f}}$.\footnote{Unfortunately, the case of fundamental hypermultiplets is not so clear.} According to the table of figure \ref{table2}, the intertwiner $\Phi$ attached to this representation should be modified into $\Phi^{P\ast}$, which produces the expected factor $\prod_\sAbox P(\chi_{\sAbox})=\CZaf(\l,v,\bar{\bmu})$ in the summands of the partition function. As we will see in the examples below, this is not the only possibility to introduce this contribution. We will comment on the equivalence between the different algebraic constructions which is reminiscent of the different brane realizations of a gauge theory using Hanany-Witten transitions.

% 
% This shift modifies the scalar product in the vertical module, introducing the hypermultiplet contributions as $\bar{n}_\l^P=\CZf(\bl,\bv,\bmu)\bar{n}_\l$ since  as we can also consider the shift of horizontal representations which leads to replace the intertwiners $\Phi$ and $\Phi^\ast$ by their shifted versions $\Phi^P$ and $\Phi^{P\ast}$. Then, the (anti)fundamental contribution \ref{CZ_f_af} appears in the new factors $t_\l^P$ and $t_\l^{P\ast}$ as shown in \ref{shift_tl}.

\subsubsection{Matter \texorpdfstring{$U(1)$}{U(1)} gauge theory}
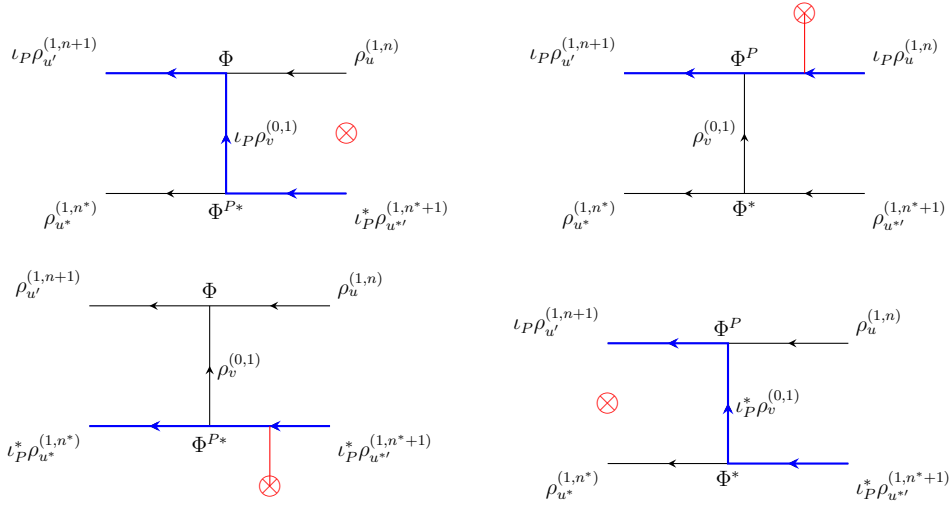
\begin{figure}
\begin{center}
\begin{tikzpicture}[scale=.8]
\draw[postaction={on each segment={mid arrow=black}}] (4,0) -- (2,0) -- (0,0);
\draw[postaction={on each segment={mid arrow=black}}] (2,0) -- (2,2);
\draw[postaction={on each segment={mid arrow=black}}] (4,2) -- (2,2) -- (0,2);
\draw[blue, thick, postaction={on each segment={mid arrow=blue}}] (4,0) -- (2,0) -- (2,2) -- (0,2);
\node[scale=.7,right] at (2,1) {$\iota_P\rho_v^{(0,1)}$};
\node[scale=.7,below right] at (4,0) {$\iota_P^\ast\rho_{u^{\ast\prime}}^{(1,n^\ast+1)}$};
\node[scale=.7,below left] at (0,0) {$\rho_{u^\ast}^{(1,n^\ast)}$};
\node[scale=.7,above right] at (4,2) {$\rho_u^{(1,n)}$};
\node[scale=.7,above left] at (0,2) {$\iota_P\rho_{u'}^{(1,n+1)}$};
\node[scale=.7,below] at (2,0) {$\Phi^{P\ast}$};
\node[scale=.7,above] at (2,2) {$\Phi$};
\node[scale=1,thick, red] at (4,1) {$\otimes$};
\end{tikzpicture}
\hspace{6mm}
\begin{tikzpicture}[scale=.8]
\draw[postaction={on each segment={mid arrow=black}}] (4,0) -- (2,0) -- (0,0);
\draw[postaction={on each segment={mid arrow=black}}] (2,0) -- (2,2);
\draw[postaction={on each segment={mid arrow=black}}] (4,2) -- (2,2) -- (0,2);
\draw[blue, thick, postaction={on each segment={mid arrow=blue}}] (4,2) -- (2,2) -- (0,2);
\node[scale=.7,left] at (2,1) {$\rho_v^{(0,1)}$};
\node[scale=.7,below right] at (4,0) {$\rho_{u^{\ast\prime}}^{(1,n^\ast+1)}$};
\node[scale=.7,below left] at (0,0) {$\rho_{u^\ast}^{(1,n^\ast)}$};
\node[scale=.7,above right] at (4,2) {$\iota_P\rho_u^{(1,n)}$};
\node[scale=.7,above left] at (0,2) {$\iota_P\rho_{u'}^{(1,n+1)}$};
\node[scale=.7,below] at (2,0) {$\Phi^\ast$};
\node[scale=.7,above] at (2,2) {$\Phi^P$};
\node[scale=1,thick, red] at (3,3) {$\otimes$};
\draw[red] (3,3) -- (3,2);
\end{tikzpicture}\\
\vspace{3mm}
\begin{tikzpicture}[scale=.8]
\draw[postaction={on each segment={mid arrow=black}}] (4,0) -- (2,0) -- (0,0);
\draw[postaction={on each segment={mid arrow=black}}] (2,0) -- (2,2);
\draw[postaction={on each segment={mid arrow=black}}] (4,2) -- (2,2) -- (0,2);
\draw[blue, thick, postaction={on each segment={mid arrow=blue}}] (4,0) -- (2,0) -- (0,0);
\node[scale=.7,right] at (2,1) {$\rho_v^{(0,1)}$};
\node[scale=.7,below right] at (4,0) {$\iota_P^\ast\rho_{u^{\ast\prime}}^{(1,n^\ast+1)}$};
\node[scale=.7,below left] at (0,0) {$\iota_P^\ast\rho_{u^\ast}^{(1,n^\ast)}$};
\node[scale=.7,above right] at (4,2) {$\rho_u^{(1,n)}$};
\node[scale=.7,above left] at (0,2) {$\rho_{u'}^{(1,n+1)}$};
\node[scale=.7,below] at (2,0) {$\Phi^{P\ast}$};
\node[scale=.7,above] at (2,2) {$\Phi$};
\node[scale=1,thick, red] at (3,-1) {$\otimes$};
\draw[red] (3,-1) -- (3,0);
\end{tikzpicture}
\hspace{6mm}
\begin{tikzpicture}[scale=.8]
\draw[postaction={on each segment={mid arrow=black}}] (4,0) -- (2,0) -- (0,0);
\draw[postaction={on each segment={mid arrow=black}}] (2,0) -- (2,2);
\draw[postaction={on each segment={mid arrow=black}}] (4,2) -- (2,2) -- (0,2);
\draw[blue, thick, postaction={on each segment={mid arrow=blue}}] (4,0) -- (2,0) -- (2,2) -- (0,2);
\node[scale=.7,right] at (2,1) {$\iota_P^\ast\rho_v^{(0,1)}$};
\node[scale=.7,below right] at (4,0) {$\iota_P^\ast\rho_{u^{\ast\prime}}^{(1,n^\ast+1)}$};
\node[scale=.7,below left] at (0,0) {$\rho_{u^\ast}^{(1,n^\ast)}$};
\node[scale=.7,above right] at (4,2) {$\rho_u^{(1,n)}$};
\node[scale=.7,above left] at (0,2) {$\iota_P\rho_{u'}^{(1,n+1)}$};
\node[scale=.7,below] at (2,0) {$\Phi^{\ast}$};
\node[scale=.7,above] at (2,2) {$\Phi^P$};
\node[scale=1,thick, red] at (0,1) {$\otimes$};
\end{tikzpicture}
\end{center}
\caption{Different algebraic constructions for the matter 5d $\CN=1$ $U(1)$ gauge theory}
\label{gauge1}
\end{figure}
% 
% \begin{figure}
% \begin{center}
% \begin{tikzpicture}[scale=.5]
% \draw[postaction={on each segment={mid arrow=black}}] (4,0) -- (2,0) -- (0,0);
% \draw[postaction={on each segment={mid arrow=black}}] (2,0) -- (2,2);
% \draw[postaction={on each segment={mid arrow=black}}] (4,2) -- (2,2) -- (0,2);
% \draw[blue, thick, postaction={on each segment={mid arrow=blue}}] (4,0) -- (2,0) -- (2,2) -- (0,2);
% \node[scale=.7,right] at (2,1) {$\iota_P^\ast\rho_v^{(0,1)}$};
% \node[scale=.7,below right] at (4,0) {$\iota_P^\ast\rho_{u^{\ast\prime}}^{(1,n^\ast+1)}$};
% \node[scale=.7,below left] at (0,0) {$\rho_{u^\ast}^{(1,n^\ast)}$};
% \node[scale=.7,above right] at (4,2) {$\rho_u^{(1,n)}$};
% \node[scale=.7,above left] at (0,2) {$\iota_P\rho_{u'}^{(1,n+1)}$};
% \node[scale=.7,below] at (2,0) {$\Phi^{P\ast}$};
% \node[scale=.7,above] at (2,2) {$\Phi^P$};
% \node[scale=1,thick, red] at (0,1) {$\otimes$};
% \end{tikzpicture}
% \end{center}
% \caption{Yet another realizations of the massive 5d $\CN=1$ $U(1)$ gauge theory}
% \label{gauge1b}
% \end{figure}

The simplest example is the case of a $U(1)$ gauge theory with $N^f$ fundamental hypermultiplets. We have identified four different algebraic constructions for the operator $\CT[U(1),N^f,0]$, they correspond to the network of representations shown on figure \ref{gauge1}. The shifts of representations involve the polynomial $P(z)=P_\bmu^f(z)$ containing the dependence in the exponentiated masses of the hypermultiplets. Similar constructions for hypermultiplets in the antifundamental representation are simply obtained by choosing instead the polynomial $P_{\bar{\bmu}}^{\bar{f}}(z)$. The modules on which acts a shifted representation have been indicated in blue on the figure. Since intertwiners always involve one incoming and one outgoing shifted representation (see the table \ref{table2}), the shifts ``propagate'' through the vertices.

The instanton partition function for the theory in the absence of hypermultiplets has been constructed in \cite{Bourgine2017b} as
\begin{equation}
\CZinst[U(1),0,0]=\la\CT[U(1),0,0]\ra,\quad \CT[U(1),0,0]=\sum_\l n_\l\ \Phi^\ast_\l[u^\ast,v,n^\ast]\otimes\Phi_\l[u,v,n],
\end{equation} 
with the identification of the instanton counting parameter as the ratio of weights $\qf=q_3^{-1/2}u/u^\ast$, and the Chern-Simons coupling as the difference of levels $\k=n^\ast-n$. This result is independent of the choice of the norm $n_\l$ for the scalar product.

Turning to the configurations shown in figure \ref{gauge1}, we observe that in each configuration one of the two intertwiners is replaced by its shifted version. The extra factor coming from the modification of $t_\l$ or $t_\l^\ast$ gives exactly the contribution of the hypermultiplets.\footnote{This factor can be displaced by taking e.g. $n_\l=\bn_\l^P$ but it is a purely cosmetic change.} It is important to emphasize that, although the networks of representations are different, the operator $\CT$ remains the same. We also note that if the representations $\rho_{u'}^{(1,n+1)}$ and $\rho_u^{(1,n)}$ are shifted simultaneously by $\iota_{P'}^\ast$ (for any $P'$ of the form \ref{def_P}), there is no modification to the intertwiners $\Phi$ or $\Phi^P$. In the same way, the intertwiners $\Phi^\ast$ or $\Phi^{P\ast}$ remain unchanged if we shift simultaneously the representations $\rho_{u\ast}^{(1,n^\ast)}$ and $\rho_{u^{\ast\prime}}^{(1,n^\ast+1)}$ by $\iota_{P'}$. This property is useful for the construction of gauge theories with gauge groups of higher rank.

% In the first case corresponding to the top left figure, the shift of the vertical representation $\iota_P\rho_v^{(0,1)}$ modifies the vertical scalar product between $\Phi$ and $\Phi^\ast$, replacing the norms $\bar{n}_\l$ by $\bar{n}_\l^P$ and thus introducing the hypermultiplet contribution $\prod_\sAbox P(\chi_{\sAbox})$ in the operator $\CT[U(1),N^f,0]$.  On the other hand, in the second realization (top right figure), the vertical representation is not shifted and $n_\l$ remains the same. However, $\Phi$ has been replaced by $\Phi^P$, and, as a result, the vev $\la\Phi_\l^P\ra=t_\l^P$ brings the extra factor $\prod_{\sAbox\in\l}P(\chi_\sAbox)$ in $\CT[U(1),N^f,0]$. The same phenomenon is observed in the third realization (bottom left), but this time $\Phi^\ast$ has been replaced by $\Phi^{P\ast}$. The last realization which represented on the bottom right figure involves a shifted vertical representation $\iota_P^\ast\rho_v^{(0,1)}$ and two shifted intertwiners. As a result, the inverse norm $\bar{n}_\l^{P\ast}$ brings a factor $\prod_{\sAbox\in\l}P(\chi_\sAbox )^{-1}$ that is canceled by one of the two extra factors $\prod_{\sAbox\in\l}P(\chi_\sAbox )$ coming from $t_\l^P$ and $t_\l^{P\ast}$. In the end, only a single factor $\prod_{\sAbox\in\l}P(\chi_\sAbox )$ remains which is indeed the contribution of the fundamental hypermultiplets.

\paragraph{qq-character} Nekrasov has introduced in \cite{Nekrasov_BPS1} a class of BPS observables which can be seen as deformations of the Frenkel-Reshetikhin q-characters \cite{Frenkel1998} defined in the context of the Bethe/gauge correspondence. On the gauge side, the observables are defined as particular combinations of chiral ring observables in such a way that their expectation values exhibit an important regularity property. This regularity property encodes an infinite set of constraints for the gauge theory called \textit{non-perturbative Dyson-Schwinger equations} \cite{Nekrasov_BPS1,Nekrasov_BPS2}. This property has been investigated from the point of view of toroidal quantum groups in \cite{BMZ,Bourgine2016}, and the qq-characters were later explicitly constructed in the algebraic engineering formalism in \cite{Bourgine2017b}. The qq-characters have also been studied using the (quiver, deformed) W-algebra formalism in \cite{Kimura2015,FJMV}. In this paper, we focus on the fundamental qq-character, but it is expected that similar results can be formulated for higher qq-characters using the method developed in \cite{Bourgine2017b}.

Following the construction presented in \cite{Bourgine2017b}, the fundamental qq-character is obtained from the operator $\CT[U(1),0,0]$ as an expectation value with a Drinfeld current inserted,
\begin{equation}
\chi^+(z)=\nu z^{n^\ast+1}\dfrac{\la \rho_\text{out}(x^+(\g^{-1}z))\CT[U(1),0,0]\ra}{\k _+ u^\ast \g^{n^\ast}\la\CT[U(1),0,0]\ra}=\nu z^{n^\ast+1}\dfrac{\la\CT[U(1),0,0]\rho_\text{in}(x^+(\g^{-1}z))\ra}{\k_+ u^\ast \g^{n^\ast}\la\CT[U(1),0,0]\ra}
\end{equation} 
where we used the shortcut notations $\g=\rho^{(1,n)}_u(\hg)=q_3^{1/2}$, $\nu=(-q_3v)^{-1}$, and denoted the incoming/outgoing representations on the external edges as $\rho_\text{in}=(\rho_{u^{\prime\ast}}^{(1,n^\ast+1)}\otimes\rho_{u}^{(1,n)})\D$ and $\rho_\text{out}=(\rho_{u^\ast}^{(1,n^\ast)}\otimes\rho_{u'}^{(1,n+1)})\D$. In the presence of hypermultiplets, the qq-character is known to have the form (see e.g. \cite{Nekrasov_BPS1,Bourgine2016})
\begin{equation}
\chi(z)=\dfrac1{\CZinst[U(1),N^f,0]}\sum_\l\qf^{|\l|}\CZv(\l,v)\CZCS(\l,v,\k)\CZf(\l,v,\bmu) \left(\nu z\CY_\l(zq_3^{-1})+\qf z^\k\dfrac{\textcolor{blue}{P(z)}}{\CY_\l(z)}\right),
\end{equation} 
with the Y-observables given in \ref{def_CYY}, and the deformation due to hypermultiplets highlighted in blue. This extra factor $\textcolor{blue}{P(z)}$ arises here from the evaluation of $x^+(z)$ in the shifted incoming/outgoing representations
\begin{equation}
\k_+^{-1}\rho(x^+(z))=\left\{
\begin{array}{ll}
u^{\ast\prime} z^{-n^\ast-1}\eta^+(z)\otimes1+\g^{n^\ast+1}\textcolor{blue}{P(\g z)}\vphi^-(\g^{1/2}z)\otimes u\g^{-n}z^{-n}\eta^+(\g z), & \rho=(\iota_P^\ast\otimes1)\rho_\text{in},\\
u^{\ast\prime} z^{-n^\ast-1}\eta^+(z)\otimes1+\g^{n^\ast+1}\vphi^-(\g^{1/2}z)\otimes u\g^{-n}z^{-n}\textcolor{blue}{P(\g z)}\eta^+(\g z), & \rho=(1\otimes\iota_P)\rho_\text{in},\\
u^{\ast} z^{-n^\ast}\eta^+(z)\otimes1+\g^{n^\ast}\textcolor{blue}{P(\g z)}\vphi^-(\g^{1/2}z)\otimes u'\g^{-n-1}z^{-n-1}\eta^+(\g z), & \rho=(\iota_P^\ast\otimes1)\rho_\text{out},\\
u^\ast z^{-n^\ast}\eta^+(z)\otimes1+\g^{n^\ast}\vphi^-(\g^{1/2}z)\otimes u'\g^{-n-1}z^{-n-1}\textcolor{blue}{P(\g z)}\eta^+(\g z), & \rho=(1\otimes\iota_P)\rho_\text{out}.
\end{array}
\right.
\end{equation} 
Thus, the algebraic construction of the fundamental qq-characters can also be extended to include the presence of fundamental hypermultiplets using the operator $\CT[U(1),N^f,0]$ built upon shifted representations. Of course, we arrive at the same conclusion if we build the qq-character from the current $x^-(z)$ instead of $x^+(z)$.

\subsubsection{Massive \texorpdfstring{$U(2)$}{U(2)} gauge theory}
\begin{figure}
\begin{center}
\begin{tikzpicture}[scale=.7]
\draw[postaction={on each segment={mid arrow=black}}] (8,0) -- (6,0) -- (2,0) -- (0,0);
\draw[postaction={on each segment={mid arrow=black}}] (2,0) -- (2,2);
\draw[postaction={on each segment={mid arrow=black}}] (6,0) -- (6,2);
\draw[postaction={on each segment={mid arrow=black}}] (8,2) -- (6,2) -- (2,2) -- (0,2);
\draw[blue, thick, postaction={on each segment={mid arrow=blue}}] (8,0) -- (6,0) -- (6,2) -- (2,2) -- (0,2);
\node[scale=.7,left] at (2,1) {$\rho_{v_2}^{(0,1)}$};
\node[scale=.7,right] at (6,1) {$\iota_P\rho_{v_1}^{(0,1)}$};
\node[scale=.7,below right] at (8,0) {$\iota_P^\ast\rho^{(1,n^\ast+2)}$};
\node[scale=.7,below] at (4,0) {$\rho^{(1,n^\ast+1)}$};
\node[scale=.7,below left] at (0,0) {$\rho^{(1,n^\ast)}$};
\node[scale=.7,above right] at (8,2) {$\rho^{(1,n)}$};
\node[scale=.7,above] at (4,2) {$\iota_P\rho^{(1,n+1)}$};
\node[scale=.7,above left] at (0,2) {$\iota_P\rho^{(1,n+2)}$};
\node[scale=.7,below] at (2,0) {$\Phi^\ast$};
\node[scale=.7,above] at (2,2) {$\Phi^P$};
\node[scale=.7,below] at (6,0) {$\Phi^{P\ast}$};
\node[scale=.7,above] at (6,2) {$\Phi$};
\node[scale=1,thick, red] at (8,1) {$\otimes$};
\end{tikzpicture}
\hspace{3mm}
\begin{tikzpicture}[scale=.7]
\draw[postaction={on each segment={mid arrow=black}}] (8,0) -- (6,0) -- (2,0) -- (0,0);
\draw[postaction={on each segment={mid arrow=black}}] (2,0) -- (2,2);
\draw[postaction={on each segment={mid arrow=black}}] (6,0) -- (6,2);
\draw[postaction={on each segment={mid arrow=black}}] (8,2) -- (6,2) -- (2,2) -- (0,2);
\draw[blue, thick, postaction={on each segment={mid arrow=blue}}] (8,0) -- (6,0) -- (2,0) -- (2,2) -- (0,2);
\node[scale=.7,left] at (2,1) {$\iota_P\rho_{v_2}^{(0,1)}$};
\node[scale=.7,right] at (6,1) {$\rho_{v_1}^{(0,1)}$};
\node[scale=.7,below right] at (8,0) {$\iota_P^\ast\rho^{(1,n^\ast+2)}$};
\node[scale=.7,below] at (4,0) {$\iota_P^\ast\rho^{(1,n^\ast+1)}$};
\node[scale=.7,below left] at (0,0) {$\rho^{(1,n^\ast)}$};
\node[scale=.7,above right] at (8,2) {$\rho^{(1,n)}$};
\node[scale=.7,above] at (4,2) {$\rho^{(1,n+1)}$};
\node[scale=.7,above left] at (0,2) {$\iota_P\rho^{(1,n+2)}$};
\node[scale=.7,below] at (2,0) {$\Phi^{P\ast}$};
\node[scale=.7,above] at (2,2) {$\Phi$};
\node[scale=.7,below] at (6,0) {$\Phi^{P\ast}$};
\node[scale=.7,above] at (6,2) {$\Phi$};
\node[scale=1,thick, red] at (4,1) {$\otimes$};
\end{tikzpicture}\\
\vspace{5mm}
\begin{tikzpicture}[scale=.7]
\draw[postaction={on each segment={mid arrow=black}}] (8,0) -- (6,0) -- (2,0) -- (0,0);
\draw[postaction={on each segment={mid arrow=black}}] (2,0) -- (2,2);
\draw[postaction={on each segment={mid arrow=black}}] (6,0) -- (6,2);
\draw[postaction={on each segment={mid arrow=black}}] (8,2) -- (6,2) -- (2,2) -- (0,2);
\draw[blue, thick, postaction={on each segment={mid arrow=blue}}] (8,0) -- (6,0) -- (2,0) -- (0,0);
\node[scale=.7,left] at (2,1) {$\rho_{v_2}^{(0,1)}$};
\node[scale=.7,right] at (6,1) {$\rho_{v_1}^{(0,1)}$};
\node[scale=.7,below right] at (8,0) {$\iota_P^\ast\rho^{(1,n^\ast+2)}$};
\node[scale=.7,below] at (4,0) {$\iota_P^\ast\rho^{(1,n^\ast+1)}$};
\node[scale=.7,below left] at (0,0) {$\iota_P^\ast\rho^{(1,n^\ast)}$};
\node[scale=.7,above right] at (8,2) {$\rho^{(1,n)}$};
\node[scale=.7,above] at (4,2) {$\rho^{(1,n+1)}$};
\node[scale=.7,above left] at (0,2) {$\rho^{(1,n+2)}$};
\node[scale=.7,below] at (2,0) {$\Phi^{P\ast}$};
\node[scale=.7,above] at (2,2) {$\Phi$};
\node[scale=.7,below] at (6,0) {$\Phi^{P\ast}$};
\node[scale=.7,above] at (6,2) {$\Phi$};
\node[scale=1,thick, red] at (4,-1) {$\otimes$};
\draw[red] (4,-1) -- (4,0);
\end{tikzpicture}
\hspace{3mm}
\begin{tikzpicture}[scale=.7]
\draw[postaction={on each segment={mid arrow=black}}] (8,0) -- (6,0) -- (2,0) -- (0,0);
\draw[postaction={on each segment={mid arrow=black}}] (2,0) -- (2,2);
\draw[postaction={on each segment={mid arrow=black}}] (6,0) -- (6,2);
\draw[postaction={on each segment={mid arrow=black}}] (8,2) -- (6,2) -- (2,2) -- (0,2);
\draw[blue, thick, postaction={on each segment={mid arrow=blue}}] (8,2) -- (6,2) -- (2,2) -- (0,2);
\node[scale=.7,left] at (2,1) {$\rho_{v_2}^{(0,1)}$};
\node[scale=.7,right] at (6,1) {$\rho_{v_1}^{(0,1)}$};
\node[scale=.7,below right] at (8,0) {$\rho^{(1,n^\ast+2)}$};
\node[scale=.7,below] at (4,0) {$\rho^{(1,n^\ast+1)}$};
\node[scale=.7,below left] at (0,0) {$\rho^{(1,n^\ast)}$};
\node[scale=.7,above right] at (8,2) {$\iota_P\rho^{(1,n)}$};
\node[scale=.7,above] at (4,2) {$\iota_P\rho^{(1,n+1)}$};
\node[scale=.7,above left] at (0,2) {$\iota_P\rho^{(1,n+2)}$};
\node[scale=.7,below] at (2,0) {$\Phi^\ast$};
\node[scale=.7,above] at (2,2) {$\Phi^P$};
\node[scale=.7,below] at (6,0) {$\Phi^\ast$};
\node[scale=.7,above] at (6,2) {$\Phi^P$};
\node[scale=1,thick, red] at (4,3) {$\otimes$};
\draw[red] (4,3) -- (4,2);
\end{tikzpicture}\\
\vspace{5mm}
\begin{tikzpicture}[scale=.7]
\draw[postaction={on each segment={mid arrow=black}}] (8,0) -- (6,0) -- (2,0) -- (0,0);
\draw[postaction={on each segment={mid arrow=black}}] (2,0) -- (2,2);
\draw[postaction={on each segment={mid arrow=black}}] (6,0) -- (6,2);
\draw[postaction={on each segment={mid arrow=black}}] (8,2) -- (6,2) -- (2,2) -- (0,2);
\draw[blue, thick, postaction={on each segment={mid arrow=blue}}] (8,0) -- (6,0) -- (6,2) -- (2,2) -- (0,2);
\node[scale=.7,left] at (2,1) {$\rho_{v_2}^{(0,1)}$};
\node[scale=.7,right] at (6,1) {$\iota_P^\ast\rho_{v_1}^{(0,1)}$};
\node[scale=.7,below right] at (8,0) {$\iota_P^\ast\rho^{(1,n^\ast+2)}$};
\node[scale=.7,below] at (4,0) {$\rho^{(1,n^\ast+1)}$};
\node[scale=.7,below left] at (0,0) {$\rho^{(1,n^\ast)}$};
\node[scale=.7,above right] at (8,2) {$\rho^{(1,n)}$};
\node[scale=.7,above] at (4,2) {$\iota_P\rho^{(1,n+1)}$};
\node[scale=.7,above left] at (0,2) {$\iota_P\rho^{(1,n+2)}$};
\node[scale=.7,below] at (2,0) {$\Phi^\ast$};
\node[scale=.7,above] at (2,2) {$\Phi^P$};
\node[scale=.7,below] at (6,0) {$\Phi^{\ast}$};
\node[scale=.7,above] at (6,2) {$\Phi^P$};
\node[scale=1,thick, red] at (4,1) {$\otimes$};
\end{tikzpicture}
\hspace{3mm}
\begin{tikzpicture}[scale=.7]
\draw[postaction={on each segment={mid arrow=black}}] (8,0) -- (6,0) -- (2,0) -- (0,0);
\draw[postaction={on each segment={mid arrow=black}}] (2,0) -- (2,2);
\draw[postaction={on each segment={mid arrow=black}}] (6,0) -- (6,2);
\draw[postaction={on each segment={mid arrow=black}}] (8,2) -- (6,2) -- (2,2) -- (0,2);
\draw[blue, thick, postaction={on each segment={mid arrow=blue}}] (8,0) -- (6,0) -- (2,0) -- (2,2) -- (0,2);
\node[scale=.7,right] at (2,1) {$\iota_P^\ast\rho_{v_2}^{(0,1)}$};
\node[scale=.7,right] at (6,1) {$\rho_{v_1}^{(0,1)}$};
\node[scale=.7,below right] at (8,0) {$\iota_P^\ast\rho^{(1,n^\ast+2)}$};
\node[scale=.7,below] at (4,0) {$\iota_P^\ast\rho^{(1,n^\ast+1)}$};
\node[scale=.7,below left] at (0,0) {$\rho^{(1,n^\ast)}$};
\node[scale=.7,above right] at (8,2) {$\rho^{(1,n)}$};
\node[scale=.7,above] at (4,2) {$\rho^{(1,n+1)}$};
\node[scale=.7,above left] at (0,2) {$\iota_P\rho^{(1,n+2)}$};
\node[scale=.7,below] at (2,0) {$\Phi^{\ast}$};
\node[scale=.7,above] at (2,2) {$\Phi^P$};
\node[scale=.7,below] at (6,0) {$\Phi^{P\ast}$};
\node[scale=.7,above] at (6,2) {$\Phi$};
\node[scale=1,thick, red] at (0,1) {$\otimes$};
\end{tikzpicture}\\
\vspace{3mm}
\end{center}
\caption{Algebraic realizations of massive 5d $\CN=1$ $U(2)$ gauge theory}
\label{gauge2}
\end{figure}
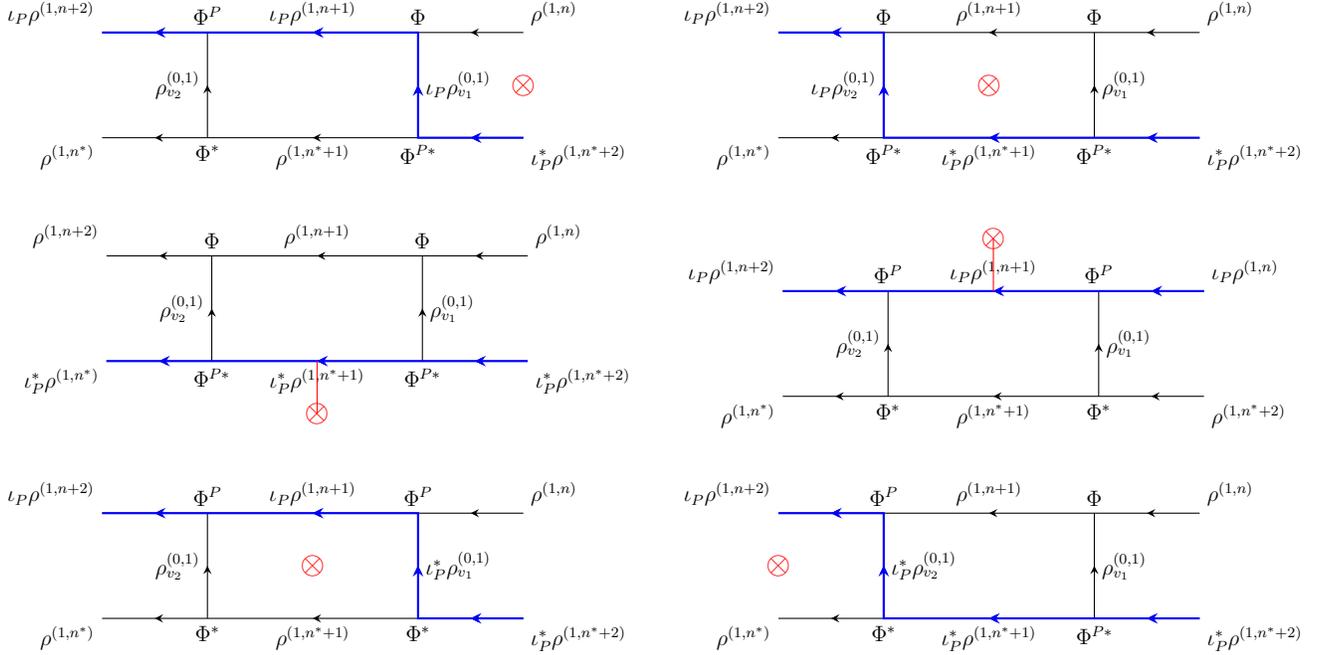

The previous analysis can be repeated for the case of the gauge group $G=U(2)$. In the absence of matter hypermultiplets, the instanton partition function $\CZinst[U(2),0,0]$ of the pure $U(2)$ gauge theory is obtained as the vev of the operator
\begin{align}
\begin{split}
% &\CZinst[U(2),0,0]=\la\CT[U(2),0,0]\ra,\\
&\CT[U(2),0,0]=\sum_{\l^{(1)},\l^{(2)}}n_{\l^{(1)}}n_{\l^{(2)}}\ \Phi_{\l^{(1)}}^\ast[u_1^\ast,v_1,n_1^\ast]\Phi^\ast_{\l^{(2)}}[u_2^\ast,v_2,n_2^\ast]\otimes\Phi_{\l^{(1)}}[u_1,v_1,n_1]\Phi_{\l^{(2)}}[u_2,v_2,n_2].
\end{split}
\end{align}
In the presence of fundamental hypermultiplets, the summands defining the operator $\CT[U(2),N^f,0]$ contain the extra factor $\prod_{\sAbox\in\l^{(1)}\cup\l^{(2)}}P(\chi_\sAbox)$. We identified six different ways of introducing the factor by shifting representations, they have been represented on the figure \ref{gauge2}. To emphasize the role of the shift, we have again drawn in blue the shifted representations. We observe that for each vertical representation, exactly one of the two intertwiners attached is shifted, carrying the expected factor $\prod_{\sAbox\in\l^{(l)}}P(\chi_\sAbox)$. Furthermore, the construction of the fundamental qq-character is the same as in the case of the $U(1)$ gauge group, since the shift of the external representations are the same: $1\otimes\iota_P$ or $\iota_P^\ast\otimes1$. The generalization of this construction to the case of $U(N)$ gauge groups is obvious.

% In the first case (above left), the first vertical representation is shifted, and one of the four intertwiners as well. As a result, $n_{\l^{(1)}}$ acquires an extra factor $\prod_{\sAbox\in\l^{(1)}}P(\chi_\sAbox)$ and $t_\l^{(2)}$ the factor $\prod_{\sAbox\in\l^{(2)}}P(\chi_\sAbox)$, so that the total factor is indeed equal to the contribution \ref{U2_hyper} of hypermultiplets. In the second case (above right), this factor comes from $n_{\l^{(2)}}$ and $t_{\l^{(1)}}^\ast$. In the third case (below left), the vertical representations are not shifted but two shifted intertwiners appear, bringing a modification to both $t_{\l^{(1)}}^\ast$ and $t_{\l^{(2)}}^\ast$. Then, in the fourth case (below right), the two shifted intertwiners leads to modify $t_{\l^{(1)}}$ and $t_{\l^{(2)}}$. Finally, in the last two cases, an extra cancellation is observed between the factor in $n_\l^{P\ast}$ and one of the two factors brought by $t_\l^P$ and $t_\l^{P\ast}$.

\subsubsection{Massive \texorpdfstring{$U(1)\times U(1)$}{U(1)xU(1)} gauge theory}
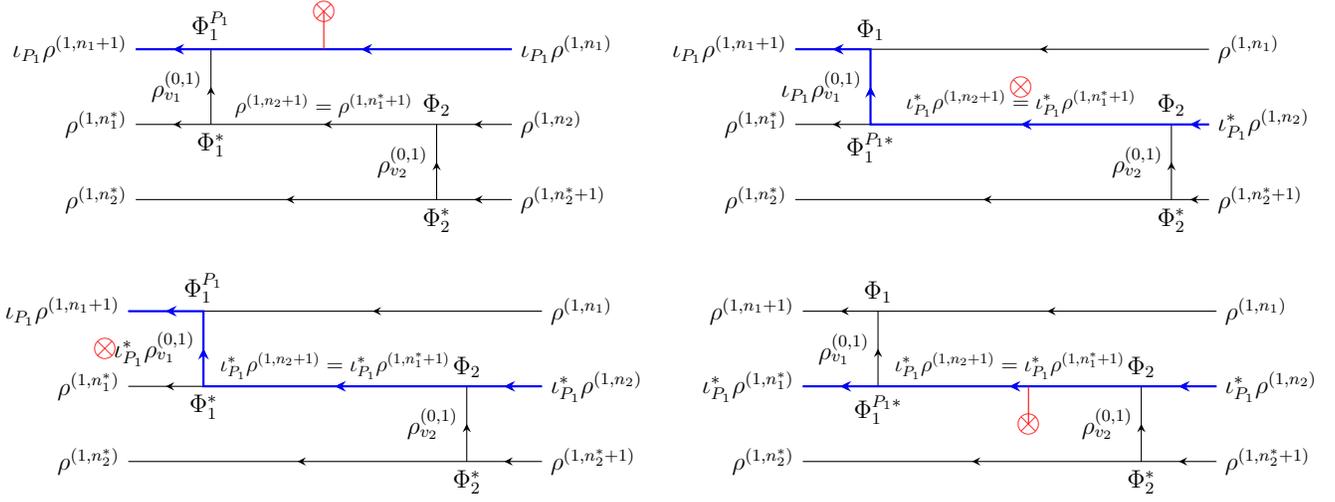
\begin{figure}
\begin{center}
\begin{tikzpicture}[scale=.5]
\draw[postaction={on each segment={mid arrow=black}}] (10,0) -- (8,0) -- (0,0);
\draw[postaction={on each segment={mid arrow=black}}] (10,2) -- (8,2) -- (2,2) -- (0,2);
\draw[postaction={on each segment={mid arrow=black}}] (10,4) -- (2,4) -- (0,4);
\draw[postaction={on each segment={mid arrow=black}}] (8,0) -- (8,2);
\draw[postaction={on each segment={mid arrow=black}}] (2,2) -- (2,4);
\draw[blue, thick, postaction={on each segment={mid arrow=blue}}] (10,4) -- (2,4) -- (0,4);
\node[scale=.8,left] at (8,1) {$\rho_{v_2}^{(0,1)}$};
\node[scale=.8,left] at (2,3) {$\rho_{v_1}^{(0,1)}$};
\node[scale=.8,right] at (10,0) {$\rho^{(1,n_2^\ast+1)}$};
\node[scale=.8,left] at (0,0) {$\rho^{(1,n_2^\ast)}$};
\node[scale=.8,right] at (10,2) {$\rho^{(1,n_2)}$};
\node[scale=.8,right] at (10,4) {$\iota_{P_1}\rho^{(1,n_1)}$};
\node[scale=.7,above] at (5,2) {$\rho^{(1,n_2+1)}=\rho^{(1,n_1^\ast+1)}$};
\node[scale=.8,left] at (0,2) {$\rho^{(1,n_1^\ast)}$};
\node[scale=.8,left] at (0,4) {$\iota_{P_1}\rho^{(1,n_1+1)}$};
\node[scale=.8,below] at (8,0) {$\Phi_2^\ast$};
\node[scale=.8,above] at (8,2) {$\Phi_2$};
\node[scale=.8,below] at (2,2) {$\Phi_1^\ast$};
\node[scale=.8,above] at (2,4) {$\Phi_1^{P_1}$};
\node[scale=1,thick, red] at (5,5) {$\otimes$};
\draw[red] (5,5) -- (5,4);
\end{tikzpicture}
\hspace{3mm}
\begin{tikzpicture}[scale=.5]
\draw[postaction={on each segment={mid arrow=black}}] (11,0) -- (10,0) -- (0,0);
\draw[postaction={on each segment={mid arrow=black}}] (11,2) -- (10,2) -- (2,2) -- (0,2);
\draw[postaction={on each segment={mid arrow=black}}] (11,4) -- (2,4) -- (0,4);
\draw[postaction={on each segment={mid arrow=black}}] (10,0) -- (10,2);
\draw[postaction={on each segment={mid arrow=black}}] (2,2) -- (2,4);
\draw[blue, thick, postaction={on each segment={mid arrow=blue}}] (11,2) -- (10,2) -- (2,2) -- (2,4) -- (0,4);
\node[scale=.8,left] at (10,1) {$\rho_{v_2}^{(0,1)}$};
\node[scale=.8,left] at (2,3) {$\iota_{P_1}\rho_{v_1}^{(0,1)}$};
\node[scale=.8,right] at (11,0) {$\rho^{(1,n_2^\ast+1)}$};
\node[scale=.8,left] at (0,0) {$\rho^{(1,n_2^\ast)}$};
\node[scale=.8,right] at (11,2) {$\iota_{P_1}^\ast\rho^{(1,n_2)}$};
\node[scale=.8,right] at (11,4) {$\rho^{(1,n_1)}$};
\node[scale=.7,above] at (6,2) {$\iota_{P_1}^\ast\rho^{(1,n_2+1)}=\iota_{P_1}^\ast\rho^{(1,n_1^\ast+1)}$};
\node[scale=.8,left] at (0,2) {$\rho^{(1,n_1^\ast)}$};
\node[scale=.8,left] at (0,4) {$\iota_{P_1}\rho^{(1,n_1+1)}$};
\node[scale=.8,below] at (10,0) {$\Phi_2^\ast$};
\node[scale=.8,above] at (10,2) {$\Phi_2$};
\node[scale=.8,below] at (2,2) {$\Phi_1^{P_1\ast}$};
\node[scale=.8,above] at (2,4) {$\Phi_1$};
\node[scale=1,thick, red] at (6,3) {$\otimes$};
\end{tikzpicture}\\
\vspace{3mm}
\begin{tikzpicture}[scale=.5]
\draw[postaction={on each segment={mid arrow=black}}] (11,0) -- (9,0) -- (0,0);
\draw[postaction={on each segment={mid arrow=black}}] (11,2) -- (9,2) -- (2,2) -- (0,2);
\draw[postaction={on each segment={mid arrow=black}}] (11,4) -- (2,4) -- (0,4);
\draw[postaction={on each segment={mid arrow=black}}] (9,0) -- (9,2);
\draw[postaction={on each segment={mid arrow=black}}] (2,2) -- (2,4);
\draw[blue, thick, postaction={on each segment={mid arrow=blue}}] (11,2) -- (9,2) -- (2,2) -- (2,4) -- (0,4);
\node[scale=.8,left] at (9,1) {$\rho_{v_2}^{(0,1)}$};
\node[scale=.8,left] at (2,3) {$\iota_{P_1}^\ast\rho_{v_1}^{(0,1)}$};
\node[scale=.8,right] at (11,0) {$\rho^{(1,n_2^\ast+1)}$};
\node[scale=.8,left] at (0,0) {$\rho^{(1,n_2^\ast)}$};
\node[scale=.8,right] at (11,2) {$\iota_{P_1}^\ast\rho^{(1,n_2)}$};
\node[scale=.8,right] at (11,4) {$\rho^{(1,n_1)}$};
\node[scale=.7,above] at (5.5,2) {$\iota_{P_1}^\ast\rho^{(1,n_2+1)}=\iota_{P_1}^\ast\rho^{(1,n_1^\ast+1)}$};
\node[scale=.8,left] at (0,2) {$\rho^{(1,n_1^\ast)}$};
\node[scale=.8,left] at (0,4) {$\iota_{P_1}\rho^{(1,n_1+1)}$};
\node[scale=.8,below] at (9,0) {$\Phi_2^\ast$};
\node[scale=.8,above] at (9,2) {$\Phi_2$};
\node[scale=.8,below] at (2,2) {$\Phi_1^{\ast}$};
\node[scale=.8,above] at (2,4) {$\Phi_1^{P_1}$};
\node[scale=1,left, red] at (0,3) {$\otimes$};
\end{tikzpicture}
\hspace{3mm}
\begin{tikzpicture}[scale=.5]
\draw[postaction={on each segment={mid arrow=black}}] (11,0) -- (9,0) -- (0,0);
\draw[postaction={on each segment={mid arrow=black}}] (11,2) -- (9,2) -- (2,2) -- (0,2);
\draw[postaction={on each segment={mid arrow=black}}] (11,4) -- (2,4) -- (0,4);
\draw[postaction={on each segment={mid arrow=black}}] (9,0) -- (9,2);
\draw[postaction={on each segment={mid arrow=black}}] (2,2) -- (2,4);
\draw[blue, thick, postaction={on each segment={mid arrow=blue}}] (11,2) -- (9,2) -- (2,2) -- (0,2);
\node[scale=.8,left] at (9,1) {$\rho_{v_2}^{(0,1)}$};
\node[scale=.8,left] at (2,3) {$\rho_{v_1}^{(0,1)}$};
\node[scale=.8,right] at (11,0) {$\rho^{(1,n_2^\ast+1)}$};
\node[scale=.8,left] at (0,0) {$\rho^{(1,n_2^\ast)}$};
\node[scale=.8,right] at (11,2) {$\iota_{P_1}^\ast\rho^{(1,n_2)}$};
\node[scale=.8,right] at (11,4) {$\rho^{(1,n_1)}$};
\node[scale=.7,above] at (5.5,2) {$\iota_{P_1}^\ast\rho^{(1,n_2+1)}=\iota_{P_1}^\ast\rho^{(1,n_1^\ast+1)}$};
\node[scale=.8,left] at (0,2) {$\iota_{P_1}^\ast\rho^{(1,n_1^\ast)}$};
\node[scale=.8,left] at (0,4) {$\rho^{(1,n_1+1)}$};
\node[scale=.8,below] at (9,0) {$\Phi_2^\ast$};
\node[scale=.8,above] at (9,2) {$\Phi_2$};
\node[scale=.8,below] at (2,2) {$\Phi_1^{P_1\ast}$};
\node[scale=.8,above] at (2,4) {$\Phi_1$};
\node[scale=1,thick, red] at (6,1) {$\otimes$};
\draw[red] (6,1) -- (6,2);
\end{tikzpicture}
\end{center}
\caption{Algebraic realizations of 5d $\CN=1$ $U(1)\times U(1)$ gauge theory with hypermultiplets attached the first node}
\label{gauge3}
\end{figure}

\begin{figure}
\begin{center}
\begin{tikzpicture}[scale=.5]
\draw[postaction={on each segment={mid arrow=black}}] (11,0) -- (8,0) -- (0,0);
\draw[postaction={on each segment={mid arrow=black}}] (11,2) -- (8,2) -- (2,2) -- (0,2);
\draw[postaction={on each segment={mid arrow=black}}] (11,4) -- (2,4) -- (0,4);
\draw[postaction={on each segment={mid arrow=black}}] (9,0) -- (9,2);
\draw[postaction={on each segment={mid arrow=black}}] (2,2) -- (2,4);
\draw[blue, thick, postaction={on each segment={mid arrow=blue}}] (10,0) -- (9,0) -- (0,0);
\node[scale=.8,left] at (9,1) {$\rho_{v_2}^{(0,1)}$};
\node[scale=.8,left] at (2,3) {$\rho_{v_1}^{(0,1)}$};
\node[scale=.8,right] at (11,0) {$\iota_{P_2}^\ast\rho^{(1,n_2^\ast+1)}$};
\node[scale=.8,left] at (0,0) {$\iota_{P_2}^\ast\rho^{(1,n_2^\ast)}$};
\node[scale=.8,right] at (11,2) {$\rho^{(1,n_2)}$};
\node[scale=.8,right] at (11,4) {$\rho^{(1,n_1)}$};
\node[scale=.8,above] at (5.5,2) {$\rho^{(1,n_2+1)}=\rho^{(1,n_1^\ast+1)}$};
\node[scale=.8,left] at (0,2) {$\rho^{(1,n_1^\ast)}$};
\node[scale=.8,left] at (0,4) {$\rho^{(1,n_1+1)}$};
\node[scale=.8,below] at (9,0) {$\Phi_2^{P_2\ast}$};
\node[scale=.8,above] at (9,2) {$\Phi_2$};
\node[scale=.8,below] at (2,2) {$\Phi_1^\ast$};
\node[scale=.8,above] at (2,4) {$\Phi_1$};
\node[scale=1,thick, red] at (5,-1) {$\otimes$};
\draw[red] (5,-1) -- (5,0);
\end{tikzpicture}
% \hspace{3mm}
\begin{tikzpicture}[scale=.5]
\draw[postaction={on each segment={mid arrow=black}}] (11,0) -- (9,0) -- (0,0);
\draw[postaction={on each segment={mid arrow=black}}] (11,2) -- (9,2) -- (2,2) -- (0,2);
\draw[postaction={on each segment={mid arrow=black}}] (11,4) -- (2,4) -- (0,4);
\draw[postaction={on each segment={mid arrow=black}}] (9,0) -- (9,2);
\draw[postaction={on each segment={mid arrow=black}}] (2,2) -- (2,4);
\draw[blue, thick, postaction={on each segment={mid arrow=blue}}] (11,0) -- (9,0) -- (9,2)-- (2,2) -- (0,2);
\node[scale=.8,left] at (9,1) {$\iota_{P_2}\rho_{v_2}^{(0,1)}$};
\node[scale=.8,left] at (2,3) {$\rho_{v_1}^{(0,1)}$};
\node[scale=.8,right] at (11,0) {$\iota_{P_2}^\ast\rho^{(1,n_2^\ast+1)}$};
\node[scale=.8,left] at (0,0) {$\rho^{(1,n_2^\ast)}$};
\node[scale=.8,right] at (11,2) {$\rho^{(1,n_2)}$};
\node[scale=.8,right] at (11,4) {$\rho^{(1,n_1)}$};
\node[scale=.7,above] at (5.5,2) {$\iota_{P_2}\rho^{(1,n_2+1)}=\iota_{P_2}\rho^{(1,n_1^\ast+1)}$};
\node[scale=.8,left] at (0,2) {$\iota_{P_2}\rho^{(1,n_1^\ast)}$};
\node[scale=.8,left] at (0,4) {$\rho^{(1,n_1+1)}$};
\node[scale=.8,below] at (9,0) {$\Phi_2^{P_2\ast}$};
\node[scale=.8,above] at (9,2) {$\Phi_2$};
\node[scale=.8,below] at (2,2) {$\Phi_1^\ast$};
\node[scale=.8,above] at (2,4) {$\Phi_1$};
\node[scale=1,thick, red] at (10,1) {$\otimes$};
\end{tikzpicture}\\
\vspace{5mm}
\begin{tikzpicture}[scale=.5]
\draw[postaction={on each segment={mid arrow=black}}] (11,0) -- (9,0) -- (0,0);
\draw[postaction={on each segment={mid arrow=black}}] (11,2) -- (9,2) -- (2,2) -- (0,2);
\draw[postaction={on each segment={mid arrow=black}}] (11,4) -- (2,4) -- (0,4);
\draw[postaction={on each segment={mid arrow=black}}] (9,0) -- (9,2);
\draw[postaction={on each segment={mid arrow=black}}] (2,2) -- (2,4);
\draw[blue, thick, postaction={on each segment={mid arrow=blue}}] (11,0) -- (9,0) -- (9,2)-- (2,2) -- (0,2);
\node[scale=.8,left] at (9,1) {$\iota_{P_2}^\ast\rho_{v_2}^{(0,1)}$};
\node[scale=.8,left] at (2,3) {$\rho_{v_1}^{(0,1)}$};
\node[scale=.8,right] at (11,0) {$\iota_{P_2}^\ast\rho^{(1,n_2^\ast+1)}$};
\node[scale=.8,left] at (0,0) {$\rho^{(1,n_2^\ast)}$};
\node[scale=.8,right] at (11,2) {$\rho^{(1,n_2)}$};
\node[scale=.8,right] at (11,4) {$\rho^{(1,n_1)}$};
\node[scale=.7,above] at (5.5,2) {$\iota_{P_2}\rho^{(1,n_2+1)}=\iota_{P_2}\rho^{(1,n_1^\ast+1)}$};
\node[scale=.8,left] at (0,2) {$\iota_{P_2}\rho^{(1,n_1^\ast)}$};
\node[scale=.8,left] at (0,4) {$\rho^{(1,n_1+1)}$};
\node[scale=.8,below] at (9,0) {$\Phi_2^{\ast}$};
\node[scale=.8,above] at (9,2) {$\Phi_2^{P_2}$};
\node[scale=.8,below] at (2,2) {$\Phi_1^\ast$};
\node[scale=.8,above] at (2,4) {$\Phi_1$};
\node[scale=1,thick, red] at (5.5,1) {$\otimes$};
\end{tikzpicture}
\hspace{3mm}
\begin{tikzpicture}[scale=.5]
\draw[postaction={on each segment={mid arrow=black}}] (11,0) -- (9,0) -- (0,0);
\draw[postaction={on each segment={mid arrow=black}}] (11,2) -- (9,2) -- (2,2) -- (0,2);
\draw[postaction={on each segment={mid arrow=black}}] (11,4) -- (2,4) -- (0,4);
\draw[postaction={on each segment={mid arrow=black}}] (9,0) -- (9,2);
\draw[postaction={on each segment={mid arrow=black}}] (2,2) -- (2,4);
\draw[blue, thick, postaction={on each segment={mid arrow=blue}}] (11,2) -- (9,2)-- (2,2) -- (0,2);
\node[scale=.8,left] at (9,1) {$\rho_{v_2}^{(0,1)}$};
\node[scale=.8,left] at (2,3) {$\rho_{v_1}^{(0,1)}$};
\node[scale=.8,right] at (11,0) {$\rho^{(1,n_2^\ast+1)}$};
\node[scale=.8,left] at (0,0) {$\rho^{(1,n_2^\ast)}$};
\node[scale=.8,right] at (11,2) {$\iota_{P_2}\rho^{(1,n_2)}$};
\node[scale=.8,right] at (11,4) {$\rho^{(1,n_1)}$};
\node[scale=.7,above] at (5.5,2) {$\iota_{P_2}\rho^{(1,n_2+1)}=\iota_{P_2}\rho^{(1,n_1^\ast+1)}$};
\node[scale=.8,left] at (0,2) {$\iota_{P_2}\rho^{(1,n_1^\ast)}$};
\node[scale=.8,left] at (0,4) {$\rho^{(1,n_1+1)}$};
\node[scale=.8,below] at (9,0) {$\Phi_2^{\ast}$};
\node[scale=.8,above] at (9,2) {$\Phi_2^{P_2}$};
\node[scale=.8,below] at (2,2) {$\Phi_1^\ast$};
\node[scale=.8,above] at (2,4) {$\Phi_1$};
\node[scale=1,thick, red] at (5.5,3) {$\otimes$};
\draw[red] (5.5,3) -- (5.5,2);
\end{tikzpicture}
\end{center}
\caption{Algebraic realizations of massive 5d $\CN=1$ $U(1)\times U(1)$ with hypermultiplets attached the second node}
\label{gauge4}
\end{figure}
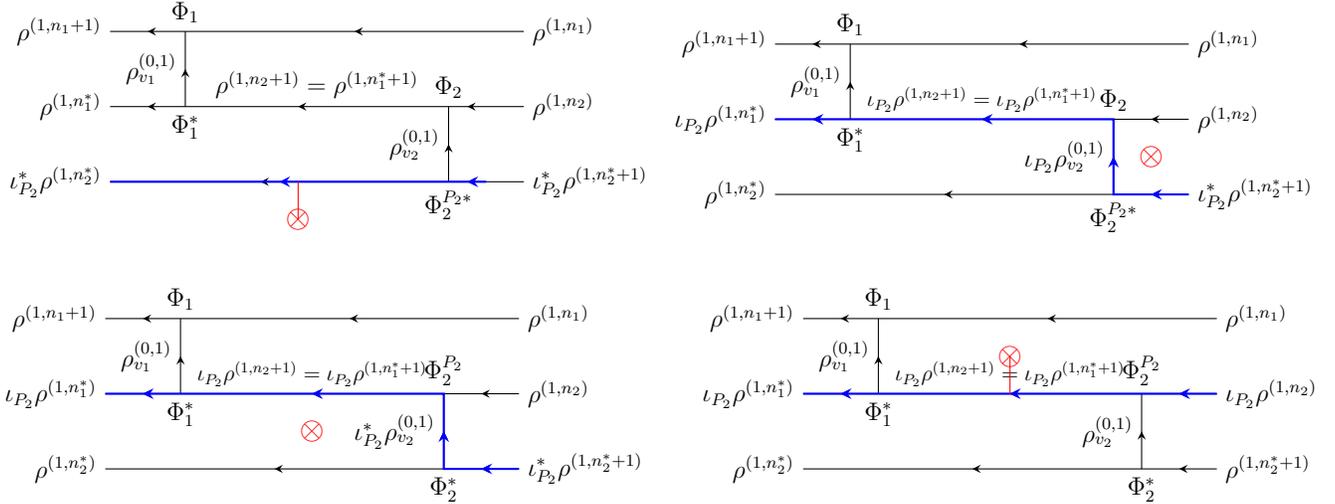

\begin{figure}
\begin{center}
\begin{tikzpicture}[scale=.5]
\draw[postaction={on each segment={mid arrow=black}}] (11,0) -- (9,0) -- (0,0);
\draw[postaction={on each segment={mid arrow=black}}] (11,2) -- (9,2) -- (2,2) -- (0,2);
\draw[postaction={on each segment={mid arrow=black}}] (11,4) -- (2,4) -- (0,4);
\draw[postaction={on each segment={mid arrow=black}}] (9,0) -- (9,2);
\draw[postaction={on each segment={mid arrow=black}}] (2,2) -- (2,4);
\draw[blue, thick, postaction={on each segment={mid arrow=blue}}] (11,4) -- (2,4) -- (0,4);
\draw[blue, thick, postaction={on each segment={mid arrow=blue}}] (11,0) -- (9,0) -- (0,0);
\node[scale=.8,left] at (9,1) {$\rho_{v_2}^{(0,1)}$};
\node[scale=.8,left] at (2,3) {$\rho_{v_1}^{(0,1)}$};
\node[scale=.7,right] at (11,0) {$\iota_{P_2}^\ast\rho^{(1,n_2^\ast+1)}$};
\node[scale=.8,left] at (0,0) {$\iota_{P_2}^\ast\rho^{(1,n_2^\ast)}$};
\node[scale=.8,right] at (11,2) {$\rho^{(1,n_2)}$};
\node[scale=.8,right] at (11,4) {$\iota_{P_1}\rho^{(1,n_1)}$};
\node[scale=.8,above] at (5.5,2) {$\rho^{(1,n_2+1)}=\rho^{(1,n_1^\ast+1)}$};
\node[scale=.8,left] at (0,2) {$\rho^{(1,n_1^\ast)}$};
\node[scale=.8,left] at (0,4) {$\iota_{P_1}\rho^{(1,n_1+1)}$};
\node[scale=.8,below] at (9,0) {$\Phi_2^{P_2\ast}$};
\node[scale=.8,above] at (9,2) {$\Phi_2$};
\node[scale=.8,below] at (2,2) {$\Phi_1^\ast$};
\node[scale=.8,above] at (2,4) {$\Phi_1^{P_1}$};
\node[scale=1,thick, red] at (5,5) {$\otimes$};
\node[scale=1,thick, green] at (5,-1) {$\otimes$};
\draw[red] (5,5) -- (5,4);
\draw[green] (5,-1) -- (5,0);
\end{tikzpicture}
% \hspace{3mm}
\begin{tikzpicture}[scale=.5]
\draw[postaction={on each segment={mid arrow=black}}] (11,0) -- (9,0) -- (0,0);
\draw[postaction={on each segment={mid arrow=black}}] (11,2) -- (9,2) -- (2,2) -- (0,2);
\draw[postaction={on each segment={mid arrow=black}}] (11,4) -- (2,4) -- (0,4);
\draw[postaction={on each segment={mid arrow=black}}] (9,0) -- (9,2);
\draw[postaction={on each segment={mid arrow=black}}] (2,2) -- (2,4);
\draw[blue, thick, postaction={on each segment={mid arrow=blue}}] (11,2) -- (9,2) -- (2,2) -- (0,2);
\draw[blue, thick, postaction={on each segment={mid arrow=blue}}] (11,0) -- (9,0) -- (9,2) -- (2,2) -- (2,4) -- (0,4);
\node[scale=.8,left] at (9,1) {$\iota_{P_2}\rho_{v_2}^{(0,1)}$};
\node[scale=.8,left] at (2,3) {$\iota_{P_1}\rho_{v_1}^{(0,1)}$};
\node[scale=.7,right] at (11,0) {$\iota_{P_2}^\ast\rho^{(1,n_2^\ast+1)}$};
\node[scale=.8,left] at (0,0) {$\rho^{(1,n_2^\ast)}$};
\node[scale=.8,right] at (11,2) {$\iota_{P_1}^\ast\rho^{(1,n_2)}$};
\node[scale=.8,right] at (11,4) {$\rho^{(1,n_1)}$};
\node[scale=.8,above] at (5.5,2) {$\iota_{P_2}\iota_{P_1}^\ast\rho^{(1,n_2+1)}$};
\node[scale=.8,below] at (5.5,2) {$=\iota_{P_2}\iota_{P_1}^\ast\rho^{(1,n_1^\ast+1)}$};
\node[scale=.8,left] at (0,2) {$\iota_{P_2}\rho^{(1,n_1^\ast)}$};
\node[scale=.8,left] at (0,4) {$\iota_{P_1}\rho^{(1,n_1+1)}$};
\node[scale=.8,below] at (9,0) {$\Phi_2^{P_2\ast}$};
\node[scale=.8,above] at (9,2) {$\Phi_2$};
\node[scale=.8,below] at (2,2) {$\Phi_1^{P_1\ast}$};
\node[scale=.8,above] at (2,4) {$\Phi_1$};
\node[scale=1,thick, red] at (5,3) {$\otimes$};
\node[scale=1,thick, green] at (11,1) {$\otimes$};
\end{tikzpicture}\\
\vspace{5mm}
\begin{tikzpicture}[scale=.5]
\draw[postaction={on each segment={mid arrow=black}}] (11,0) -- (9,0) -- (0,0);
\draw[postaction={on each segment={mid arrow=black}}] (11,2) -- (9,2) -- (2,2) -- (0,2);
\draw[postaction={on each segment={mid arrow=black}}] (11,4) -- (2,4) -- (0,4);
\draw[postaction={on each segment={mid arrow=black}}] (9,0) -- (9,2);
\draw[postaction={on each segment={mid arrow=black}}] (2,2) -- (2,4);
\draw[blue, thick,  postaction={on each segment={mid arrow=blue}}] (11,2) -- (9,2) -- (2,2) -- (0,2);
\draw[blue, thick, postaction={on each segment={mid arrow=blue}}] (11,0) -- (9,0) -- (9,2) -- (2,2) -- (2,4) -- (0,4);
\node[scale=.8,left] at (9,1) {$\iota_{P_2}\rho_{v_2}^{(0,1)}$};
\node[scale=.8,right] at (2,3) {$\iota_{P_1}^\ast\rho_{v_1}^{(0,1)}$};
\node[scale=.7,right] at (11,0) {$\iota_{P_2}^\ast\rho^{(1,n_2^\ast+1)}$};
\node[scale=.8,left] at (0,0) {$\rho^{(1,n_2^\ast)}$};
\node[scale=.8,right] at (11,2) {$\iota_{P_1}^\ast\rho^{(1,n_2)}$};
\node[scale=.8,right] at (11,4) {$\rho^{(1,n_1)}$};
\node[scale=.7,above] at (5.6,2) {$\iota_{P_2}\iota_{P_1}^\ast\rho^{(1,n_2+1)}$};
\node[scale=.7,below] at (5.6,2) {$=\iota_{P_2}\iota_{P_1}^\ast\rho^{(1,n_1^\ast+1)}$};
\node[scale=.8,left] at (0,2) {$\iota_{P_2}\rho^{(1,n_1^\ast)}$};
\node[scale=.8,left] at (0,4) {$\iota_{P_1}\rho^{(1,n_1+1)}$};
\node[scale=.8,below] at (9,0) {$\Phi_2^{P_2\ast}$};
\node[scale=.8,above] at (9,2) {$\Phi_2$};
\node[scale=.8,below] at (2,2) {$\Phi_1^{\ast}$};
\node[scale=.8,above] at (2,4) {$\Phi_1^{P_1}$};
\node[scale=1,thick, red] at (0,3) {$\otimes$};
\node[scale=1,thick, green] at (11,1) {$\otimes$};
\end{tikzpicture}
% \hspace{3mm}
\begin{tikzpicture}[scale=.5]
\draw[postaction={on each segment={mid arrow=black}}] (11,0) -- (9,0) -- (0,0);
\draw[postaction={on each segment={mid arrow=black}}] (11,2) -- (9,2) -- (2,2) -- (0,2);
\draw[postaction={on each segment={mid arrow=black}}] (11,4) -- (2,4) -- (0,4);
\draw[postaction={on each segment={mid arrow=black}}] (9,0) -- (9,2);
\draw[postaction={on each segment={mid arrow=black}}] (2,2) -- (2,4);
\draw[blue, thick,  postaction={on each segment={mid arrow=blue}}] (11,2) -- (9,2) -- (2,2) -- (0,2);
\draw[blue, thick, postaction={on each segment={mid arrow=blue}}] (11,0) -- (9,0) -- (9,2) -- (2,2) -- (2,4) -- (0,4);
\node[scale=.8,right] at (9,1) {$\iota_{P_2}^\ast\rho_{v_2}^{(0,1)}$};
\node[scale=.8,left] at (2,3) {$\iota_{P_1}\rho_{v_1}^{(0,1)}$};
\node[scale=.7,right] at (11,0) {$\iota_{P_2}^\ast\rho^{(1,n_2^\ast+1)}$};
\node[scale=.8,left] at (0,0) {$\rho^{(1,n_2^\ast)}$};
\node[scale=.8,right] at (11,2) {$\iota_{P_1}^\ast\rho^{(1,n_2)}$};
\node[scale=.8,right] at (11,4) {$\rho^{(1,n_1)}$};
\node[scale=.8,above] at (5.5,2) {$\iota_{P_2}\iota_{P_1}^\ast\rho^{(1,n_2+1)}$};
\node[scale=.8,below] at (5.5,2) {$=\iota_{P_2}\iota_{P_1}^\ast\rho^{(1,n_1^\ast+1)}$};
\node[scale=.8,left] at (0,2) {$\iota_{P_2}\rho^{(1,n_1^\ast)}$};
\node[scale=.8,left] at (0,4) {$\iota_{P_1}\rho^{(1,n_1+1)}$};
\node[scale=.8,below] at (9,0) {$\Phi_2^{\ast}$};
\node[scale=.8,above] at (9,2) {$\Phi_2^{P_2}$};
\node[scale=.8,below] at (2,2) {$\Phi_1^{P_1\ast}$};
\node[scale=.8,above] at (2,4) {$\Phi_1$};
\node[scale=1,thick, red] at (5,3) {$\otimes$};
\node[scale=1,thick, green] at (5,1) {$\otimes$};
\end{tikzpicture}\\
\vspace{3mm}
\begin{tikzpicture}[scale=.5]
\draw[postaction={on each segment={mid arrow=black}}] (11,0) -- (9,0) -- (0,0);
\draw[postaction={on each segment={mid arrow=black}}] (11,2) -- (9,2) -- (2,2) -- (0,2);
\draw[postaction={on each segment={mid arrow=black}}] (11,4) -- (2,4) -- (0,4);
\draw[postaction={on each segment={mid arrow=black}}] (9,0) -- (9,2);
\draw[postaction={on each segment={mid arrow=black}}] (2,2) -- (2,4);
\draw[blue, thick,  postaction={on each segment={mid arrow=blue}}] (11,2) -- (9,2) -- (2,2) -- (0,2);
\draw[blue, thick, postaction={on each segment={mid arrow=blue}}] (11,0) -- (9,0) -- (9,2) -- (2,2) -- (2,4) -- (0,4);
\node[scale=.8,right] at (9,1) {$\iota_{P_2}^\ast\rho_{v_2}^{(0,1)}$};
\node[scale=.8,right] at (2,3) {$\iota_{P_1}^\ast\rho_{v_1}^{(0,1)}$};
\node[scale=.7,right] at (11,0) {$\iota_{P_2}^\ast\rho^{(1,n_2^\ast+1)}$};
\node[scale=.8,left] at (0,0) {$\rho^{(1,n_2^\ast)}$};
\node[scale=.8,right] at (11,2) {$\iota_{P_1}^\ast\rho^{(1,n_2)}$};
\node[scale=.8,right] at (11,4) {$\rho^{(1,n_1)}$};
\node[scale=.7,above] at (5.6,2) {$\iota_{P_2}\iota_{P_1}^\ast\rho^{(1,n_2+1)}$};
\node[scale=.7,below] at (5.6,2) {$=\iota_{P_2}\iota_{P_1}^\ast\rho^{(1,n_1^\ast+1)}$};
\node[scale=.8,left] at (0,2) {$\iota_{P_2}\rho^{(1,n_1^\ast)}$};
\node[scale=.8,left] at (0,4) {$\iota_{P_1}\rho^{(1,n_1+1)}$};
\node[scale=.8,below] at (9,0) {$\Phi_2^{\ast}$};
\node[scale=.8,above] at (9,2) {$\Phi_2^{P_2}$};
\node[scale=.8,below] at (2,2) {$\Phi_1^{\ast}$};
\node[scale=.8,above] at (2,4) {$\Phi_1^{P_1}$};
\node[scale=1,thick, red] at (0,3) {$\otimes$};
\node[scale=1,thick, green] at (5,1) {$\otimes$};
\end{tikzpicture}
% \hspace{3mm}
\begin{tikzpicture}[scale=.5]
\draw[postaction={on each segment={mid arrow=black}}] (11,0) -- (9,0) -- (0,0);
\draw[postaction={on each segment={mid arrow=black}}] (11,2) -- (9,2) -- (2,2) -- (0,2);
\draw[postaction={on each segment={mid arrow=black}}] (11,4) -- (2,4) -- (0,4);
\draw[postaction={on each segment={mid arrow=black}}] (9,0) -- (9,2);
\draw[postaction={on each segment={mid arrow=black}}] (2,2) -- (2,4);
\draw[blue, thick,  postaction={on each segment={mid arrow=blue}}] (11,2) -- (9,2) -- (2,2) -- (0,2);
\draw[blue, thick, postaction={on each segment={mid arrow=blue}}] (11,0) -- (9,0) -- (9,2) -- (2,2) -- (0,2);
\node[scale=.8,right] at (9,1) {$\iota_{P_2}^\ast\rho_{v_2}^{(0,1)}$};
\node[scale=.8,right] at (2,3) {$\rho_{v_1}^{(0,1)}$};
\node[scale=.7,right] at (11,0) {$\iota_{P_2}^\ast\rho^{(1,n_2^\ast+1)}$};
\node[scale=.8,left] at (0,0) {$\rho^{(1,n_2^\ast)}$};
\node[scale=.8,right] at (11,2) {$\iota_{P_1}^\ast\rho^{(1,n_2)}$};
\node[scale=.8,right] at (11,4) {$\rho^{(1,n_1)}$};
\node[scale=.8,above] at (5.5,2) {$\iota_{P_2}\iota_{P_1}^\ast\rho^{(1,n_2+1)}$};
\node[scale=.8,below] at (5.5,2) {$=\iota_{P_2}\iota_{P_1}^\ast\rho^{(1,n_1^\ast+1)}$};
\node[scale=.7,left] at (0,2) {$\iota_{P_2}\iota_{P_1}^\ast\rho^{(1,n_1^\ast)}$};
\node[scale=.8,left] at (0,4) {$\rho^{(1,n_1+1)}$};
\node[scale=.8,below] at (9,0) {$\Phi_2^{\ast}$};
\node[scale=.8,above] at (9,2) {$\Phi_2^{P_2}$};
\node[scale=.8,below] at (2,2) {$\Phi_1^{P_1\ast}$};
\node[scale=.8,above] at (2,4) {$\Phi_1$};
\node[scale=1,thick, red] at (4,1) {$\otimes$};
\draw[red] (4,1) -- (4,2);
\node[scale=1,thick, green] at (6,1) {$\otimes$};
\end{tikzpicture}
\end{center}
\caption{Different realizations of massive 5d $\CN=1$ $U(1)\times U(1)$ gauge theory with hypermultiplets attached to both nodes}
\label{gauge5}
\end{figure}
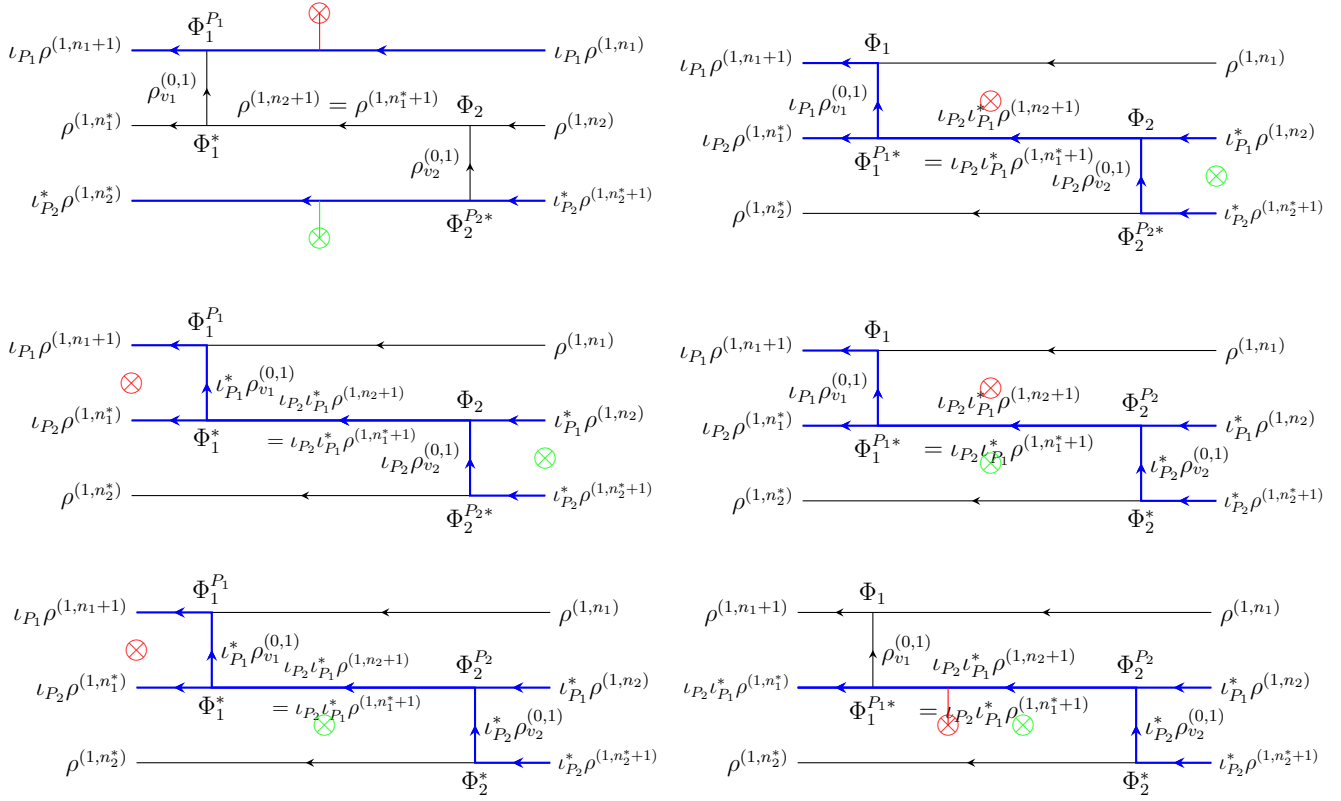

% \begin{figure}
% \begin{center}
% \begin{tikzpicture}[scale=.5]
% \draw[postaction={on each segment={mid arrow=black}}] (10,0) -- (8,0) -- (0,0);
% \draw[postaction={on each segment={mid arrow=black}}] (10,2) -- (8,2) -- (2,2) -- (0,2);
% \draw[postaction={on each segment={mid arrow=black}}] (10,4) -- (2,4) -- (0,4);
% \draw[postaction={on each segment={mid arrow=black}}] (8,0) -- (8,2);
% \draw[postaction={on each segment={mid arrow=black}}] (2,2) -- (2,4);
% % \draw[blue, thick, postaction={on each segment={mid arrow=blue}}] (8,0) -- (6,0) -- (6,2) -- (2,2) -- (0,2);
% \node[scale=.7,left] at (8,1) {$\rho_{v_2}^{(0,1)}$};
% \node[scale=.7,left] at (2,3) {$\rho_{v_1}^{(0,1)}$};
% \node[scale=.7,below right] at (10,0) {$\rho^{(1,n_2^\ast+1)}$};
% \node[scale=.7,below left] at (0,0) {$\rho^{(1,n_2^\ast)}$};
% \node[scale=.7,above right] at (10,2) {$\rho^{(1,n_2)}$};
% \node[scale=.7,above right] at (10,4) {$\rho^{(1,n_1)}$};
% \node[scale=.7,above] at (5,2) {$\rho^{(1,n_2+1)}=\rho^{(1,n_1^\ast+1)}$};
% \node[scale=.7,above left] at (0,2) {$\rho^{(1,n_1^\ast)}$};
% \node[scale=.7,above left] at (0,4) {$\rho^{(1,n_1+1)}$};
% \node[scale=.7,below] at (8,0) {$\Phi_2^\ast$};
% \node[scale=.7,above] at (8,2) {$\Phi_2$};
% \node[scale=.7,below] at (2,2) {$\Phi_1^\ast$};
% \node[scale=.7,above] at (2,4) {$\Phi_1$};
% \end{tikzpicture}
% \hspace{3mm}
% \end{center}
% \caption{Algebraic realizations of massive 5d $\CN=1$ $U(1)\times U(1)$ gauge theory I}
% \label{fig4}
% \end{figure}

The partition function of the pure $U(1)\times U(1)$ gauge theory coincides with the vev of the operator
\begin{align}
\begin{split}
\sum_{\l^{(1)},\l^{(2)}}n_{\l^{(1)}}n_{\l^{(2)}}\ \Phi^\ast_{\l^{(2)}}[u_2^\ast,v_2,n_2^\ast]\otimes\Phi_{\l^{(1)}}^\ast[u_1^\ast,v_1,n_1^\ast]\Phi_{\l^{(2)}}[u_2,v_2,n_2]\otimes\Phi_{\l^{(1)}}[u_1,v_1,n_1].
\end{split}
\end{align}
The introduction of hypermultiplets for only one of the two gauge groups can be realized with one of the four diagrams of figures \ref{gauge3} (first node) or \ref{gauge4} (second node). These configurations are easily combined to describe a gauge theory with hypermultiplets for both nodes, and some examples are shown on the figure \ref{gauge5}. The representations associated to the external edges carry one of the four following shifts: $(1\otimes\iota_{P_2}\otimes \iota_{P_1})$, $(\iota_{P_2}^\ast\otimes1\otimes \iota_{P_1})$, $(\iota_{P_2}^\ast\otimes \iota_{P_1}^\ast\otimes1)$, $(1\otimes\iota_{P_2}\iota_{P_1}^\ast\otimes 1)$. They are responsible for bringing the extra factors $P_1(\g z)$ and $P_1(\g^2 z)P_2(\g z)$ in the evaluation of the coproduct $(\D\otimes 1)\D(x^+(z))$ which also enter in the expression of the fundamental qq-characters.

\subsubsection{Hanany-Witten transitions}
\begin{figure}
\begin{center}
\begin{subfigure}[c]{0.2\textwidth}
\begin{tikzpicture}[scale=.5]
\draw (-.7,-.7) -- (0,0) -- (2,0) -- (2.7,-.7) ;
\draw (-.7,-.7) -- (0,0) -- (0,2) -- (-.7,2.7)-- (-.7,3.7);
\draw (-.7,2.7) -- (0,2) -- (2,2) -- (2.7,2.7);
\draw[red] (-.7,2.7) -- (-1.7,2.7);
\draw (2,0) -- (2,2);
\draw[red] (0,0) -- (2,0);
\draw[red] (0,2) -- (2,2);
% \node[scale=.6,left] at (0,1) {NS5};
\node[scale=.7,below] at (1,0) {D5};
% \node[scale=.6,right] at (2,1) {NS5};
\node[scale=.7,below] at (1,2) {D5};
\node[scale=.7,below] at (-1.3,2.7) {D5};
% \node[scale=.6,above right] at (2.7,2.7) {$(1,1)$};
% \node[scale=.6,below left] at (-.7,-.7) {$(1,1)$};
% \node[scale=.6,above left] at (-.7,2.7) {$(-1,1)$};
% \node[scale=.6,below right] at (2.7,-.7) {$(-1,1)$};
\end{tikzpicture}
\end{subfigure}
% \hspace{15mm}
\begin{subfigure}[c]{0.2\textwidth}
\begin{tikzpicture}[scale=.5]
\draw (-.7,-.7) -- (0,0) -- (2,0) -- (2.7,-.7) ;
\draw (-.7,-.7) -- (0,0) -- (0,2) -- (-.7,2.7)-- (-.7,3.7);
\draw (-.7,2.7) -- (0,2) -- (2,2) -- (2.7,2.7);
\draw[red] (-.7,2.7) -- (-1.7,2.7);
\draw (2,0) -- (2,2);
\draw[red] (0,0) -- (2,0);
\draw[red] (0,2) -- (2,2);
% \node[scale=.6,left] at (0,1) {NS5};
\node[scale=.7,below] at (1,0) {D5};
% \node[scale=.6,right] at (2,1) {NS5};
\node[scale=.7,below] at (1,2) {D5};
\node[scale=.7,below] at (-1.3,2.7) {D5};
\node at (-1.7,2.7) {$\otimes$};
% \node[scale=.6,above right] at (2.7,2.7) {$(1,1)$};
% \node[scale=.6,below left] at (-.7,-.7) {$(1,1)$};
% \node[scale=.6,above left] at (-.7,2.7) {$(-1,1)$};
% \node[scale=.6,below right] at (2.7,-.7) {$(-1,1)$};
\end{tikzpicture}
\end{subfigure}
\begin{subfigure}[c]{0.2\textwidth}
\begin{tikzpicture}[scale=.5]
\draw (-.7,-.7) -- (0,0) -- (2,0) -- (2.7,-.7) ;
\draw (-.7,-.7) -- (0,0) -- (0,2) -- (-.7,2.7);
\draw (-.7,2.7) -- (0,2) -- (2,2) -- (2.7,2.7);
% \draw[red] (-.7,2.7) -- (-1.7,2.7);
\draw (2,0) -- (2,2);
\draw[red] (0,0) -- (2,0);
\draw[red] (0,2) -- (2,2);
% \node[scale=.6,left] at (0,1) {NS5};
\node[scale=.7,below] at (1,0) {D5};
% \node[scale=.6,right] at (2,1) {NS5};
\node[scale=.7,below] at (1,2) {D5};
\node at (1,2.7) {$\otimes$};
% \node[scale=.6,below] at (-1.3,2.7) {D5};
% \node[scale=.6,above right] at (2.7,2.7) {$(1,1)$};
% \node[scale=.6,below left] at (-.7,-.7) {$(1,1)$};
% \node[scale=.6,above left] at (-.7,2.7) {$(-1,1)$};
% \node[scale=.6,below right] at (2.7,-.7) {$(-1,1)$};
\end{tikzpicture}
\end{subfigure}
\begin{subfigure}[c]{0.2\textwidth}
\begin{tikzpicture}[scale=.5]
\draw (-.7,-.7) -- (0,0) -- (2,0) -- (2.7,-.7) ;
\draw (-.7,-.7) -- (0,0) -- (0,2) -- (-.7,2.7);
\draw (-.7,2.7) -- (0,2) -- (2,2) -- (2.7,2.7) -- (2.7,3.7);
\draw[red] (2.7,2.7) -- (3.7,2.7);
\draw (2,0) -- (2,2);
\draw[red] (0,0) -- (2,0);
\draw[red] (0,2) -- (2,2);
% \node[scale=.6,left] at (0,1) {NS5};
\node[scale=.7,below] at (1,0) {D5};
% \node[scale=.6,right] at (2,1) {NS5};
\node[scale=.7,below] at (1,2) {D5};
\node[scale=.7,below] at (3.3,2.7) {D5};
\node at (3.7,2.7) {$\otimes$};
% \node[scale=.6,above right] at (2.7,2.7) {$(1,1)$};
% \node[scale=.6,below left] at (-.7,-.7) {$(1,1)$};
% \node[scale=.6,above left] at (-.7,2.7) {$(-1,1)$};
% \node[scale=.6,below right] at (2.7,-.7) {$(-1,1)$};
\end{tikzpicture}
\end{subfigure}
\end{center}
\caption{Different brane realizations of the $U(2)$ gauge theory with $N^f=1$}
\label{U2_Nf}
\end{figure}
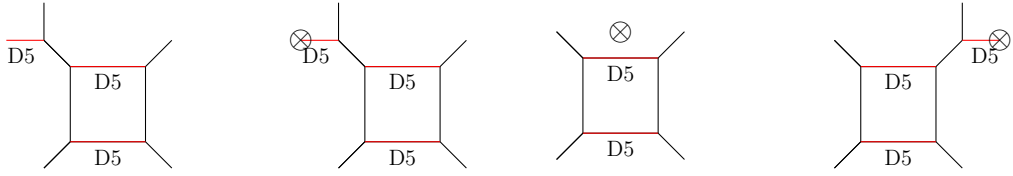

It is tempting to associate the equivalent networks of shifted representations to different brane realizations of the gauge theories. Hypermultiplets are usually introduced using semi-infinite D5-branes ending on a (dressed) NS5-brane. Instead of extending to infinity, the D5-branes can also end on a transverse D7-brane extending in the directions 01234789. Using the Hanany-Witten transition which states that a D5-brane is created whenever a D7-brane crosses a NS5-branes, we find another brane realization in which the D7-branes are free and lie between two NS5-branes. These different possibilities are represented on the figure \ref{U2_Nf} for the case of a $U(2)$ gauge theory with one fundamental hypermultiplet. The transverse D7-brane is represented by the symbol $\otimes$. These brane configurations can be compared with the representation networks drawn on figure \ref{gauge2}.\footnote{Recall that the brane web is flattened and rotated by 90$^\circ$, so that D5-branes become vertical lines and dressed NS5-branes horizontal lines.} We represented our conjecture for the position of the D7-brane by the red symbol $\otimes$. Hence, the first two and the last two networks of representations in figure \ref{gauge2} correspond to free D7-branes between NS5-branes (at different positions with respect to the D5-branes), while the two networks in the middle have the D7-branes attached to a dressed NS5-branes through a D5-brane (in red). Similar conjectures are made for the networks of figures \ref{gauge1}, \ref{gauge3}, \ref{gauge4} and \ref{gauge5}.

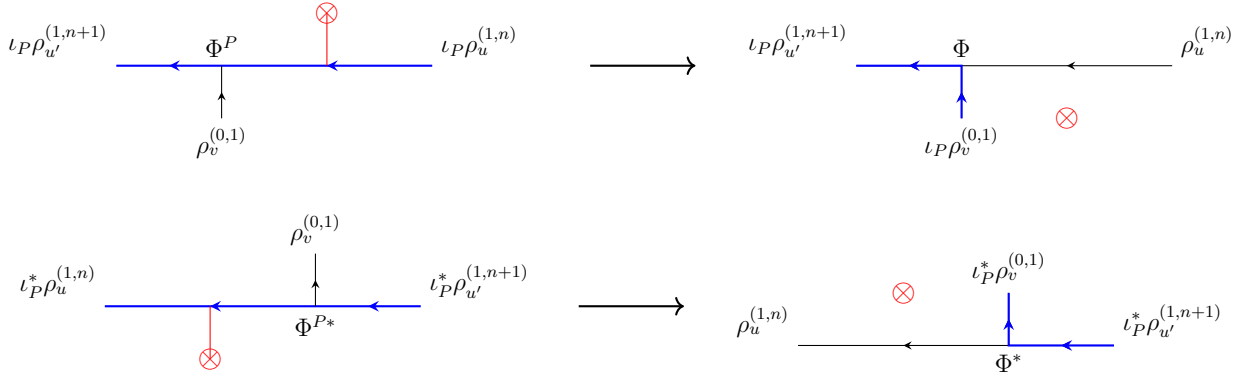
\begin{figure}
\begin{center}
\begin{tikzpicture}[scale=.7]
\draw[blue, thick,postaction={on each segment={mid arrow=blue}}] (6,0) -- (2,0) -- (0,0);
\draw[postaction={on each segment={mid arrow=black}}] (2,-1) -- (2,0) ;
\node[scale=.8,below] at (2,-1) {$\rho_{v}^{(0,1)}$};
\node[scale=.8,above left] at (0,0) {$\iota_P\rho_{u'}^{(1,n+1)}$};
\node[scale=.8,above right] at (6,0) {$\iota_P\rho_{u}^{(1,n)}$};
\node[scale=.8,above] at (2,0) {$\Phi^P$};
\node[scale=1,red] at (4,1) {$\otimes$};
\draw[red] (4,1) -- (4,0);
\draw[->,thick] (9,0) -- (11,0);
\end{tikzpicture}
% \hspace{3mm}
% \begin{tikzpicture}[scale=.7]
% \draw[->,thick] (0,3) -- (2,3);
% \end{tikzpicture}
\hspace{3mm}
\begin{tikzpicture}[scale=.7]
\draw[postaction={on each segment={mid arrow=black}}] (6,0) -- (2,0) -- (0,0);
\draw[postaction={on each segment={mid arrow=black}}] (2,-1) -- (2,0) ;
\draw[blue, thick,postaction={on each segment={mid arrow=blue}}] (2,-1) -- (2,0) -- (0,0);
\node[scale=.8,below] at (2,-1) {$\iota_P\rho_{v}^{(0,1)}$};
\node[scale=.8,above left] at (0,0) {$\iota_P\rho_{u'}^{(1,n+1)}$};
\node[scale=.8,above right] at (6,0) {$\rho_{u}^{(1,n)}$};
\node[scale=.8,above] at (2,0) {$\Phi$};
\node[scale=1,red] at (4,-1) {$\otimes$};
\end{tikzpicture}\\
\vspace{5mm}
\begin{tikzpicture}[scale=.7]
\draw[blue, thick,postaction={on each segment={mid arrow=blue}}] (6,0) -- (4,0) -- (0,0);
\draw[postaction={on each segment={mid arrow=black}}] (4,0) -- (4,1) ;
\node[scale=.8,above] at (4,1) {$\rho_{v}^{(0,1)}$};
\node[scale=.8,above left] at (0,0) {$\iota_P^\ast\rho_{u}^{(1,n)}$};
\node[scale=.8,above right] at (6,0) {$\iota_P^\ast\rho_{u'}^{(1,n+1)}$};
\node[scale=.8,below] at (4,0) {$\Phi^{P\ast}$};
\node[scale=1,red] at (2,-1) {$\otimes$};
\draw[red] (2,-1) -- (2,0);
\draw[->,thick] (9,0) -- (11,0);
\end{tikzpicture}
% \hspace{3mm}
% \begin{tikzpicture}[scale=.7]
% \draw[->,thick] (0,1) -- (2,1);
% \end{tikzpicture}
\hspace{3mm}
\begin{tikzpicture}[scale=.7]
\draw[postaction={on each segment={mid arrow=black}}] (6,0) -- (4,0) -- (0,0);
\draw[postaction={on each segment={mid arrow=black}}] (4,0) -- (4,1) ;
\draw[blue, thick,postaction={on each segment={mid arrow=blue}}] (6,0) -- (4,0) -- (4,1);
\node[scale=.8,above] at (4,1) {$\iota_P^\ast\rho_{v}^{(0,1)}$};
\node[scale=.8,above left] at (0,0) {$\rho_{u}^{(1,n)}$};
\node[scale=.8,above right] at (6,0) {$\iota_P^\ast\rho_{u'}^{(1,n+1)}$};
\node[scale=.8,below] at (4,0) {$\Phi^{\ast}$};
\node[scale=1,red] at (2,1) {$\otimes$};
\end{tikzpicture}
\end{center}
\caption{Hanany-Witten transitions}
\label{figHW}
\end{figure}

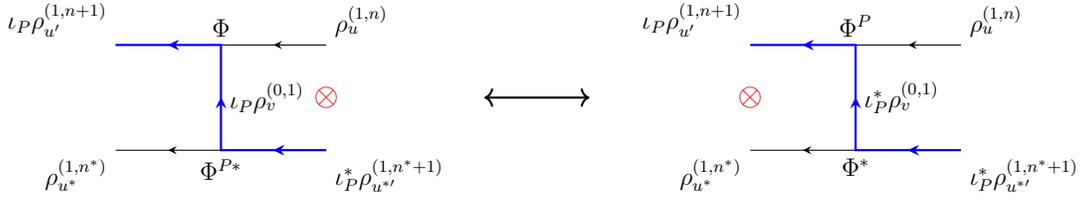
\begin{figure}
\begin{center}
\begin{tikzpicture}[scale=.7]
\draw[postaction={on each segment={mid arrow=black}}] (4,0) -- (2,0) -- (0,0);
\draw[postaction={on each segment={mid arrow=black}}] (2,0) -- (2,2);
\draw[postaction={on each segment={mid arrow=black}}] (4,2) -- (2,2) -- (0,2);
\draw[blue, thick, postaction={on each segment={mid arrow=blue}}] (4,0) -- (2,0) -- (2,2) -- (0,2);
\node[scale=.8,right] at (2,1) {$\iota_P\rho_v^{(0,1)}$};
\node[scale=.8,below right] at (4,0) {$\iota_P^\ast\rho_{u^{\ast\prime}}^{(1,n^\ast+1)}$};
\node[scale=.8,below left] at (0,0) {$\rho_{u^\ast}^{(1,n^\ast)}$};
\node[scale=.8,above right] at (4,2) {$\rho_u^{(1,n)}$};
\node[scale=.8,above left] at (0,2) {$\iota_P\rho_{u'}^{(1,n+1)}$};
\node[scale=.8,below] at (2,0) {$\Phi^{P\ast}$};
\node[scale=.8,above] at (2,2) {$\Phi$};
\node[scale=1,thick, red] at (4,1) {$\otimes$};
\draw[<->,thick] (7,1) -- (9,1);
\end{tikzpicture}
\hspace{3mm}
\begin{tikzpicture}[scale=.7]
\draw[postaction={on each segment={mid arrow=black}}] (4,0) -- (2,0) -- (0,0);
\draw[postaction={on each segment={mid arrow=black}}] (2,0) -- (2,2);
\draw[postaction={on each segment={mid arrow=black}}] (4,2) -- (2,2) -- (0,2);
\draw[blue, thick, postaction={on each segment={mid arrow=blue}}] (4,0) -- (2,0) -- (2,2) -- (0,2);
\node[scale=.8,right] at (2,1) {$\iota_P^\ast\rho_v^{(0,1)}$};
\node[scale=.8,below right] at (4,0) {$\iota_P^\ast\rho_{u^{\ast\prime}}^{(1,n^\ast+1)}$};
\node[scale=.8,below left] at (0,0) {$\rho_{u^\ast}^{(1,n^\ast)}$};
\node[scale=.8,above right] at (4,2) {$\rho_u^{(1,n)}$};
\node[scale=.8,above left] at (0,2) {$\iota_P\rho_{u'}^{(1,n+1)}$};
\node[scale=.8,below] at (2,0) {$\Phi^{\ast}$};
\node[scale=.8,above] at (2,2) {$\Phi^P$};
\node[scale=1,thick, red] at (0,1) {$\otimes$};
\end{tikzpicture}
\end{center}
\caption{Vertical transition}
\label{figV}
\end{figure}

Our conjecture for the position of the D7-branes follows from the assumption that the Hanany-Witten transition is realized algebraically as shown on figure \ref{figHW}. In addition, we also assumed another transition when a D7-brane crosses a D5-brane which brings the modification in the choice of representations drawn on figure \ref{figV}. Unfortunately, since all these algebraic descriptions give the same instanton partition function and qq-characters, it appears very difficult to check our conjectures.

\subsection{Matter vertex operators}\label{sec_mass}
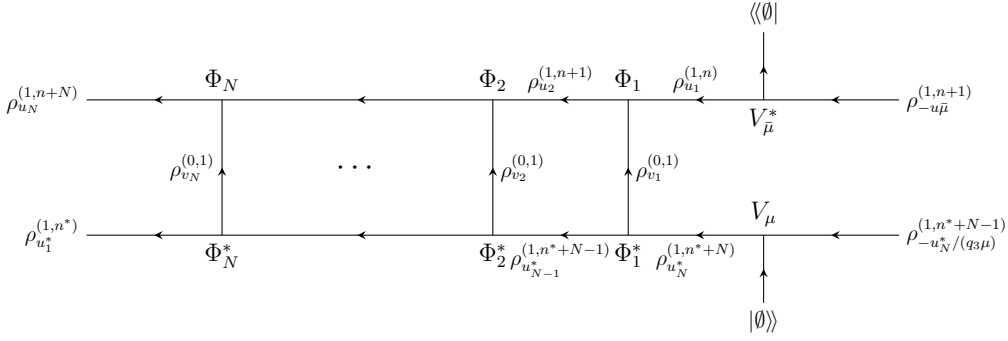
\begin{figure}
\begin{center}
\begin{tikzpicture}[scale=.9]
\draw[postaction={on each segment={mid arrow=black}}] (12,0) -- (10,0) -- (8,0) -- (6,0) -- (2,0) -- (0,0);
\draw[postaction={on each segment={mid arrow=black}}] (2,0) -- (2,2);
\draw[postaction={on each segment={mid arrow=black}}] (6,0) -- (6,2);
\draw[postaction={on each segment={mid arrow=black}}] (8,0) -- (8,2);
\draw[postaction={on each segment={mid arrow=black}}] (12,2) -- (10,2) -- (8,2) -- (6,2) -- (2,2) -- (0,2);
\draw[postaction={on each segment={mid arrow=black}}] (10,-1) -- (10,0);
\draw[postaction={on each segment={mid arrow=black}}] (10,2) -- (10,3);
\node[scale=.7,above] at (10,3) {$\dbra{\vac}$};
\node[scale=.7,below] at (10,-1) {$\dket{\vac}$};
\node[scale=.7,left] at (2,1) {$\rho_{v_N}^{(0,1)}$};
\node[scale=.7,right] at (8,1) {$\rho_{v_1}^{(0,1)}$};
\node[scale=.7,right] at (6,1) {$\rho_{v_2}^{(0,1)}$};
\node[scale=.7,below] at (9,0) {$\rho_{u_N^\ast}^{(1,n^\ast+N)}$};
\node[scale=.7,right] at (12,0) {$\rho_{-u_N^\ast/(q_3\mu)}^{(1,n^\ast+N-1)}$};
\node[scale=.7,below] at (7,0) {$\rho_{u_{N-1}^\ast}^{(1,n^\ast+N-1)}$};
\node[scale=.7,left] at (0,0) {$\rho_{u_1^\ast}^{(1,n^\ast)}$};
\node[scale=.7,above] at (9,2) {$\rho_{u_1}^{(1,n)}$};
\node[scale=.7,right] at (12,2) {$\rho_{-u\bar{\mu}}^{(1,n+1)}$};
\node[scale=.7,above] at (7,2) {$\rho_{u_2}^{(1,n+1)}$};
\node[scale=.7,left] at (0,2) {$\rho_{u_N}^{(1,n+N)}$};
\node[scale=.8,below] at (2,0) {$\Phi_N^\ast$};
\node[scale=.8,above] at (2,2) {$\Phi_N$};
\node[scale=.8,below] at (8,0) {$\Phi_1^\ast$};
\node[scale=.8,above] at (8,2) {$\Phi_1$};
\node[scale=.8,below] at (6,0) {$\Phi_2^\ast$};
\node[scale=.8,above] at (6,2) {$\Phi_2$};
\node[scale=.8,below] at (10,2) {$V_{\bar{\mu}}^\ast$};
\node[scale=.8,above] at (10,0) {$V_{\mu}$};
\node at (4,1) {$\cdots$};
\end{tikzpicture}
\end{center}
\caption{Representation network for the 5d $\CN=1$ $U(N)$ gauge theory with $N^f=N^{\bar{f}}=1$}
\label{fig_matter}
\end{figure}

When hypermultiplets are introduced in the brane web as semi-infinite D5-branes, attached either to the right-most (fundamental) or left-most (antifundamental) dressed NS5-branes, they can be realized in the algebraic description as an insertion of \text{matter vertex operators} in the Fock module attached to the dressed NS5-brane,
\begin{align}
\begin{split}
&V_\mu=\Phi[u,q_3^{1/2}\mu,n](\dket{\vac}\otimes1)=\Phi_\vac[u,q_3^{1/2}\mu,n],\\
&V_{\bar{\mu}}^\ast=(\dbra{\vac}\otimes1)\Phi^\ast[u,q_3^{-1/2}\bar{\mu},n]=\Phi_\vac^\ast[u,q_3^{-1/2}\bar{\mu},n].
\end{split}
\end{align}
The parameters $u$ and $n$ are fixed to match the representation of $\CE$ acting on this module \cite{Bourgine2017b}. To compare this procedure with the shift of representations introduced previously, we restrict ourselves to the gauge group $U(N)$ and consider the network of representations of figure \ref{fig_matter}. For generic values of the masses and Coulomb branch vevs (i.e. $\bar{\mu}\neq vq_1^kq_2^l$ $\forall k,l\in\mZ$), the mass vertex operator $V_{\bar\mu}^\ast$ commutes with the intertwiner $\Phi$ and the position of insertion is irrelevant,
\begin{center}
\begin{tikzpicture}[scale=.7]
\draw[postaction={on each segment={mid arrow=black}}] (8,0) -- (6,0) -- (2,0) -- (0,0);
\draw[postaction={on each segment={mid arrow=black}}] (2,-2) -- (2,0);
\draw[postaction={on each segment={mid arrow=black}}] (6,0) -- (6,1);
\node[scale=.7,left] at (0,0) {$\rho_{u_0}^{(1,n+1)}$};
\node[scale=.7,right] at (8,0) {$\rho_{u_1}^{(1,n+1)}$};
\node[scale=.7,below] at (2,-2) {$\rho_{v}^{(0,1)}$};
\node[scale=.7,above] at (4,0) {$\rho_{-u_0/(\g v)}^{(1,n)}$};
\node[scale=.7, above] at (6,1) {$\dbra{\vac}$};
\node[scale=.8,above] at (2,0) {$\Phi$};
\node[scale=.8,below] at (6,0) {$V_{\bar{\mu}}^\ast$};
\node[scale=1] at (11,0) {$=$};
\end{tikzpicture}
\hspace{5mm}
\begin{tikzpicture}[scale=.7]
\draw[postaction={on each segment={mid arrow=black}}] (8,0) -- (6,0) -- (2,0) -- (0,0);
\draw[postaction={on each segment={mid arrow=black}}] (6,-2) -- (6,0);
\draw[postaction={on each segment={mid arrow=black}}] (2,0) -- (2,1);
\node[scale=.7,left] at (0,0) {$\rho_{u_0}^{(1,n+1)}$};
\node[scale=.7,right] at (8,0) {$\rho_{u_1}^{(1,n+1)}$};
\node[scale=.7,below] at (6,-2) {$\rho_{v}^{(0,1)}$};
\node[scale=.7,above] at (4,0) {$\rho_{-\g u_1 v}^{(1,n+2)}$};
\node[scale=.7,above] at (2,1) {$\dbra{\vac}$};
\node[scale=.8,above] at (6,0) {$\Phi$};
\node[scale=.8,below] at (2,0) {$V_{\bar{\mu}}^\ast$};
\end{tikzpicture}
\end{center}
with $u_0\bar{\mu}=\g vu_1$.\footnote{This equality requires to adjust the weights and level according to the intermediate representation which is different. It is easy to see that the operator part matches on both sides, and the result follows from comparing the vevs
\begin{align}
\begin{split}
&\la\Phi_{\l_1}[-u_0/(\g v),v,n]\Phi_{\l_2}^\ast[u_1,\g^{-1}\bar\mu,n]\ra=\dfrac{N_{\l_2,\l_1}(\bar\mu/v)}{\CG(q_3^{-1}\bar\mu/v)}u_0^{|\l_1|}(\g u_1)^{-|\l_2|}\prod_{\sAbox\in\l_1}\chi_{\sAbox}^{-n-1}\prod_{\sAbox\in\l_2}\chi_{\sAbox}^n,\\
&\la\Phi_{\l_2}^\ast[-\g u_1v,\g^{-1}\bar\mu,n+1]\Phi_{\l_1}[u_1,v,n+1]\ra=\dfrac{N_{\l_1,\l_2}(q_3v/\bar\mu)}{\CG(v/\bar\mu)}(-\g u_1v)^{|\l_1|}(-q_3u_1\bar\mu)^{-|\l_2|}\prod_{\sAbox\in\l_1}\chi_{\sAbox}^{-n-2}\prod_{\sAbox\in\l_2}\chi_{\sAbox}^{n+1},
\end{split}
\end{align}
using the reflection formula for the Nekrasov factor (see e.g. footnote 6 in \cite{Bourgine2017b}). The two expressions match up to the one-loop factor involving the function $\CG(z)$, and the result follows from the specialization at $\l_2=\vac$. We refer to the appendix B of \cite{Bourgine2017b} for the definition of the function $\CG(z)$.} The same is true for the operator $V_{\mu}$ and the intertwiner $\Phi^\ast$. Comparing the two formalisms, we observe that the factors obtained from normal ordering the product of operators $\Phi_\l$ and $V_{\bar{\mu}}^\ast$ gives exactly the quantity $\prod_{\sAbox\in\l}P_{\bar{\mu}}^{\bar{f}}(\chi_\sAbox)$ introduced in the definition of the shifted intertwiner $\Phi^{P}$ with $P=P_{\bar{\mu}}^{\bar{f}}$,
\begin{equation}
\Phi_{\l}[u,v,n]V_{\bar{\mu}}^\ast=\CG(\bar{\mu}/(q_3v))^{-1}\ :\Phi_{\l}^P[u,v,n]V_{\bar{\mu}}^\ast:,
\end{equation} 
up to a $\l$-independent one-loop contribution that can be absorbed in an overall rescaling of $t_\l^{P\ast}$. Upon evaluation of the vacuum expectation value, the presence of $V_{\bar{\mu}}^\ast$ in the normal-ordered product brings no further contribution. However, when $V_{\bar{\mu}}^\ast$ is inserted inside a product $\Phi_1\Phi_2\cdots\Phi_N$, all intertwiners are replaced by their shifted versions. This is equivalent to the replacement of all the horizontal representations $\rho_u^{(1,n)}$ by their shifted versions $\iota_P\rho_u^{(1,n)}$. In the same way, the insertion of $V_\mu$ inside a product $\Phi_1^\ast\Phi_2^\ast\cdots\Phi_N^\ast$ is equivalent to the replacement of the intertwiners $\Phi_\a^\ast$ by their shifted version $\Phi_\a^{P\ast}$ with $P=P_\mu^f$, effectively replacing the horizontal representations $\rho_{u^\ast}^{(1,n^\ast)}$ with  $\iota^\ast_P\rho_{u^\ast}^{(1,n^\ast)}$.

The matter vertex operators carry a vertical representation $\rho_v^{(0,1)}$ of $\CE$, and so they are interpreted as the insertion of a D5-brane attached to a transverse D7-brane. Replacing these operators by shifted intertwiners as explained above, we end up with the type of configurations represented on the left of figure \ref{figHW} which were indeed associated to the insertion of hypermultiplets using D5-D7 branes, in contrast with the configurations on the right involving free D7-branes. Thus, this interpretation is in agreement with our conjecture on the position of D7-branes.

\begin{figure}
\begin{center}
\begin{tikzpicture}[scale=.7]
\draw[postaction={on each segment={mid arrow=black}}] (4,0) -- (2,0);
\draw[blue, thick, postaction={on each segment={mid arrow=blue}}] (2,-2) -- (2,0) -- (0,0);
\node[scale=.8,left] at (0,0) {$\iota_P\rho_{-q_3u\mu}^{(1,n+1)}$};
\node[scale=.8,below] at (2,-2) {$\rho_{\tilde{P}}$};
\node[scale=.8,right] at (4,0) {$\rho_u^{(1,n)}$};
\node[scale=.8,above] at (2,0) {$V_\mu$};
\end{tikzpicture}
\hspace{10mm}
\begin{tikzpicture}[scale=.7]
\draw[postaction={on each segment={mid arrow=black}}] (2,0) -- (0,0);
\draw[blue, thick, postaction={on each segment={mid arrow=blue}}] (4,0) -- (2,0) -- (2,2);
\node[scale=.8,left] at (0,0) {$\rho_u^{(1,n)}$};
\node[scale=.8,above] at (2,2) {$\rho_{\tilde{P}}$};
\node[scale=.8,right] at (4,0) {$\iota_P^\ast\rho_{-u\bar\mu}^{(1,n+1)}$};
\node[scale=.8,below] at (2,0) {$V_{\bar\mu}^\ast$};
\end{tikzpicture}
\end{center}
\caption{Network of representations involving the matter vertex operators $V_\mu$ (left) and $V_{\bar\mu}^\ast$ (right).}
\label{fig_PhiP}
\end{figure}
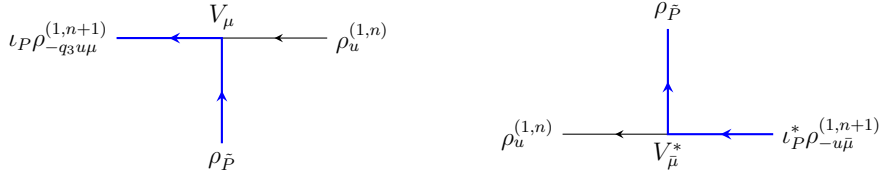

\paragraph{Matter operators as shifted intertwiners} Conversely, matter vertex operators can be realized as shifted intertwiners by tuning the zeros of the polynomials so that $P(\chi_\sAbox)=0$ for the first box $\Abox=(1,1)\in\l$. As a result $\prod_{\sAbox\in\l}P(\chi_\sAbox)$ vanishes unless $\l=\vac$, and the summations over $\l$ in the definition \ref{def_Phi} of the AFS intertwiners reduce to a single term,
\begin{align}
\begin{split}
&\Phi^P[u,q_3^{1/2}\mu,n]=\dbra{\vac}\otimes V_\mu,\quad P(z)=P_\mu^f(q_3^{1/2}z),\\
&\Phi^{P\ast}[u,q_3^{-1/2}\bar{\mu},n]=\dket{\vac}\otimes V_{\bar{\mu}}^\ast,\quad P(z)=P_{\bar{\mu}}^{\bar{f}}(q_3^{1/2}z).
\end{split}
\end{align}
For this choice of polynomial, the shifted vertical representation $\iota_P\rho_v^{(0,1)}$ reduces to the one dimensional representation $\rho_{\tilde{P}}$ used to build shifted representations, with $\tilde{P}(z)=q_3^{1/2}P(q_3^{-1}z)$ for $\Phi^P$ and $\tilde{P}(z)=q_3^{-1/2}P(q_3^{-1}z)$ for $\Phi^{P\ast}$. Matter vertex operators can also be seen as intertwiners involving this particular representation.\footnote{Note that when $\iota_P\rho_v^{(0,1)}$ is of dimension one, there is no difference between the operators $\Phi$ and $\Phi^P$ since $\prod_{\sAbox}P(\chi_\sAbox)=1$.} Using the intertwining properties summarized in the table of figure \ref{table2} and selecting the ones involving $\iota_P\rho_v^{(0,1)}$, we find a new interpretation of matter vertex operators which has been represented on figure \ref{fig_PhiP}. It is clear from this realization that the insertion of matter vertex operators has the effect of shifting the horizontal representation $\rho_u^{(1,n)}$ into either $\iota_P\rho_{-q_3u\mu}^{(1,n+1)}$ or $\iota_P^\ast\rho_{-u\bar\mu}^{(1,n+1)}$. This shift will then propagate through the network.

\section{Vortex partition functions and Higgsing}
The intertwiners $\Phi$ and $\Phi^\ast$ of the shifted quantum affine $\sl(2)$ algebra constructed in the section \ref{sec_3d_intw} have been used in \cite{Bourgine2021d} to reconstruct the partition function and qq-characters of a set of 3d $\CN=2$ gauge theories. These theories can be obtained from the massive 5d $\CN=1$ gauge theories considered in the previous section through a two steps procedure. The first step is called Higgsing, and consists in adjusting the mass of some hypermultiplets to a critical value. In this way, we obtain a 3d $\CN=2$ gauge theory with an adjoint chiral multiplet of mass $m_\Phi=-\e_1$ \cite{Hellerman2011,Nedelin2017,Aprile2018}. The second step is to send the omega background parameter $\e_1\to\infty$ in order to decouple the fields of this multiplet.

We start this section with a short review of the vortex partition functions for 3d $\CN=2$ gauge theories. Then, we briefly recall the engineering technique presented in \cite{Bourgine2021d} and based on the shifted quantum affine $\sl(2)$ algebras $\CU^\bmu$. We present the analysis of a pure gauge theory with $G=U(1)\times U(1)$ gauge group as an example of application to (A-type) quiver theories. Finally, we discuss the Higgsing procedure from an algebraic point of view using the relations between the shifted algebras $\CE^\bmu$ and $\CU^\bmu$ observed in the section \ref{sec_SQA_sl2}.

\subsection{3d \texorpdfstring{$\CN=2$}{N=2} gauge theories}\label{sec_COHA}
The supersymmetric gauge theories with four supercharges ($\CN=2$) considered in this section are obtained as massive deformations of 3d $\CN=4$ gauge theories. The Lagrangian of $\CN=4$ theories is built out of vector and hyper multiplets, it is fully specified by the choice of gauge group $G$ and representation of hypermultiplets. Here, we restrict ourselves to the gauge group $G=\times U(N_\a)$ of a linear quiver gauge theory. Turning on the omega-background in dimensions transverse to the spacetime breaks the $\CN=4$ supersymmetry to $\CN=2$, splitting the $\CN=4$ vector multiplet into a $\CN=2$ vector multiplet and an adjoint chiral multiplet to which it gives the real mass $m_\Phi=-\e_1$ \cite{Hellerman2011}.\footnote{In contrast with the `complex masses' entering in the superpotential, `real masses' correspond to a deformation of the supersymmetry algebra.} We are interested in the limit $\e_1\to\infty$ in which this adjoint chiral multiplet fully decouples. The $\CN=4$ hypermultiplets split into two $\CN=2$ chiral multiplets $(q_l,\tilde{q}^l)$ with $q_l$ in the fundamental and $\tilde{q}^l$ in the antifundamental representation of the gauge group. In fact, a different mass can be chosen for each chiral multiplet in the $\CN=2$ Lagrangian, and we can further decouple some of them by sending their mass to infinity. Thus, in general, to each node $\a$ of the quiver is attached $N_\a^f\geq N$ chiral multiplets in the fundamental representation with exponentiated masses $(\nu_1^{(\a)},\cdots,\nu_{N_\a^f}^{(\a)})$, and $N_\a^{\bar{f}}$ chiral multiplets in the antifundamental representations with exponentiated masses $(\bar{\nu}_1^{(\a)},\cdots,\bar{\nu}_{N_\a^{\bar{f}}}^{(\a)})$.

In the case of a single node, i.e. of the gauge group $G=U(N)$, the partition function has been evaluated by Higgs branch localization on the omega-deformed background $\mC_\e\times S^1$ in \cite{Yoshida2011,Fujitsuka2013,Chen2013}. This spacetime defines the quantum group parameter $q^2=e^{-R\e}$ where $R$ is the radius of $S^1$. Up to permutation, the vacuum can be chosen such that the first $N$ chiral multiplets have a non-zero vacuum expectation value $\z^{1/2}$, while all other fields vanish. The $N$ masses $\bnu=(\nu_1,\cdots,\nu_{N})$ play the role of the 5d Coulomb branch vevs in the localization formulas, and we denote the vector of remaining fundamental chiral masses as $\bnu^f=(\nu_{N+1},\cdots,\nu_{N^f})$ and the antifundamental chiral masses as $\bnu^{\bar{f}}=(\bar{\nu}_1,\cdots,\bar{\nu}_{N^{\bar{f}}})$. The partition function receives factorized contributions from classical, one-loop and non-perturbative vortex corrections, and we focus here on the latter. The vortex partition function is a sum over the $N$-tuples of vortex numbers $\bk=(k_1,\cdots,k_N)\in(\mZ^{\geq0})^N$ where $k_i$ is the vorticity for the subgroup $U(1)\subset U(N)$ on the $i$th diagonal, 
\begin{equation}\label{CZ_V}
Z_V[U(N),N^f,N^{\bar{f}}]=\sum_{\bk}\qf^{\sum_lk_l}\Zv(\bk,\bnu)\ZCS(\bk,\bnu,\k)\Zf(\bk,\bnu,\bnu^f)\Zaf(\bk,\bnu,\bnu^{\bar{f}}).
\end{equation} 
Each configuration is weighted by the vortex counting parameter $\qf$, and the summands factorize into contributions from the various $\CN=2$ multiplets and the Chern-Simons term (of level $\k$),
\begin{align}\label{vortex_UN}
\begin{split}
&\Zv(\bk,\bnu)=\prod_{l,l'=1}^NN_{k_l,k_{l'}}(\nu_l/\nu_{l'})^{-1},\quad \ZCS(\bk,\bnu,\k)=\prod_{l=1}^N (\nu_lq^{k_l(k_l-1)})^{\k},\quad \Zf(\bk,\bnu,\bnu^f)=\prod_{l=1}^N\prod_{j=1}^{k_l}p_{\bnu^f}(\nu_lq^{2j-2})^{-1},\\
&\Zaf(\bk,\bnu,\bnu^{\bar{f}})=\prod_{l=1}^N\prod_{j=1}^{k_l}p_{\bnu^{\bar{f}}}(\nu_lq^{2j-2}),\quad \text{with}\quad p_{\bnu}(z)=\prod_a(1-z/\nu_a).
\end{split}
\end{align}
The notation $\Zv(\bk,\bnu)$ is a little abusive here as it includes the contribution from the first $N$ fundamental chiral multiplets with masses $\bnu$. Its expression involves a 3d version of the Nekrasov factor expressed using the q-Pochhammer symbol  (defined in \ref{def_qPoch}),
\begin{equation}\label{def_Nkk}
N_{k,k'}(\a)=(\a q^{2k-2k'+2};q^2)_{k'}.
\end{equation} 

\subsection{Algebraic engineering}
The algebraic engineering of 3d $\CN=2$ theories is based on the fact that the normal-ordering of the intertwiners $\Phi$ and $\Phi^\ast$ of $\CU^\bmu$ reproduce the 3d Nekrasov factor \ref{def_Nkk},\footnote{In these expressions, we used the shortcut notation $A(z)B(w)::f(z/w)$ for $A(z)B(w)=f(z/w):A(z)B(w):$. To say it differently,
\begin{equation}
A(z)B(w)::f(z/w)\quad \Leftrightarrow\quad f(z/w)=\dfrac{\la A(z)B(w)\ra}{\la A(z)\ra\la B(w)\ra}.
\end{equation} Normal-ordering relations of this type appear often in this context, and this trick shortens many formulas.}
\begin{align}\label{NO_3d}
\begin{split}
&\Phi_{k_1}[u_1,\nu_1,n_1]\Phi_{k_2}[u_2,\nu_2,n_2]::q^{-2k_1k_2}N_{k_2,k_1}(\nu_2/\nu_1)^{-1},\\
&\Phi_{k_1}[u_1,\nu_1,n_1]\Phi_{k_2}^\ast[u_2,\nu_2,n_2]::\dfrac{N_{k_1,k_2}(\nu_1/\nu_2)}{N_{0,k_2}(\nu_1/\nu_2)},\\
&\Phi_{k_1}^\ast[u_1,\nu_1,n_1]\Phi_{k_2}[u_2,\nu_2,n_2]::(q^2\nu_2/\nu_1;q^2)_\infty\dfrac{N_{k_2,k_1}(\nu_2/\nu_1)}{(q^2\nu_2/\nu_1;q^2)_{k_2}},\\
% &\Phi_{k_1}^\ast[u_1,\nu_1,n_1]\Phi_{k_2}[u_2,\nu_2,n_2]::(-\nu_1)^{k_2}\nu_2^{-k_2}q^{2k_1k_2-k_2(k_2+1)}\dfrac{(q^{2-2k_1}\nu_2/\nu_1;q^2)_\infty}{N_{k_1,k_2}(q^{-2}\nu_1/\nu_2)},\\
&\Phi_{k_1}^\ast[u_1,\nu_1,n_1]\Phi_{k_2}^\ast[u_2,\nu_2,n_2]::(-\nu_1)^{k_2}\nu_2^{-k_2}q^{2k_1k_2-k_2(k_2-1)}N_{k_1,k_2}(\nu_1/\nu_2)^{-1}.
\end{split}
\end{align}

\paragraph{Remark} In the section \ref{sec_SQA_sl2}, the intertwiners $\tPhi$ and $\tPhi^\ast$ were also introduced. Their normal-ordering relations
\begin{align}
\begin{split}\label{NO_3d_2}
&\tPhi_{k_1}[u_1,\nu_1,n_1]\tPhi_{k_2}[u_2,\nu_2,n_2]::q^{-2k_1k_2}\tN_{k_2,k_1}(\nu_2/\nu_1)^{-1},\\
&\tPhi_{k_1}[u_1,\nu_1,n_1]\tPhi_{k_2}^\ast[u_2,\nu_2,n_2]::q^{2k_1k_2}\tN_{k_2,k_1}(q^{-2}\nu_2/\nu_1)^{-1},\\
&\tPhi_{k_1}^\ast[u_1,\nu_1,n_1]\tPhi_{k_2}[u_2,\nu_2,n_2]::\tN_{k_2,k_1}(\nu_2/\nu_1),\\
&\tPhi_{k_1}^\ast[u_1,\nu_1,n_1]\tPhi_{k_2}^\ast[u_2,\nu_2,n_2]::\tN_{k_2,k_1}(q^{-2}\nu_2/\nu_1),
\end{split}
\end{align}
involve a different quantity $\tN_{k,k'}(\a)=(\a q^{2k-2k'+2};q^2)_\infty$. It is related to the 3d Nekrasov factor \ref{def_Nkk} as follows,
\begin{equation}\label{def_tNkk}
\tN_{k,k'}(\a)=(\a q^{2k+2},q^2)_\infty N_{k,k'}(\a).
\end{equation}
We note the reflection properties
\begin{equation}\label{prop_Nkk}
\dfrac{N_{k,k'}(\a)}{N_{0,k'}(\a)}=q^{2kk'}\dfrac{N_{0,k}(q^{-2}\a^{-1})}{N_{k',k}(q^{-2}\a^{-1})},\quad \dfrac{\tN_{k,k'}(\a)}{\tN_{0,k'}(\a)}=(-\a)^{k}q^{2kk'-k(k+1)}\dfrac{\tN_{k',0}(q^{-2}\a^{-1})}{\tN_{k',k}(q^{-2}\a^{-1})}.
\end{equation}

At the moment, the string theory interpretation of the intertwiners $\tPhi$ and $\tPhi^\ast$ is not fully clear, but rewriting the product
\begin{equation}
\prod_{l,l'=1}^N\tN_{k_l,k_{l'}}(\nu_l/\nu_{l'})^{-1}=\prod_{l,l'=1}^{N}(q^2\nu_l/\nu_{l'};q^2)_\infty^{-1} \times\Zv(\bk,\bnu)\Zaf(\bk,\bnu,q^{-2}\bnu),
\end{equation} 
hints toward an explanation involving the extra insertion of $N$ antifundamental chiral multiplets of mass $q^{-2}\nu_1,\cdots,q^{-2}\nu_N$ (the $\bk$-independent prefactors $(\nu_l/\nu_{l'};q^2)_\infty$ are one-loop contributions). Thus, the corresponding theory should contain $N$ $\CN=4$ hypermultiplets broken down to $N$ chiral fundamental and $N$ chiral antifundamental multiplets by the omega-background. But the situation is not entirely clear due to the presence of spurious $\th_{q^2}$-functions that can be seen e.g. by rewriting
\begin{equation}
\tPhi_{k_1}^\ast[u_1,\nu_1,n_1]\tPhi_{k_2}^\ast[u_2,\nu_2,n_2]::(-\nu_1)^{k_1}(\nu_2)^{-k_1}q^{2k_1k_2-k_1(k_1+1)}\th_{q^2}(q^{2k_2}\nu_2/\nu_1) \tN_{k_1,k_2}(\nu_1/\nu_2)^{-1},
\end{equation} 
and enter in the expression of the partition functions.\footnote{We recall the definition
\begin{equation}
\th_{q^2}(z)=\prod_{j=0}^\infty(1-zq^{2j})(1-z^{-1}q^{2j+2})=(z;q^2)_\infty(z^{-1}q^2;q^2)_\infty.
\end{equation}}
In fact, similar factors are also encountered in the Higgs network calculus proposed in \cite{Zenkevich2018}.

\subsubsection{Gauge theories with \texorpdfstring{$U(N)$}{U(N)} gauge group}
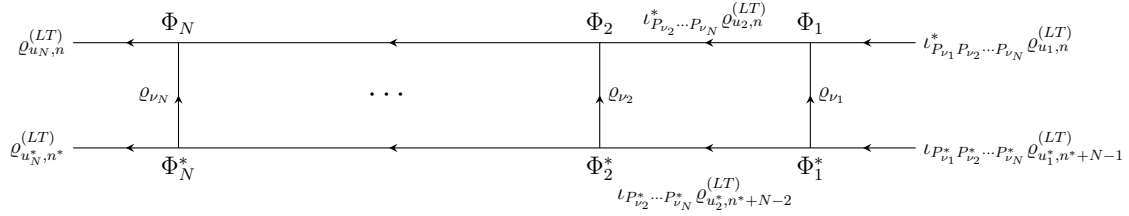
\begin{figure}
\begin{center}
\begin{tikzpicture}[scale=1.4]
\draw[postaction={on each segment={mid arrow=black}}] (9,0) -- (8,0) -- (6,0) -- (2,0) -- (1,0);
\draw[postaction={on each segment={mid arrow=black}}] (2,0) -- (2,1);
\draw[postaction={on each segment={mid arrow=black}}] (6,0) -- (6,1);
\draw[postaction={on each segment={mid arrow=black}}] (8,0) -- (8,1);
\draw[postaction={on each segment={mid arrow=black}}] (9,1) -- (8,1) -- (6,1) -- (2,1) -- (1,1);
\node[scale=.7,left] at (2,.5) {$\vrho_{\nu_N}$};
\node[scale=.7,right] at (8,.5) {$\vrho_{\nu_1}$};
\node[scale=.7,right] at (6,.5) {$\vrho_{\nu_2}$};
\node[scale=.7,right] at (9,0) {$\iota_{P_{\nu_1}^\ast P_{\nu_2}^\ast\cdots P_{\nu_N}^\ast}\vrho_{u_1^\ast,n^\ast+N-1}^{(LT)}$};
\node[scale=.7,below] at (7,-.2) {$\iota_{P_{\nu_2}^\ast\cdots P_{\nu_N}^\ast}\vrho_{u_2^\ast,n^\ast+N-2}^{(LT)}$};
\node[scale=.7,left] at (1,0) {$\vrho_{u_N^\ast,n^\ast}^{(LT)}$};
\node[scale=.7,right] at (9,1) {$\iota^\ast_{P_{\nu_1}P_{\nu_2}\cdots P_{\nu_N}}\vrho_{u_1,n}^{(LT)}$};
\node[scale=.7,above] at (7,1) {$\iota^\ast_{P_{\nu_2}\cdots P_{\nu_N}}\vrho_{u_2,n}^{(LT)}$};
\node[scale=.7,left] at (1,1) {$\vrho_{u_N,n}^{(LT)}$};
\node[scale=.8,below] at (2,0) {$\Phi_N^\ast$};
\node[scale=.8,above] at (2,1) {$\Phi_N$};
\node[scale=.8,below] at (8,0) {$\Phi_1^\ast$};
\node[scale=.8,above] at (8,1) {$\Phi_1$};
\node[scale=.8,below] at (6,0) {$\Phi_2^\ast$};
\node[scale=.8,above] at (6,1) {$\Phi_2$};
\node at (4,.5) {$\cdots$};
\end{tikzpicture}
\end{center}
\caption{Network of representations for the 3d $\CN=2$ $U(N)$ gauge theory with $N$ fundamental chiral multiplets.}
\label{fig_UN_3d}
\end{figure}

The pure $U(N)$ 3d $\CN=2$ gauge theory has been associated to the network of representations of figure \ref{fig_UN_3d} with $u_{l}=u$, $u_{l}^\ast=u^\ast\prod_{l'>l}(-\nu_{l'})$, $P_\nu(z)=1-\nu/z$ and $P_\nu^\ast(z)=-\nu^{-1}z(1-q^2z/\nu)$.\footnote{One should not confuse the transformation $\iota_{P_\nu^\ast}$, that is $\iota_P$ with the polynomial $P_\nu^\ast$, and $\iota_{P_\nu}^\ast$ which corresponds to $\iota_P^\ast$ with the polynomial $P_\nu$. In the same way, $\Phi^{P_\nu\ast}$ denotes the intertwiner $\Phi^\ast$ shifted by the polynomial $P_\nu$ while $\Phi^{P_\nu^\ast}$ is the intertwiner $\Phi$ shifted by $P_\nu^\ast$.} It defines the operator
\begin{align}
\begin{split}
T[U(N),0,0]=\sum_{k_1,k_2,\cdots,k_N=0}^\infty \prod_{l=1}^N n_{k_l}\ &\Phi^\ast_{k_N}[u^\ast_N,\nu_N,n^\ast_N]\cdots\Phi^\ast_{k_2}[u^\ast_2,\nu_2,n^\ast_2]\Phi^\ast_{k_1}[u^\ast_1,\nu_1,n^\ast_1] \\
&\otimes\Phi_{k_N}[u_N,\nu_N,n_N]\cdots\Phi_{k_2}[u_2,\nu_2,n_2]\Phi_{k_1}[u_1,\nu_1,n_1].
\end{split}
\end{align}
We recover indeed the vortex partition function $Z[U(N),0,0]$ by evaluating the vev $\la T[U(N),0,0]\ra$, upon the identifications $\k=n^\ast-n$ and $\qf=q^2 u/u^\ast$. Chiral multiplets in the antifundamental representations can be introduced using shifted representations just like in the case of 5d $\CN=1$ hypermultiplets. On the other hand, it appears difficult to introduce fundamental chiral multiplets in this formalism since $p_{\bnu^f}$ enters in $\Zf(\bk,\bnu,\bnu^f)$ with a negative power.

\subsubsection{Quiver gauge theory}
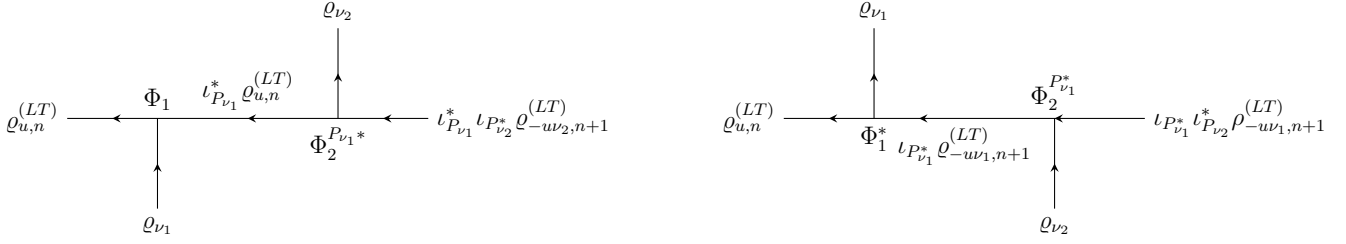
\begin{figure}
\begin{center}
\begin{tikzpicture}[scale=1.2]
\draw[postaction={on each segment={mid arrow=black}}] (4,0) -- (3,0) -- (1,0) -- (0,0);
\draw[postaction={on each segment={mid arrow=black}}] (3,0) -- (3,1);
\draw[postaction={on each segment={mid arrow=black}}] (1,-1) -- (1,0);
\node[above,scale=0.8] at (1,0) {$\Phi_1$};
\node[below,scale=0.8] at (3,0) {$\Phi_2^{P_{\nu_1}\ast}$};
\node[below,scale=0.8] at (1,-1) {$\vrho_{\nu_1}$};
\node[above,scale=0.8] at (3,1) {$\vrho_{\nu_2}$};
\node[left,scale=0.8] at (0,0) {$\vrho_{u,n}^{(LT)}$};
\node[above,scale=0.8] at (2,0) {$\iota_{P_{\nu_1}}^\ast\vrho_{u,n}^{(LT)}$};
\node[right,scale=0.8] at (4,0) {$\iota_{P_{\nu_1}}^\ast\iota_{P_{\nu_2}^\ast}\vrho_{-u\nu_2,n+1}^{(LT)}$};
\end{tikzpicture}
\hspace{10mm}
\begin{tikzpicture}[scale=1.2]
\draw[postaction={on each segment={mid arrow=black}}] (3,0) -- (1,0) -- (0,0) -- (-1,0);
\draw[postaction={on each segment={mid arrow=black}}] (0,0) -- (0,1);
\draw[postaction={on each segment={mid arrow=black}}] (2,-1) -- (2,0);
\node[below,scale=0.8] at (0,0) {$\Phi_1^\ast$};
\node[above,scale=0.8] at (2,0) {$\Phi_2^{P_{\nu_1}^\ast}$};
\node[above,scale=0.8] at (0,1) {$\vrho_{\nu_1}$};
\node[below,scale=0.8] at (2,-1) {$\vrho_{\nu_2}$};
\node[left,scale=0.8] at (-1,0) {$\vrho_{u,n}^{(LT)}$};
\node[below,scale=0.8] at (1,0) {$\iota_{P_{\nu_1}^\ast}\vrho_{-u\nu_1,n+1}^{(LT)}$};
\node[right,scale=0.8] at (3,0) {$\iota_{P_{\nu_1}^\ast}\iota_{P_{\nu_2}}^\ast\rho_{-u\nu_1,n+1}^{(LT)}$};
\end{tikzpicture}
\end{center}
\caption{Network of representations for the horizontal couplings of $\Phi\Phi^\ast$ and $\Phi^\ast\Phi$}
\label{fig6}
\end{figure}

The case of quiver gauge theories has not been discussed in \cite{Bourgine2021d}. We present here the treatment of the simplest gauge group $G=U(1)\times U(1)$, and the generalization to any A-type quiver is straightforward. D-type quivers might also be addressed along the same lines provided that an equivalent of the reflection states used in \cite{BFMZ} is introduced.

The main new phenomenon in quiver gauge theories is the ``propagation'' of the shifts which requires to shift all the consecutive intertwiners along the dressed NS5-brane's module. For instance, the intertwiner $\Phi$ introduces the shifted representation $\iota_{P_\nu}^\ast$ along the NS5-brane, which does not affect other intertwiners $\Phi$ since they also intertwine between shifted horizontal representations $\iota_P^\ast\rho^{(LT)}$ according to the table of figure \ref{table3}. On the other hand, the intertwiners $\Phi^\ast$ have to be replaced by $\Phi^{P\ast}$ when inserted between shifted horizontal representations with $\iota_P^\ast$. This replacement leads to the cancellation of unwanted factors in the normal ordering \ref{NO_3d}.

The simplest examples of horizontal couplings have been represented on figure \ref{fig6}. In the first configuration (left), the intertwiner $\Phi_2^\ast$ has been replaced by the shifted intertwiner $\Phi_2^{P_{\nu_1}\ast}$, thus bringing the extra factor
\begin{equation}
\prod_{j=1}^{k_2}P_{\nu_1}(\nu_2q^{2j-2})=N_{0,k_2}(\nu_1/\nu_2).
\end{equation} 
This factor can be interpreted as an extra contribution from a fundamental chiral multiplet of mass $\nu_1$, but it cancels with a similar factor coming from the normal-ordering of $\Phi_{k_1}\Phi_{k_2}^\ast$ in \ref{NO_3d}. A similar cancellation is observed in the second configuration (figure \ref{fig6}, right) where $\Phi_2$ is replaced by $\Phi_2^{P_{\nu_1}^\ast}$, thus bringing the extra factor
\begin{equation}
\prod_{j=1}^{k_2}P_{\nu_1}^\ast(\nu_2q^{2j-2})=(-\nu_2)^{k_2}\nu_1^{-k_2}q^{k_2(k_2-1)}(q^2\nu_2/\nu_1;q^2)_{k_2}
\end{equation} 
that cancels the unwanted factor $(q^2\nu_2/\nu_1;q^2)_{k_2}$ in the normal-ordering of $\Phi_{k_1}^\ast\Phi_{k_2}$.

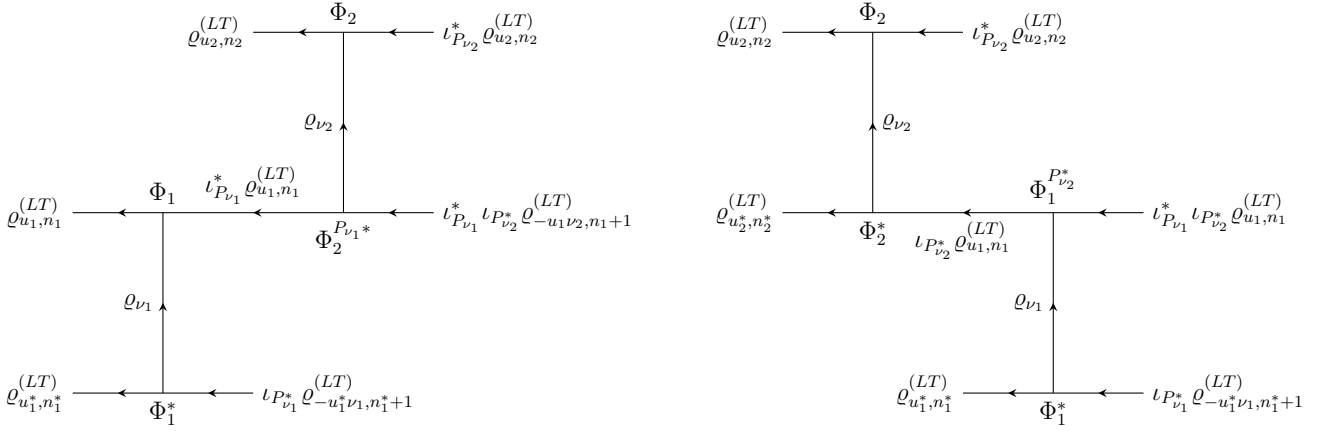
\begin{figure}
\begin{center}
\begin{tikzpicture}[scale=1.2]
\draw[postaction={on each segment={mid arrow=black}}] (4,0) -- (3,0) -- (1,0) -- (0,0);
\draw[postaction={on each segment={mid arrow=black}}] (3,0) -- (3,2);
\draw[postaction={on each segment={mid arrow=black}}] (1,-2) -- (1,0);
\draw[postaction={on each segment={mid arrow=black}}] (4,2) -- (3,2) -- (2,2);
\draw[postaction={on each segment={mid arrow=black}}] (2,-2) -- (1,-2) -- (0,-2);
\node[above,scale=0.8] at (1,0) {$\Phi_1$};
\node[below,scale=0.8] at (3,0) {$\Phi_2^{P_{\nu_1}\ast}$};
\node[below,scale=0.8] at (1,-2) {$\Phi_1^\ast$};
\node[above,scale=0.8] at (3,2) {$\Phi_2$};
\node[left,scale=0.8] at (1,-1) {$\vrho_{\nu_1}$};
\node[left,scale=0.8] at (3,1) {$\vrho_{\nu_2}$};
\node[left,scale=0.8] at (0,0) {$\vrho_{u_1,n_1}^{(LT)}$};
\node[above,scale=0.8] at (2,0) {$\iota_{P_{\nu_1}}^\ast\vrho_{u_1,n_1}^{(LT)}$};
\node[right,scale=0.8] at (4,0) {$\iota_{P_{\nu_1}}^\ast\iota_{P_{\nu_2}^\ast}\vrho_{-u_1\nu_2,n_1+1}^{(LT)}$};
\node[left,scale=0.8] at (0,-2) {$\vrho_{u_1^\ast,n_1^\ast}^{(LT)}$};
\node[right,scale=0.8] at (2,-2) {$\iota_{P_{\nu_1}^\ast}\vrho_{-u_1^\ast\nu_1,n_1^\ast+1}^{(LT)}$};
\node[right,scale=0.8] at (4,2) {$\iota_{P_{\nu_2}}^\ast\vrho_{u_2,n_2}^{(LT)}$};
\node[left,scale=0.8] at (2,2) {$\vrho_{u_2,n_2}^{(LT)}$};
\end{tikzpicture}
\hspace{6mm}
\begin{tikzpicture}[scale=1.2]
% \draw[postaction={on each segment={mid arrow=black}}] (2.7,-0.7) -- (2,0) -- (0,0) -- (-0.7,-0.7);
\draw[postaction={on each segment={mid arrow=black}}] (2,0) -- (1,0) -- (0,0);
\draw[postaction={on each segment={mid arrow=black}}] (1,0) -- (1,2);
\draw[postaction={on each segment={mid arrow=black}}] (2,2) -- (1,2) -- (-1,2) -- (-2,2);
\draw[postaction={on each segment={mid arrow=black}}] (-1,2) -- (-1,4);
\draw[postaction={on each segment={mid arrow=black}}] (0,4) -- (-1,4) -- (-2,4);
\node[below,scale=0.8] at (1,0) {$\Phi_1^\ast$};
\node[above,scale=0.8] at (1,2) {$\Phi_1^{P_{\nu_2}^\ast}$};
\node[above,scale=0.8] at (-1,4) {$\Phi_2$};
\node[below,scale=0.8] at (-1,2) {$\Phi_2^\ast$};
\node[left,scale=0.8] at (0,0) {$\vrho_{u_1^\ast,n_1^\ast}^{(LT)}$};
\node[right,scale=0.8] at (2,0) {$\iota_{P_{\nu_1}^\ast}\vrho_{-u_1^\ast\nu_1,n_1^\ast+1}^{(LT)}$};
\node[right,scale=0.8] at (2,2) {$\iota_{P_{\nu_1}}^\ast\iota_{P_{\nu_2}^\ast}\vrho_{u_1,n_1}^{(LT)}$};
\node[below,scale=0.8] at (0,2) {$\iota_{P_{\nu_2}^\ast}\vrho_{u_1,n_1}^{(LT)}$};
\node[left,scale=0.8] at (-2,2) {$\vrho_{u_2^\ast,n_2^\ast}^{(LT)}$};
\node[right,scale=0.8] at (0,4) {$\iota_{P_{\nu_2}}^\ast\vrho_{u_2,n_2}^{(LT)}$};
\node[left,scale=0.8] at (-2,4) {$\vrho_{u_2,n_2}^{(LT)}$};
\node[left,scale=0.8] at (1,1) {$\vrho_{\nu_1}$};
\node[right,scale=0.8] at (-1,3) {$\vrho_{\nu_2}$};
\end{tikzpicture}
\end{center}
\caption{Networks of representations for a $U(1)\times U(1)$ gauge theory}
\label{fig5}
\end{figure}

These considerations apply to the $U(1)\times U(1)$ quiver for which two equivalent networks of representations have been drawn on figure \ref{fig5}. Starting with the diagram on the right, we compute the operator
\begin{equation}
T[U(1)\times U(1)]=\sum_{k_1,k_2=0}^\infty n_{k_1}n_{k_2}\ \Phi_{k_1}^\ast[u^\ast_1,\nu_1,n^\ast_1]\otimes\Phi_{k_1}[u_1,\nu_1,n_1]\Phi_{k_2}^{P_{\nu_1}\ast}[u^\ast_2,\nu_2,n^\ast_2]\otimes\Phi_{k_2}[u_2,\nu_2,n_2],
\end{equation} 
with $u_2^\ast=u_1$, $n_2^\ast=n_1$. Its vacuum expectation value gives
\begin{align}
\begin{split}
\sum_{k_1,k_2=0}^\infty \prod_{i=1,2}\qf_i^{k_i}\Zv(k_i,\nu_i)\ZCS(k_i,\nu_i,\k_i)\times N_{k_1,k_2}(\nu_1/\nu_2),
% \sum_{k_1,k_2=0}^\infty \prod_{i=1,2}\left(-\dfrac{q^2u_i}{\nu_i u_i^\ast}\right)^{k_i}\ZCS(k_i,\nu_i,n_i^\ast-n_i+1)N_{k_i,k_i}(1)^{-1}\times (-\nu_1/\nu_2)^{k_2}q^{-k_2(k_2-1)} q^{2k_1k_2}N_{k_1,k_2}(\nu_1/\nu_2)\\
\end{split}
\end{align}
with the identification of the gauge theory parameters $\qf_1=q^2 u_1/u_1^\ast$, $\qf_2=q^2 u_2/u_1$, $\k_1=n_1^\ast-n_1$ and $\k_2=n_1-n_2$. From this expression, we read the contribution $N_{k_1,k_2}(\nu_1/\nu_2)$ for the bifundamental fields.

We now turn to the second diagram on figure \ref{fig6} in which $\Phi_1$ has been replaced by $\Phi_1^{P_{\nu_2}^\ast}$. Notice that the outer representations match those of the first diagram up to the shifts $n_1\to n_1-1$ and $u_1\to-u_1/\nu_2$, and the constraints are also different: $u_1=-u_2^\ast\nu_2$ and $n_1=n_2^\ast+1$. The vev of the corresponding operator $T$ reads
\begin{align}
\begin{split}
&\sum_{k_1,k_2=0}^\infty n_{k_1}n_{k_2}\ \la\Phi_{k_1}^\ast[u^\ast_1,\nu_1,n^\ast_2]\ra\la\Phi^\ast_{k_2}[u^\ast_2,\nu_2,n^\ast_2]\Phi_{k_1}^{P_{\nu_2}^\ast}[u_1,\nu_1,n_1]\ra\la\Phi_{k_2}[u_2,\nu_2,n_2]\ra\\
% &=(q^2\nu_2/\nu_1;q^2)_\infty\sum_{k_1,k_2=0}^\infty \prod_{i=1,2}\left(-\dfrac{q^2u_i}{\nu_i u_i^\ast}\right)^{k_i}\ZCS(k_i,\nu_i,n_i^\ast-n_i+1)N_{k_i,k_i}(1)^{-1}\\
% &\qquad\qquad\qquad\times(-\nu_1/\nu_2)^{k_1+k_2}q^{k_1(k_1-1)-k_2(k_2-1)}q^{2k_1k_2}N_{k_1,k_2}(\nu_1/\nu_2)\\
&=(q^2\nu_1/\nu_2;q^2)_\infty\sum_{k_1,k_2=0}^\infty \prod_{i=1,2}\qf_i^{k_i}\Zv(k_i,\nu_i)\ZCS(k_i,\nu_i,\k_i)\times N_{k_1,k_2}(\nu_1/\nu_2)
\end{split}
\end{align}
with $\qf_1=-q^2 u_1/(\nu_2u_1^\ast)$, $\qf_2=-q^2\nu_2 u_2/u_1$, $\k_1=n_1^\ast-n_1+1$ and $\k_2=n_1-1-n_2$. Thus, we recover the same expression for these parameters, up to the replacement $n_1\to n_1-1$ and $u_1\to-u_1/\nu_2$. We also find the same bifundamental contribution as before, up to an extra one-loop factor $(q^2\nu_1/\nu_2;q^2)_\infty$.

\subsection{Higgsing from the algebraic perspective}
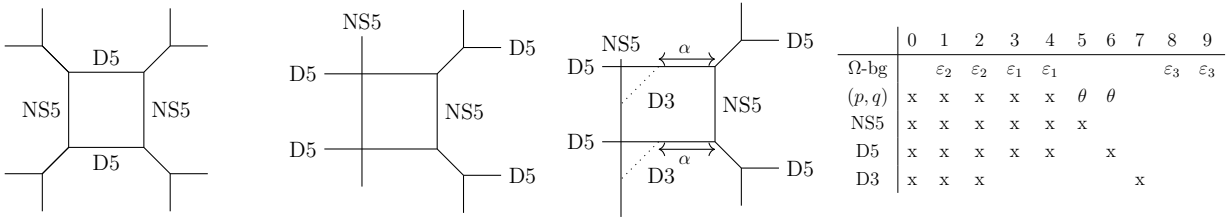
\begin{figure}
% \begin{center}
\begin{subfigure}[c]{0.2\textwidth}
\begin{tikzpicture}[scale=.5]
\draw (-1.7,-.7) -- (-.7,-.7) -- (0,0) -- (2,0) -- (2.7,-.7) -- (3.7,-.7);
\draw (-.7,-1.7) -- (-.7,-.7) -- (0,0) -- (0,2) -- (-.7,2.7) -- (-1.7,2.7);
\draw (-.7,3.7) -- (-.7,2.7) -- (0,2) -- (2,2) -- (2.7,2.7) -- (2.7,3.7);
\draw (2,0) -- (2,2);
\draw (2.7,2.7) -- (3.7,2.7);
\draw (2.7,-.7) -- (2.7,-1.7);
\node[scale=.7,left] at (0,1) {NS5};
\node[scale=.7,below] at (1,0) {D5};
\node[scale=.7,right] at (2,1) {NS5};
\node[scale=.7,above] at (1,2) {D5};
\end{tikzpicture}
\end{subfigure}
% \hspace{5mm}
\begin{subfigure}{0.2\textwidth}
\begin{tikzpicture}[scale=.5]
\draw (-1,0) -- (2,0) -- (2.7,-.7) -- (3.7,-.7);
\draw (-1,2) -- (2,2) -- (2.7,2.7) -- (2.7,3.7);
\draw (2,0) -- (2,2);
\draw (0,-1) -- (0,3);
\draw (2.7,2.7) -- (3.7,2.7);
\draw (2.7,-.7) -- (2.7,-1.7);
\node[scale=.7,above] at (0,3) {NS5};
\node[scale=.7,left] at (-1,2) {D5};
\node[scale=.7,left] at (-1,0) {D5};
\node[scale=.7,right] at (2,1) {NS5};
\node[scale=.7,right] at (3.7,2.7) {D5};
\node[scale=.7,right] at (3.7,-.7) {D5};
\end{tikzpicture}
\end{subfigure}
% \hspace{5mm}
\begin{subfigure}{0.2\textwidth}
\begin{tikzpicture}[scale=.5]
\draw (-1,0) -- (2,0) -- (2.7,-.7) -- (3.7,-.7);
\draw (-1,2) -- (2,2) -- (2.7,2.7) -- (2.7,3.7);
\draw (2,0) -- (2,2);
\draw (-.5,-2) -- (-.5,2.2);
\draw (2.7,2.7) -- (3.7,2.7);
\draw (2.7,-.7) -- (2.7,-1.7);
\draw[dotted] (0.5,0) -- (-.5,-1);
\draw[dotted] (0.5,2) -- (-.5,1);
\draw[<->] (0.5,2.2) -- (2,2.2);
\draw[<->] (0.5,-.2) -- (2,-.2);
\node[scale=.6,above] at (1.2,2.2) {$\a$};
\node[scale=.6,below] at (1.2,-.2) {$\a$};
\node[scale=.7,above] at (-.5,2.2) {NS5};
\node[scale=.7,left] at (-1,2) {D5};
\node[scale=.7,left] at (-1,0) {D5};
\node[scale=.7,below right] at (0,-.5) {D3};
\node[scale=.7,below right] at (0,1.5) {D3};
\node[scale=.7,right] at (2,1) {NS5};
\node[scale=.7,right] at (3.7,2.7) {D5};
\node[scale=.7,right] at (3.7,-.7) {D5};
\end{tikzpicture}
\end{subfigure}
% \hspace{5mm}
\begin{subfigure}{0.2\textwidth}
\scalebox{.6}{\begin{tabular}[b]{c|cccccccccc}
& 0 & 1 & 2 & 3 & 4 & 5 & 6 & 7 & 8 & 9\\
\hline
$\Omega$-bg & & $\e_2$ & $\e_2$ & $\e_1$ & $\e_1$ & & & & $\e_3$ & $\e_3$\\
$(p,q)$ & x & x & x & x & x & $\th$ & $\th$ & & & \\
NS5 & x & x & x & x & x & x & & & & \\
D5 & x & x & x & x & x & & x & & & \\ 
D3 & x & x & x & & & & & x & &\\
\end{tabular}}
\end{subfigure}
\caption{Higgsing procedure on the $(p,q)$-brane web describing the 5d $\CN=1$ U(2) gauge theory with $N^f=4$ flavors.}
\label{figHiggs}
\end{figure}

The string theory description of the Higgsing procedure is well known, it is represented on the figure \ref{figHiggs} in the case of a $U(2)$ gauge theory with four hypermultiplets. More generally, one can start from a $U(N)$ gauge theory with exponentiated Coulomb branch vevs $v_l$ and at least $N$ (anti)fundamental hypermultiplets of exponentiated mass $\mu_l$. The Coulomb branch vevs indicate the position of the internal D5-branes in the $(p,q)$-brane web, and the hypermultiplets masses the position of semi-infinite D5-branes. Adjusting the masses to the critical value $\mu_l\sim v_l$ corresponds classically to the reconstruction of a broken NS5-brane orthogonal to the D5-branes. Then, the NS5-brane can be pulled out in the transverse direction, creating a number of D3-branes attached to both the D5 and NS5 branes. The number of D3-branes created depends on the precise relation between masses and Coulomb branch vevs. We study here the case $q_3\mu_l=q_1v_l$ for which only a single D3-brane is created for each internal D5. Thus, the theory in the worldvolume of D3-branes is a 3d $\CN=4$ gauge theory with $U(N)$ gauge group broken to $\CN=2^\ast$ by the omega-background deformation which gives a mass $m_\Phi=-\e_1$ to the $\CN=2$ chiral multiplet of the $\CN=4$ vector multiplet. Then, it only remains to send the parameter $\e_1\to\infty$ to recover the 3d $\CN=2$ gauge theories considered in this paper.

On the Coulomb branch of the 5d theory, the vev of the scalar in the vector multiplet break the gauge group to $U(1)^N$. The separation of the internal D5-branes induces a separations of D3-branes corresponding to the mass of $N$ chiral multiplets breaking $U(N)\to U(1)^N$ on the Higgs branch. Instantons of the 5d theories correspond to a D1-D5 system, and these D1-branes introduce vortices on the worldvolume of D3 branes \cite{Hanany1996}. As a result, the 5d instanton partition functions and the 3d vortex partition functions are related in the limit $q_1\to\infty$, and we would like to interpret this relation from the point of view of quantum groups, exploiting the relation between the representations of the algebras $\CU^\bmu$ and $\CE^\bmu$ observed in the section \ref{sec_SQA_sl2} of this paper. For this purpose, we focus on a 5d $\CN=1$ $U(N)$ gauge theory with $N$ hypermultiplets either in the fundamental or antifundamental representation of the gauge group. Extra hypermultiplets can be introduced by further shift of the quantum group representations as explained in the previous sections, but they are spectators in this procedure.

\begin{figure}
\begin{center}
\begin{tikzpicture}[scale=1.4]
\draw[postaction={on each segment={mid arrow=black}}] (9,0) -- (8,0) -- (6,0) -- (2,0) -- (1,0);
\draw[postaction={on each segment={mid arrow=black}}] (2,0) -- (2,1);
\draw[postaction={on each segment={mid arrow=black}}] (6,0) -- (6,1);
\draw[postaction={on each segment={mid arrow=black}}] (8,0) -- (8,1);
\draw[postaction={on each segment={mid arrow=black}}] (9,1) -- (8,1) -- (6,1) -- (2,1) -- (1,1);
\node[scale=.8,left] at (2,.5) {$\iota_{P_N}\rho_{v_N}^{(0,1)}$};
\node[scale=.8,right] at (8,.5) {$\iota_{P_1}\rho_{v_1}^{(0,1)}$};
\node[scale=.8,right] at (6,.5) {$\iota_{P_2}\rho_{v_2}^{(0,1)}$};
\node[scale=.8,right] at (9,0) {$\iota^\ast_{P_1P_2\cdots P_N}\rho_{u_1^\ast}^{(1,n^\ast+N)}$};
\node[scale=.7,below] at (7,-.25) {$\iota^\ast_{P_2\cdots P_N}\rho_{u_2^\ast}^{(1,n^\ast+N-1)}$};
\node[scale=.8,left] at (1,0) {$\rho_{u_N^\ast}^{(1,n^\ast)}$};
\node[scale=.8,right] at (9,1) {$\rho_{u_1}^{(1,n)}$};
\node[scale=.8,above] at (7,1) {$\iota_{P_1}\rho_{u_2}^{(1,n+1)}$};
\node[scale=.8,left] at (1,1) {$\iota_{P_1P_2\cdots P_N}\rho_{u_N}^{(1,n+N)}$};
\node[scale=.8,below] at (2,0) {$\Phi_N^\ast$};
\node[scale=.8,above] at (2,1) {$\Phi_N^{P_1\cdots P_{N-1}}$};
\node[scale=.8,below] at (8,0) {$\Phi_1^{P_2P_3\cdots P_N\ast}$};
\node[scale=.8,above] at (8,1) {$\Phi_1$};
\node[scale=.8,below] at (6,0) {$\Phi_2^{P_3\cdots P_N\ast}$};
\node[scale=.8,above] at (6,1) {$\Phi_2^{P_1}$};
\node at (4,.5) {$\cdots$};
\end{tikzpicture}
\end{center}
\caption{Representation network for a 5d $\CN=1$ $U(N)$ gauge theory with $N$ hypermultiplets}
\label{fig_UN_N_5d}
\end{figure}
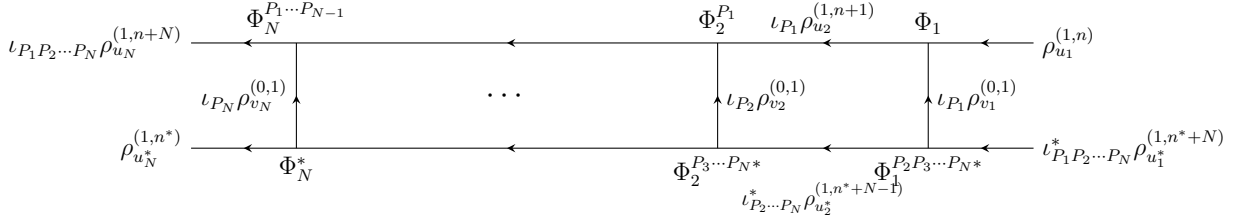

We associate to the gauge theory the network of representations drawn on figure \ref{fig_UN_N_5d} where each internal D5-brane carries a shifted representation $\iota_{P_l}\rho_{v_l}^{(0,1)}$ with $P_l(z)=1-q_1q_2z/\mu_l$ (resp. $P_l(z)=1-\bar{\mu}_l/z$) for fundamental (resp. antifundamental) hypermultiplets. Adjusting the mass to $q_3\mu_l=q_1v_l$ (or $\bar{\mu}_l=q_1v_l$), the vertical representations $\iota_{P_l}\rho_{v_l}^{(0,1)}$ can be restricted to the submodule $\CL$ spanned by the states $\dket{k}$ corresponding to Young diagrams which consist of a single column of $k$ boxes. We note that other algebraic realizations of this gauge theory leads to similar restrictions that manifest themselves as the vanishing of summands in the summations \ref{def_Phi} defining the shifted AFS intertwiners when $\l$ has more than one column. Taking further the limit $\e_1\to -\infty$, which corresponds to $q_1\to\infty$, the vertical action on $\CL$ reduces to the prefundamental representation of $\CU^{(0,-1)}$. These are indeed the representations attached to D3-branes in the network of representations of figure \ref{fig_UN_3d}, upon the identification of $\nu_l=v_l$ and $q^2=q_2$.

It has been argued in the section \ref{sec_limit_horiz} that the twisted Fock representation $\vrho^{(LT)}$ appears in the limit of the horizontal vertex representations $\rho_u^{(1,n)}$ as $q_1\to\infty$ with $q_2$ fixed. This is indeed what we observe when comparing the network of representations \ref{fig_UN_N_5d} and \ref{fig_UN_3d} for the 5d and 3d gauge theories, provided that we discard the shifts by the functors $\iota_P$ and $\iota_P^\ast$. Trying to understand directly how the various shifted representations arise appears difficult. Naively, the shifts by $\iota_{P_l}$ or $\iota_{P_l}^\ast$ should vanish in the limit $q_1\to\infty$ as $P_l(\chi_k)\sim 1$ (resp. $P_l(\chi_k)\sim q_1v_l/\chi_k$ in the antifundamental case) brings only a trivial factor. On the other hand, the extra shifts by $\iota_{P_\nu}^\ast$ or $\iota_{P_\nu^\ast}$ are difficult to justify in this picture, apart from the fact that they are required by the insertion of the 3d intertwiners.

In fact, it is easier to justify the limit of the intertwiners rather than horizontal representations. Indeed, when $\l$ is restricted to a column of $k$ boxes, the vertical component of the 5d intertwiner $\Phi^\ast$ read
\begin{align}
\begin{split}
\Phi_k^{(5d)\ast}[u,v,n]&=t_k^\ast[u,v,n]e^{\sum_{l>0}\frac{v^l}{l}q_2^{-l}\left[q_1^{-l}+(1-q_1^{-l})(1-q_2^{kl})\right]J_{-l}}e^{-\sum_{l>}\frac{v^{-l}}{l(1-q_2^{-l})}\left[\frac1{1-q_1^{-l}}-(1-q_2^{-kl})\right]J_l},\\
&\sim_{LII} (-)^k\Phi_k^{(3d)\ast}[u,v,n],
\end{split}
\end{align}
thus justifying the replacement by its 3d equivalent in the network \ref{fig_UN_3d}. This is in agreement with the limit of the normal-ordering relation\footnote{While the limit LII of $\CG(q_3^{-1}z)$ (with $z$ finite) is well defined and produces the factor $(zq^2;q^2)_\infty^{-1}$, the limit of $\CG(z)$ requires a regularization that we simply denote $\CG_{LII}(z)$ since we are not interested in the perturbative contributions.} 
\begin{align}
\begin{split}
\Phi_{k_1}^{(5d)\ast}[u_1,v_1,n_1]\Phi_{k_2}^{(5d)\ast}[u_2,v_2,n_2]&::\dfrac{\CG(v_2/v_1)}{\CN_{k_2,k_1}(q_3v_2/v_1)}=\dfrac{\CG(v_2/v_1)}{(q_1^{-1}q_2^{k_2-k_1}v_2/v_1;q_2)_{k_1}(q_2^{-k_1}v_2/v_1;q_2)_{k_2}}\\
&\sim_{LII}\CG_{LII}(v_2/v_1)N_{k_1,k_2}(v_1/v_2)^{-1}\times(-v_1/v_2)^{k_2} q^{2k_1k_2-k_2(k_2-1)}
\end{split}
\end{align}
that reproduces the normal-ordering \ref{NO_3d} provided that we disregard the $k$-independent (one-loop) factor. Although the direct limit of the 5d intertwiner $\Phi$ is ill-defined, the limit of its normal-ordering relations justifies its replacement by the 3d equivalent,
\begin{align}
\begin{split}
\Phi_{k_1}^{(5d)}[u_1,v_1,n_1]\Phi_{k_2}^{(5d)}[u_2,v_2,n_2]&::\dfrac{\CG(q_3^{-1}v_2/v_1)}{\CN_{k_2,k_1}(v_2/v_1)}=\dfrac{\CG(q_3^{-1}v_2/v_1)}{(q_1q_2^{1-k_1}v_2/v_1;q_2)_{k_1}N_{k_2,k_1}(v_2/v_1)}\\
&\sim_{LII}q_1^{-k_1}(q^2v_2/v_1;q^2)_\infty^{-1} (-v_1/v_2)^{k_1}q^{k_1(k_1-1)}N_{k_2,k_1}(v_2/v_1)^{-1}
\end{split}
\end{align}
up to spurious factors that can be absorbed in the modification of the factors $\t_k[u,v,n]$. Upon this replacement of intertwiners, we find indeed the network of representations of figure \ref{fig_UN_3d}.

\section{Discussion}
In this paper, we have presented several new results on the representation theory of the shifted quantum groups $\CE^\bmu$ and $\CU^\bmu$, and discussed their application to the algebraic engineering of supersymmetric gauge theories. On the mathematical side, we have extended the notion of shifted representations which led to the definition of pit subrepresentations for highest $\ell$-weight representations. In particular, we have obtained a series of finite dimensional highest weight representations labeled by a Young diagrams for $\CE^{\bmu}$ with the dominant shift $\bmu\geq2$. We also studied the relations between representations of $\CE^{\bmu}$ in the limit LII (i.e. $q_1\to\infty$ and $q_2$ fixed) and $\CU^\bmu$. We have introduced a notion of \textit{formal limit} to solve a problem of order of limits in the Drinfeld presentation, which specifies that the limit of $\ell$-weights should be taken before the modes expansion. As a result, we found several relations between various representations, they have been summarized in the diagram of figure \ref{fig_limit}.

New representations called left/right twisted Fock representations were also introduced for the infinitely shifted quantum groups $\CE^{(-\infty,0)}$ and $\CU^{(-\infty,0)}$ (or $\CE^{(0,-\infty)}$ and $\CU^{(0,-\infty)}$). These representations bear some similarities with the Frenkel-Jing vertex representations since the Drinfeld currents are also expressed as vertex operators acting on a 2d boson Fock space. The twisted representations of $\CU^{(-\infty,0)}$ can be understood as the limit LII of the horizontal Fock representation of $\CE$. They define a natural action on Hall-Littlewood polynomials.

From the physics perspective, we extended the algebraic engineering formalism to include (anti)fundamental hypermultiplets (5d) or antifundamental chiral multiplets using shifted representations. We reproduced successfully the known partition functions and qq-characters of matter gauge theories. We also extended the construction of 3d $\CN=2$ theories presented in \cite{Bourgine2021d} to A-type quivers, and gave an algebraic description of Higgsing using the limiting procedure on the representations studied earlier.

Our limit LII is somewhat related to the Nekrasov-Shatashvili (NS) limit $(q_1,q_2)\to(q,1)$. Identifying the quantum group parameters as $q_\a=e^{-R\e_a}$ with $\a=1,2$, the limit LII is obtained by sending $\e_1\to-\infty$ keeping $R$ and $\e_2$ fixed, while the NS limit sends $\e_2\to0$ with $R$, $\e_1$ fixed. Taking first the limit of small radius $R\to0$, we obtain a 4D $\CN=2$ theory on $\mC_{\e_1}\times\mC_{\e_2}$ for which the BPS quantities only depend the ratio $\b=-\e_1/\e_2$. In this case, the two limits appear equivalent as they both send $\b\to\infty$. However, a rescaling of the masses and Coulomb branch parameters is hidden in this formulation, which effectively leads to different limits. Alternatively, this difference can be understood in the light of the AGT correspondence. While the CFT seems to depend only on $\b$, e.g. through the central charge $c$, the product $\hbar\sim\sqrt{\e_1\e_2}$ defines a scale which is usually set to one. However, this scale will diverge in the limit LII, and tend to zero in the NS limit. As a result, vertex operators $V_P(z)$ of charge $\a=Q/2+P/\hbar$ that would be considered light in one limit, will appear as heavy in the other.

In our description of Higgsing, we focused on the simplest case in which only a single D3-brane is created. By considering a more general position for the pit of the shifted vertical representation, it is possible to access more general brane configurations which have been recently studied in \cite{Kimura2021}. In this way, we could obtain an algebraic construction of two 3d $\CN=2$ gauge theories, defined respectively on $\mC_{\e_1}\times S^1$ and $\mC_{\e_2}\times S^1$, and coupled on the common $S^1$. It would be interesting to explore further the connection between quantum group representations and W-algebras in this context.

Miki's automorphism \cite{Miki2007} induces a map between vertical and horizontal Fock representations of $\CE$ that plays an important role in the algebraic description of type IIB S-duality \cite{Bourgine2018a,Fukuda2019}. In the limit LII, we expect to find a similar map between the prefundamental action $\bigoplus_{\l_2\geq0}\bigoplus_{\mu\diagup\mu_1\leq\l_2}\vrho_{vq^{2\l_2}}$ of $\CU^{(0,-1)}$ on $\CF_0$ and the left twisted Fock representation of $\CU^{(-\infty,0)}$,
% \tikzexternaldisable
\begin{equation}
\begin{tikzcd}
\rho_v^{(0,1)}\arrow{r}{\CS} \arrow[swap]{d}{LII} & \rho_u^{(1,0)} \arrow{d}{LII} \\
 \bigoplus_{\superp{\l_2\geq0}{\mu\diagup\mu_1\leq\l_2}}\vrho_{vq^{2\l_2}}\arrow{r}{???} & \vrho^{(LT)}_{u,0}
\end{tikzcd}
\end{equation} 
% \tikzexternalenable
Indeed, we have observed that the subalgebra $\CU_-^{(0,-1)}\subset\CU^{(0,-1)}$ spanned by the negative modes $X_k^\pm$, $\Psi_{k}^-$ with $k<0$ can be mapped to a subalgebra of $\CU^{(-\infty,0)}$ generated by the modes $X_k^-$ with $k=0,\pm1$ of $\CU^{(-\infty,0)}$. This map might be useful in the context of 3d mirror symmetry \cite{Intriligator1996,deBoer1996,Nieri2018,Cheng2021} and bispectral duality of quantum integrable systems \cite{Gaiotto2013}. We plan to come back to this question in a future work.

Finally, it seems important to explore further the possible relations between our algebraic construction, the finite AGT correspondence \cite{Braverman2010}, and quantum integrable systems \cite{Gadde2013}. For instance, it is natural to ask if the proof of the q-deformed AGT correspondence obtained in \cite{Fukuda2019} can be extended to the finite version. In this case, an equivalent of the Kimura-Pestun quiver W-algebras \cite{Kimura2015} could be obtained from a tensor product of left twisted Fock representations that would play the role of the horizontal Fock representations of $\CE$ in \cite{Feigin2009a}. This problem is currently under investigation.

\section*{Acknowledgements}
The author would like to thank Sasha Garbali, David Hernandez, Taro Kimura and Gufang Zhao for discussions.

\appendix

\section{Example of a finite dimensional shifted representation}\label{AppA}
In this appendix, we provide some explicit formulas for the action of the Drinfeld currents defining the algebra $\CE^{(0,3)}$ on the five dimensional module $\CM_\mu$ indexed by the Young diagram $\mu=21$. The representation is obtained as $\iota_{P_\mu}\rho_v^{(0,1)}$ with 
\begin{equation}
P_\mu(z)=(1-vq_1^2/z)(1-vq_2^2/z)(1-vq_1q_2/z)
\end{equation} 
In the basis $e_0=\dket{\vac}$, $e_1=\dket{1}$, $e_2=\dket{2}$, $e_3=\dket{1^2}$ and $e_4=\dket{21}$, the currents $x^\pm(z)$ have the form of $5\times 5$ matrices
\begin{equation}
x^+(z)=
\begin{pmatrix}
0 & \a_1 & 0 & 0 & 0\\
0 & 0 & \a_2 & \a_3 & 0\\
0 & 0 & 0 & 0 & \a_4\\
0 & 0 & 0 & 0 & \a_5\\
0 & 0 & 0 & 0 & 0
\end{pmatrix},\quad
x^-(z)=\begin{pmatrix}
0 & 0 & 0 & 0 & 0\\
\b_1 & 0 & 0 & 0 & 0\\
0 & \b_2 & 0 & 0 & 0\\
0 & \b_3 & 0 & 0 & 0\\
0 & 0 & \b_4 & \b_5 & 0
\end{pmatrix}.
\end{equation} 
The expression of the coefficients shortens if we introduce the notation $p_{i,j}=1-q_1^iq_2^j$,
\begin{align}
\begin{split}
\a_1&=\d(v/z)q_3^{-1/2}p_{-1,-1}p_{2,0}p_{0,2}p_{1,1},\quad \b_1=\d(v/z)p_{1,1}p_{-1,0}p_{0,-1}\\
\a_2&=\d(vq_2/z)q_3^{-1/2}p_{-1,-1}p_{2,-1}p_{0,1}p_{1,0}^2p_{1,-1}^{-1},\quad \b_2=\d(vq_2/z)p_{1,1}p_{-1,0}p_{0,-2}\\
\a_3&=\d(vq_1/z)q_3^{-1/2}p_{-1,-1}p_{-1,2}p_{0,1}^2p_{1,0}p_{-1,1}^{-1},\quad \b_3=\d(vq_1/z)p_{1,1}p_{-2,0}p_{0,-1}\\
\a_4&=\d(vq_1/z)q_3^{-1/2}p_{-1,-1}p_{0,1}p_{1,0}p_{0,2},\quad \b_4=\d(vq_1/z)p_{1,1}p_{-1,0}p_{0,-1}p_{-2,1}p_{-1,1}^{-1}\\
\a_5&=\d(vq_2/z)q_3^{-1/2}p_{-1,-1}p_{0,1}p_{1,0}p_{2,0}\quad \b_5=\d(vq_2/z)p_{1,1}p_{-1,0}p_{0,-1}p_{1,-2}p_{1,11}^{-1}.
\end{split}
\end{align}

The action of the Cartan currents on the state $e_i$ is diagonal, with the eigenvalue $q_3^{-1/2}[\tilde{\psi}_i(z)]_\pm$ obtained by expanding the following rational functions,
\begin{align}
\begin{split}
&\tpsi_0(z)=\dfrac{(z-vq_1^2)(z-vq_2^2)(z-vq_1q_2)(z-vq_1^{-1}q_2^{-1})}{z^3(z-v)},\\
&\tpsi_1(z)=\dfrac{(z-vq_1^2)(z-vq_2^2)(z-vq_1q_2)^2(z-vq_1^{-1})(z-vq_2^{-1})}{z^3(z-v)(z-vq_1)(z-vq_2)},\\
&\tpsi_2(z)=\dfrac{(z-vq_1^2)(z-vq_1q_2)(z-vq_1q_2^2)(z-vq_2^{-1})(z-vq_1^{-1}q_2)}{z^3(z-vq_1)(z-vq_2)},\\
&\tpsi_3(z)=\dfrac{(z-vq_2^2)(z-vq_1q_2)(z-vq_1^2q_2)(z-vq_1^{-1})(z-vq_1q_2^{-1})}{z^3(z-vq_1)(z-vq_2)},\\
&\tpsi_4(z)=\dfrac{(z-v)(z-vq_1^2q_2)(z-vq_1q_2^2)(z-vq_1q_2^{-1})(z-vq_1^{-1}q_2)}{z^3(z-vq_1)(z-vq_2)}.
\end{split}
\end{align}

\section{Equivariant characters}\label{AppCOHA}
\begin{figure}
\begin{center}
\begin{tikzpicture}[scale=.5]
\node[scale=.7] at (0,5) {$N$};
\node[scale=.7] at (0,0) {$K$};
\draw (-1,4) rectangle (1,6);
\draw (0,0) circle (1);
\node[scale=.7,right] at (.1,2.5) {$I$};
\draw[->,>=stealth] (.1,4) -- (.1,2.5);
\draw (.1,2.5) -- (.1,1);
\node[scale=.7,left] at (-.1,2.5) {$J$};
\draw[->,>=stealth] (-.1,1) -- (-.1,2.5);
\draw (-.1,2.5) -- (-.1,4);
\node[scale=.7,above right] at (1.8,0) {$B_1$};
\draw[->,>=stealth] (1,-.7) arc (-100:100:.7);
\node[scale=.7,above left] at (-2,0) {$B_2$};
\draw[->,>=stealth] (-1,-.7) arc (-80:-280:.7);
\end{tikzpicture}
\end{center}
\caption{Quiver variety describing the moduli space of instantons}
\label{Quiver_Inst}
\end{figure}
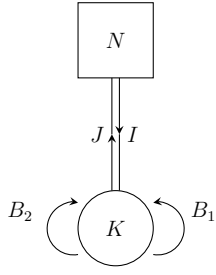

In this appendix, we analyze the 5d hypermultiplets and the Higgsing procedure from the point of view of the equivariant characters of instantons and vortex moduli spaces. The moduli space of $U(n)$ framed instantons of charge $k$ on the $\O$-deformed non-commutative $\mR^4$ follows from the ADHM construction \cite{Atiyah1978,Nakajima1994,Nekrasov1998}.\footnote{In this appendix, we denote $n$ the rank of the gauge group to avoid a conflict of notation with the vector space $N=\mC^n$.} It is a quiver variety with a single node $K=\mC^k$ framed by $N=\mC^n$, 
\begin{equation}
\CM^I_{k,n}=\{B_1,B_2\in\End(K), I\in\Hom(N,K), J\in\Hom(K,N)\diagup \mu_\mR=\z\mathbbm{1}_k,\quad \mu_\mC=0\}\diagup U(k),
\end{equation} 
with the moment maps
\begin{equation}
2\mu_\mR=II^\dagger-J^\dagger J+[B_1,B_1^\dagger]+[B_2,B_2^\dagger],\quad \mu_\mC=[B_1,B_2]+IJ.
\end{equation}
The corresponding quiver is represented on figure \ref{Quiver_Inst}. When $\z>0$, the real moment map can be replaced by the stability condition $\mC[B_1,B_2]IN=K$ provided that we quotient by the action of $GL(k)$ instead of $U(k)$. The torus $\mathfrak{t}=U(1)_{\e_1}\times U(1)_{\e_2}\times U(1)^n$ acts on this moduli space, and the fixed point are labeled by $n$-tuples of partitions $\bl$ with $|\bl|=k$ boxes. At the fixed point $\bl$, we have the following decomposition of $N$ and $K$ into one-dimensional subspaces
\begin{equation}
N=\bigoplus_{l=1}^n N_l,\qquad K=\bigoplus_{(l,i,j)\in\bl}B_1^{i-1}B_2^{j-1}IN_l.
\end{equation} 
As a result, the characters for the action of the torus $\mathfrak{t}$ on the spaces $N$ and $K$ are given by
\begin{equation}
\CN=\sum_{l=1}^nv_l,\qquad \CK_\l=\sum_{\sAbox\in\bl}\chi_\sAbox.
\end{equation}
The characters $v_l$ and $\chi_\sAbox$ are resp. identified with the exponentiated Coulomb branches and the box contents $\chi_{(l,i,j)}=v_lq_1^{i-1}q_2^{j-1}$.

The Nekrasov instanton partition function is a sum over the fixed points $\bl$, the summands are obtained from the character $\chi_{\bl,\bl}$ of the torus action on the tangent space $T_\bl\CM^I_{k,N}$ at the fixed point, with\footnote{When the two $n$-tuples of Young diagrams $\bl$ and $\bl'$ differ, the quantity $\chi_{\bl,\bl'}$ is in fact the equivariant character of a fiber defined using the Hecke correspondence introduced in \cite{Schiffmann2012} as a generalization of the moduli space of nested Hilbert scheme to $n>1$.}
\begin{equation}
\chi_{\bl,\bl'}=-(1-q_1)(1-q_2)\CK_\bl\CK_{\bl'}^\vee+\CN\CK_{\bl'}^\vee+q_1q_2\CK_\bl\CN^\vee,
\end{equation} 
with the linear operation $v_l^\vee=v_l^{-1}$, $q_\a^\vee=q_\a^{-1}$. The vector contribution is obtained by applying the index functor to this expression,
\begin{equation}\label{def_mI}
\mathbb{I}\left[\sum_{i\in I_+}e^{w_i}-\sum_{i\in I_-}e^{w_i}\right]=\dfrac{\prod_{i\in I_+} 1-e^{w_i}}{\prod_{i\in I_-} 1-e^{w_i}}\implies \mI[\chi_{\bl,\bl}]=\CZv(\bl,\bv)^{-1}, % \mI[\pm \sum_Ie^{w_I}]=\prod_I(1-e^{w_I})^{\pm1}.
\end{equation} 
and the Nekrasov factor corresponds to the case $n=1$: $\CN_{\l,\l'}(v/v')=\mI[\chi_{\l,\l'}]$.

\subsection{Massive hypermultiplets}
In the presence of $N^f$ fundamental and $N^{\bar{f}}$ antifundamental hypermultiplets, the equivariant character is replaced by,
\begin{equation}\label{def_chi}
\chi_{\bl,\bl'}=-(1-q_1)(1-q_2)\CK_\bl\CK_{\bl'}^\vee+(\CN-\bar \CM)\CK_{\bl'}^\vee+q_1q_2\CK_\bl(\CN^\vee-\CM^\vee)
\end{equation} 
with $\CM=\sum_{a=1}^{N^f}\mu_a$ and $\bCM=\sum_{a=1}^{N^{\bar{f}}}\bar\mu_a$ \cite{NPS} (neglecting the one-loop terms). The summands of the instanton partition function are again obtained as
\begin{equation}
\mI[\chi_{\bl,\bl}]^{-1}=\CZv(\bl,\bv)\CZf(\bl,\bv,\bmu)\CZaf(\bl,\bv,\bar{\bmu}).
\end{equation} 

The extra terms $-\bar \CM\CK_{\bl'}^\vee-q_1q_2\CK_\bl\CM^\vee$ in the character also induce an extra factor in the $\CY$-observables. These observables were introduced by Nekrasov in \cite{Nekrasov_BPS1}, they can be obtained as a variation of the equivariant character (see \cite{Bourgine:2019phm}),\footnote{The presence of the extra factor $q_3^{-1}$ in the argument of the function $\CY^\ast_\bl(z)$ is a convention related to the property $\chi_{\bl,\bl'}^\vee=q_3\chi_{\bl',\bl}$ in the absence of hypermultiplets (symplectic quiver) which leads to the relation $\CY_\bl^\ast(z)=\prod_l(-z/v_l)\ \CY_\bl(z)$.}
\begin{equation}
\tilde{\CY}_\bl(\chi_\sAbox)=\mI[\chi_{\bl,\bl'+\sAbox}-\chi_{\bl,\bl'}],\quad \tilde{\CY}_{\bl'}^\ast(q_3^{-1}\chi_\sAbox)=\mI[\chi_{\bl+\sAbox,\bl'}-\chi_{\bl,\bl'}].
\end{equation} 
From the bilinear expression of the character \ref{def_chi}, we deduce that the r.h.s. of these formulas is indeed independent of the second Young diagram. The $\CY$-observables are the unique rational functions satisfying these constraints for the infinite set of points $\chi_{(l,i,j)}=v_lq_1^{i-1}q_2^{j-1}$. The extra factors in the $\CY$-observables coming from the deformation of the character in the presence of hypermultiplets are simply the inverse of the polynomials defined in \ref{def_Pmu} and involved in the shift of representations,
\begin{align}
\begin{split}
&\tilde{\CY}_\bl(z)=P^{\bar{f}}_{\bar{\bmu}}(z)^{-1}\CY_\bl(z),\quad \tilde{\CY}_\bl^\ast(z)=\prod_{l=1}^n(-z/v_l)\ P^f_\bmu(q_3z)^{-1}\CY_\bl(z),\\
\text{with}\quad &\CY_\bl(z)=\prod_l\CY_{\l^{(l)}}(z),\quad \CY_\l(z)=(1-v/z)\prod_{\sAbox\in \l}\dfrac{(1-q_1\chi_{\sAbox}/z)(1-q_2\chi_{\sAbox}/z)}{(1-\chi_{\sAbox}/z)(1-q_1q_2\chi_{\sAbox}/z)}.
\end{split}
\end{align}
Using the shell formula, it is seen that the $\CY$-observable $\CY_\l(z)$ coincides with the function \ref{def_CYY} defining the vertical Fock representation. Thus, restricting ourselves to $U(1)$ instantons, the algebra $\CE$ acts on the fixed point $\dket{\l}$ of the torus action by the vertical Fock representation given in \ref{def_vert_rep}. This action is the (K-theoretic) (double) Cohomological Hall algebra of the instanton moduli space. Matrix elements of the currents are expressed in terms of the $\CY$-observables, and so they are expected to be modified in the presence of hypermultiplets as follows,
\begin{align}
\begin{split}
&\tilde{x}^+(z)\dket{\l}=\k_+\sum_{\sAbox\in A(\l)}\delta(z/\chi_\sAbox)\res_{w=\chi_\sAbox}\dfrac1{w\tilde{\CY}_{\lambda}(w)}\dket{\l+\Abox},\\
&\tilde{x}^-(z)\dket{\l}=-q_3^{-1/2}\k_-v_lz^{-1}\sum_{\sAbox\in R(\l)}\delta(z/\chi_\sAbox)\res_{w=\chi_\sAbox}w^{-1}\tilde{\CY}^\ast_{\l}(q_3^{-1}w)\dket{\l-\Abox},\\
&\tilde{\psi}^\pm(z)\dket{\l}=-q_3^{-1/2}v_lz^{-1}\left[\dfrac{\tilde{\CY}^\ast_\l(q_3^{-1}z)}{\tilde{\CY}_\l(z)}\right]_\pm\dket{\l}.
\end{split}
\end{align}
Comparing with \ref{def_vert_rep}, we deduce that antifundamental hypermultiplets lead to the shift $\iota_P\rho_v^{(0,1)}$ of the vertical representation with the polynomial $P=P_{\bar{\bmu}}^{\bar{f}}$. Naively, fundamental hypermultiplets would introduce a similar shift $\iota_{P^{-1}}^\ast\rho_v^{(0,1)}$ with $P=P_\bmu^f$. However, in this case $P(z)^{-1}$ is no longer a finite Laurent series, it introduce extra poles in the action of the Cartan current that cannot be accounted for by the commutator $[x^+(z),x^-(w)]$. Thus, it is not clear how to define properly the vertical action there. This is a limitation of our simplified approach, and this issue could be resolved using the more involved geometric techniques employed e.g. in \cite{Schiffmann2012,Rapcak2020}.

\subsection{Higgsing and the vortex moduli space}
\begin{figure}
\begin{center}
\begin{tikzpicture}[scale=.5]
\node[scale=.7] at (0,5) {$N$};
\node[scale=.7] at (0,0) {$k$};
%\draw (0,5) circle (1);
\draw (-1,4) rectangle (1,6);
\draw (0,0) circle (1);
\node[scale=.7,right] at (0,2.5) {$I$};
\draw[->,>=stealth] (0,4) -- (0,2.5);
\draw (0,2.5) -- (0,1);
% \node[scale=.7,left] at (-.1,2.5) {$I^\dagger$};
% \draw[->,>=stealth] (-.1,1) -- (-.1,2.5);
% \draw (-.1,2.5) -- (-.1,4);
\node[scale=.7,above right] at (1.8,0) {$B$};
\draw[->,>=stealth] (1,-.7) arc (-100:100:.7);
\end{tikzpicture}
\end{center}
\caption{Quiver variety describing the vortex moduli space}
\label{Quiver_Vort}
\end{figure}
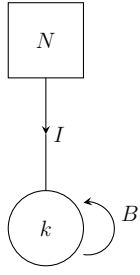

The relation between the instanton moduli space and the vortex moduli space has been investigated in \cite{Hanany2003} from the point of view of the $\CN=(2,2)$ quantum mechanics on the worldvolume of D1-branes. By weakly gauging a $U(1)$ subgroup of the total symmetry group, it is possible to give a large mass to some of the fields and decouple them. It corresponds to imposing $B_1=J=0$ in the ADHM construction, and we find the vortex moduli space given by the quiver variety of figure \ref{Quiver_Vort},
\begin{equation}
\CM_{k,n}^V=\{B_2\in\End(K), I\in\Hom(N,K)\diagup\mu_\mR=\z\mathbbm{1}_k\}/U(k),
\end{equation} 
with the real moment map $2\mu_\mR=II^\dagger+[B_2,B_2^\dagger]$. The torus $U(1)_{\e_2}\times U(1)^n$ acts on $\CM_{k,n}^V$ and the fixed point are labeled by $n$-tuple of positive integers $\bk=(k_1,\cdots,k_n)$ such that $|\bk|=\sum_lk_l=k$. At each fixed point, we have the following decomposition of $K$ and $N$ and the corresponding characters $\CK_{\bk}$ and $\CN$,
\begin{equation}
N=\bigoplus_{l=1}^n N_l,\quad K=\bigoplus_{l=1}^n\bigoplus_{j=1}^{k_l}B_2^{j-1}IN_l\implies \CN=\sum_{l=1}^nv_l,\quad \CK_\bk=\sum_{l=1}^nv_l\sum_{j=1}^{k_l}q_2^{j-1}.
\end{equation} 
The equivariant character of the tangent space of the vortex moduli space has been computed in \cite{Yoshida2011}, it corresponds to 
$\chi^\text{3d}_{\bk,\bk}$ with
\begin{equation}
\chi^\text{3d}_{\bk,\bk'}=-(1-q^{2})\CK_{\bk}\CK_{\bk'}^\vee+\CN\CK_{\bk'}^\vee,\quad \text{with}\quad q^2=q_2.
\end{equation}
Taking the index functor, this character produces the 3d version of the Nekrasov factor defined in \ref{def_Nkk}, $N_{k,k'}(v/v')=\mI[\chi_{k,k'}]$.

The vortex moduli space studied here corresponds to a non-commutative Hilbert scheme \cite{Franzen2013}, but unfortunately we were unable to find a derivation of the K-theoretic COHA for this variety. A possible approach could be to start from the K-theoretic COHA derivation of the finite dimensional module of quantum affine algebras performed in \cite{Ginzburg1995,Vasserot1998,Nakajima1999}. Specializing to the Kirillov-Reshetikhin representation of dimension $N$ for the quantum affine $\sl(2)$ algebra, it might be possible to consider the limit $N\to\infty$ of this geometric construction. Here, again, we will be more modest and simply consider the variation of the equivariant character. In this way, we define 3d analog of Nekrasov's $\CY$-observables,
\begin{equation}
Y_\bk(v_lq^{2k_l})=\mI[\chi^\text{3d}_{\bk,\bk'+\d_l}-\chi^\text{3d}_{\bk,\bk'}],\quad Y_{\bk'}^\ast(v_lq^{2k_l+2})=\mI[\chi^\text{3d}_{\bk+\d_l,\bk'}-\chi^\text{3d}_{\bk,\bk'}],
\end{equation} 
where $\d_l=(0,\cdots,1,0,\cdots)\in\mZ^n$ is a vector with a $1$ in the $l$th row and zero elsewhere. It gives indeed the functions defined in \cite{Bourgine2021d},
\begin{align}
\begin{split}\label{def_Y_obs}
Y_{\bk}(z)=\prod_{l=1}^n\left(1-\dfrac{v_lq^{2k_l}}{z}\right),\quad Y_{\bk}^\ast(q^2z)=\prod_{l=1}^nq^{2k_l}\dfrac{z-v_lq^{-2}}{z-v_lq^{2k_l-2}}.
\end{split}
\end{align}
In the case of $U(1)$ vortices, the prefundamental representation acts on a module spanned by states $\dket{k}$ labeled by the fixed point of the equivariant action on the vortex moduli space, its action can be expressed using $\CY$-observables as
\begin{align}\label{COHA_vertex}
\begin{split}
&\tX^+(z)\dket{k}=\d(v q^{2k}/z)\res_{z=v q^{2k}}\dfrac1{z Y_k(z)}\dket{k+1},\\%=\d(v q^{2k}/z)\dket{k+1},\\
&\tX^-(z)\dket{k}=\d(v q^{2k-2}/z)\res_{z=v q^{2k-2}}z^{-1}Y_k^\ast(q^2z)\dket{k-1},\\%=-\d(v q^{2k-2}/z)(1-q^{2k})\dket{k-1},\\
&\tPsi^\pm(z)\dket{k}=\left[\dfrac{Y_k^\ast(q^2z)}{Y_k(z)}\right]_\pm\dket{k}.%= q^{2k} \left[\dfrac{z(z-v q^{-2})}{(z-v q^{2k})(z-vq^{2k-2})}\right]_\pm\dket{k}.
\end{split}
\end{align}
It reproduces the prefundamental action \ref{prefund} up to a change in the normalization of the states and a rescaling of the currents by harmless factors.

\paragraph{Higgsing} Recall that the 3d gauge theories are obtained from a two steps procedure that consists in 1) adjusting the mass of $n$ hypermultiplets to a critical value and 2) taking the limit $q_1\to\infty$ to decouple the adjoint chiral multiplet. Once the mass takes its critical value, the set of fixed points is restricted to $n$-tuples of single column Young diagrams identified with the $n$ positive integers $\bk=(k_1,\cdots,k_n)$, and the characters $\CK_\bl$ should be replaced by $\CK_\bk$. Furthermore, adjusting the mass of $n$ antifundamental hypermultiplets to $\bar{\mu}_l=q_1v_l$ corresponds to setting $\bar\CM=q_1\CN$ and $\CM=0$ in the expression \ref{def_chi} of the equivariant character, and we find
\begin{equation}
\chi_{\bk,\bk'}^\text{5d}=(1-q_1)\chi_{\bk,\bk'}^\text{3d}+q_1q_2\CN^\vee\CK_{\bk}.
\end{equation} 
In the limit $q_1\to\infty$, only the finite terms in $\chi_{\bk,\bk'}^\text{5d}$ will contribute once the index functor is applied. Examining the term of power $q_1^0$, we find indeed $\left[\chi_{\bk,\bk'}^\text{5d}\right]_0=\chi_{\bk,\bk'}^\text{3d}$. We would arrive at the same conclusion if we had adjusted instead the mass of fundamental hypermultiplet, i.e. setting $\CM=q_1^2q_2\CN$ and $\bar\CM=0$ in \ref{def_chi}. This relation between the equivariant characters justifies our approach of the limit of highest $\ell$-weight representation, and we can check that the $\CY$-observables have indeed the proper limit (the factor $q^{-2k}$ can be absorbed in a renormalization of the states)
\begin{align}
\begin{split}
&\CY_k(z)\tox_{LII} q^{-2k}Y_k(z),\quad \CY_k(q_3^{-1}z)\tox_{LII} q^{-2k}Y_k^\ast(q^2z)\implies \psi_k(z)\tox_{LII} q^{2k}\dfrac{z(z-vq^{-2})}{(z-vq^{2k})(z-vq^{2k-2})}.
\end{split}
\end{align}

% 
% It is worth mentioning that the limit of the $\CY$-functions obtained in \ref{CY_k} agree with the $Y$-observables defined from the equivariant character of the vortex moduli space in appendix \ref{AppCOHA},

% We observe that $(\chi_{\bk,\bk'}^\text{5d})^\vee$ tends to $(\chi_{\bk,\bk'}^\text{3d})^\vee$ in the limit $q_1\to\infty$. Choosing $\chi$ or $\chi^\vee$ is merely a convention as it gives the same Nekrasov factor.\footnote{We have also argued that the limit $q_1\to\infty$ is equivalent to $q_1\to0$ from the point of view of the algebras. In the latter, we find $\chi_{\bk,\bk'}^\text{5d}\to \chi_{\bk,\bk'}^\text{3d}$.}

\section*{Declarations}
\paragraph{Fundings}  This research was partly supported by the Basic Science Research Program through the National Research Foundation of Korea (NRF) funded by the Ministry of Education through the Center for Quantum Spacetime (CQUeST) of Sogang University (NRF-2020R1A6A1A03047877). The author also gratefully acknowledges support from the Australian Research Council Centre of Excellence for Mathematical and Statistical Frontiers (ACEMS).

\paragraph{Competing interests} The author have no competing interests to declare that are relevant to the content of this article.

\paragraph{Data}  Data sharing not applicable to this article as no datasets were generated or analysed during the current study.

% \bibliographystyle{../utphys}
% \bibliography{DIM_crystal}

\end{document}